\documentclass[12pt]{article}
\usepackage{latexsym}
\usepackage{amssymb}
\catcode`\@=11
\newcount\hour
\newcount\minute
\newtoks\amorpm \hour=\time\divide\hour by 60\minute
=\time{\multiply\hour by 60 \global\advance\minute by-\hour}
\edef\standardtime{{\ifnum\hour<12 \global\amorpm={am}%
        \else\global\amorpm={pm}\advance\hour by-12 \fi
        \ifnum\hour=0 \hour=12 \fi
        \number\hour:\ifnum\minute<10
        0\fi\number\minute\the\amorpm}}
\edef\militarytime{\number\hour:\ifnum\minute<10
0\fi\number\minute}
\def\draftlabel#1{{\@bsphack\if@filesw {\let\thepage\relax
   \xdef\@gtempa{\write\@auxout{\string
      \newlabel{#1}{{\@currentlabel}{\thepage}}}}}\@gtempa
   \if@nobreak \ifvmode\nobreak\fi\fi\fi\@esphack}
        \gdef\@eqnlabel{#1}}
\def\@eqnlabel{}
\def\@vacuum{}
\def\marginnote#1{}
\def\draftmarginnote#1{\marginpar{\raggedright\scriptsize\tt#1}}
\def\draft{
        \pagestyle{plain}
        \overfullrule=2pt
        \oddsidemargin -.1truein
        \def\@oddhead{\sl \phantom{\today\quad\militarytime} \hfil
        \smash{\Large\sl DRAFT} \hfil \today\quad\militarytime}
        \let\@evenhead\@oddhead
        \let\label=\draftlabel
        \let\marginnote=\draftmarginnote
        \def\ps@empty{\let\@mkboth\@gobbletwo
        \def\@oddfoot{\hfil \smash{\Large\sl DRAFT} \hfil}
        \let\@evenfoot\@oddhead}
        \def\@eqnnum{(\theequation)\rlap{\kern\marginparsep\tt\@eqnlabel}%
        \global\let\@eqnlabel\@vacuum}  }

\renewcommand{\theequation}{\thesection.\arabic{equation}}
\renewcommand{\thefootnote}{\fnsymbol{footnote}}
\newcommand{\newsection}{    
\setcounter{equation}{0}\section}
\def\appendix#1{\addtocounter{section}{1}\setcounter{equation}{0}
\renewcommand{\thesection}{\Alph{section}}
\section*{Appendix \thesection\protect\indent \parbox[t]{11.15cm}{#1}}
\addcontentsline{toc}{section}{Appendix \thesection\ \ \ #1}}

\def \lc {{light-cone}}

\def \bi{\bibitem}
\def \la {\label}

\def \b {\beta}
\def \Om {\Omega}

\def \d {\partial}

\def\be{\begin{equation}}
\def\ee{\end{equation}}

\catcode`\@=11

\def\bea{\begin{eqnarray}}
\def\eea{\end{eqnarray}}
\def\beann{\begin{eqnarray*}}
\def\eeann{\end{eqnarray*}}
\def\beq{\begin{equation}}
\def\eeq{\end{equation}}
\def\ba{\begin{array}}
\def\ea{\end{array}}
\def\ben{\begin{enumerate}}
\def\een{\end{enumerate}}

 \def \la {\label}
 \def\be{\begin{equation}}
\def\ee{\end{equation}}

\def \la {\label}

\font\mybb=msbm10 at 11pt

\def\bb#1{\hbox{\mybb#1}}

\def\bR {\bb{R}}

\def\bC {\bb{C}}

 \def\ep {\epsilon}

\def \ee {\epsilon}

\def \g {\gamma}
\def \bi{\bibitem}
\def\a{\alpha }

\def \ep {\epsilon}

\def \d {\delta}

\def \g {\gamma}

\def \b {\beta}

\def\lc{\lrcorner}

\def \xone {G_{-+p}}
\def \xtwo {G_{-1p}}
\def \xthree {G_{- \bar{1} p}}
\def \xfour {G_{+1p}}
\def \xfive {G_{+ \bar{1} p}}
\def \xsix {G_{1 \bar{1} p}}
\def \xseven {G_{pq}{}^q}
\def \xeight {G_{- {\bar{q}}_1 {\bar{q}}_2} \epsilon^{{\bar{q}}_1 {\bar{q}}_2}{}_p}
\def \xnine {G_{+ {\bar{q}}_1 {\bar{q}}_2} \epsilon^{{\bar{q}}_1 {\bar{q}}_2}{}_p}
\def \xten {G_{1 {\bar{q}}_1 {\bar{q}}_2} \epsilon^{{\bar{q}}_1 {\bar{q}}_2}{}_p}
\def \xeleven {G_{\bar{1} {\bar{q}}_1 {\bar{q}}_2} \epsilon^{{\bar{q}}_1 {\bar{q}}_2}{}_p}

\def \yone {F_{-+1 \bar{1} p}}
\def \ytwo {F_{-+pq}{}^q}
\def \ythree {F_{-1pq}{}^q}
\def \yfour {F_{- \bar{1} pq}{}^q}
\def \yfive {F_{+1pq}{}^q}
\def \ysix {F_{+ \bar{1} pq}{}^q}
\def \yseven {F_{-+1 {\bar{q}}_1 {\bar{q}}_2} \epsilon^{{\bar{q}}_1 {\bar{q}}_2}{}_p}
\def \yeight {F_{-+\bar{1} {\bar{q}}_1 {\bar{q}}_2} \epsilon^{{\bar{q}}_1 {\bar{q}}_2}{}_p}
\def \ynine {F_{-1 \bar{1} {\bar{q}}_1 {\bar{q}}_2} \epsilon^{{\bar{q}}_1 {\bar{q}}_2}{}_p}
\def \yten {F_{+ 1 \bar{1} {\bar{q}}_1 {\bar{q}}_2} \epsilon^{{\bar{q}}_1 {\bar{q}}_2}{}_p}

\def \aone {\Omega_{p,q}{}^q}
\def \atwo {\Omega_{p, 1 \bar{1}}}
\def \athree {\Omega_{p,-+}}
\def \afour {\Omega_{p,1-}}
\def \afive {\Omega_{p,\bar{1}-}}
\def \asix {\Omega_{p,+\bar{1}}}
\def \aseven {\Omega_{p,+1}}

\def \bone {\Omega_{1, {\bar{q}}_1 {\bar{q}}_2}  \epsilon^{{\bar{q}}_1 {\bar{q}}_2}{}_p}
\def \btwo {\Omega_{1,p-}}
\def \bthree {\Omega_{1,p+}}
\def \bfour {\Omega_{1,p \bar{1}}}
\def \bfive {\Omega_{1,1p}}

\def \cone {\Omega_{{\bar{q}}_1 , {\bar{q}}_2 \bar{1}}  \epsilon^{{\bar{q}}_1 {\bar{q}}_2}{}_p}
\def \ctwo {\Omega_{\bar{q},}{}^{\bar{q}}{}_p}
\def \cthree {\Omega_{{\bar{q}}_1 , {\bar{q}}_2 -}  \epsilon^{{\bar{q}}_1 {\bar{q}}_2}{}_p}
\def \cfour {\Omega_{{\bar{q}}_1 , {\bar{q}}_2 +}  \epsilon^{{\bar{q}}_1 {\bar{q}}_2}{}_p}
\def \cfive {\Omega_{{\bar{q}}_1 , {\bar{q}}_2 1}  \epsilon^{{\bar{q}}_1 {\bar{q}}_2}{}_p}

\def \done {\Omega_{\bar{1} , {\bar{q}}_1 {\bar{q}}_2}  \epsilon^{{\bar{q}}_1 {\bar{q}}_2}{}_p}
\def \dtwo {\Omega_{\bar{1},1p}}
\def \dthree {\Omega_{\bar{1},p-}}
\def \dfour {\Omega_{\bar{1},p+}}
\def \dfive {\Omega_{\bar{1},p\bar{1}}}

\def \eone {\Omega_{-, {\bar{q}}_1 {\bar{q}}_2}  \epsilon^{{\bar{q}}_1 {\bar{q}}_2}{}_p}
\def \etwo {\Omega_{-,1p}}
\def \ethree {\Omega_{-,p-}}
\def \efour {\Omega_{-,+p}}
\def \efive {\Omega_{-,p\bar{1}}}

\def \tone {\Omega_{+, {\bar{q}}_1 {\bar{q}}_2}  \epsilon^{{\bar{q}}_1 {\bar{q}}_2}{}_p}
\def \ttwo {\Omega_{+,1p}}
\def \tthree {\Omega_{+,p-}}
\def \tfour {\Omega_{+,+p}}
\def \tfive {\Omega_{+,\bar{1}p}}

\def \df {D_p f}
\def \dg {D_p g}

\def \xonexx {G_{-+\bar{p}}}
\def \xtwoxx {G_{-1{\bar{p}}}}
\def \xthreexx {G_{- \bar{1} {\bar{p}}}}
\def \xfourxx {G_{+1{\bar{p}}}}
\def \xfivexx {G_{+ \bar{1} {\bar{p}}}}
\def \xsixxx {G_{1 \bar{1} {\bar{p}}}}
\def \xsevenxx {G_{{\bar{p}}q}{}^q}
\def \xeightxx {G_{- q_1 q_2} \epsilon^{q_1 q_2}{}_{\bar{p}}}
\def \xninexx {G_{+ q_1 q_2} \epsilon^{q_1 q_2}{}_{\bar{p}}}
\def \xtenxx {G_{1 q_1 q_2} \epsilon^{q_1 q_2}{}_{\bar{p}}}
\def \xelevenxx {G_{\bar{1} q_1 q_2} \epsilon^{q_1 q_2}{}_{\bar{p}}}

\def \yonexx {F_{-+1 \bar{1} {\bar{p}}}}
\def \ytwoxx {F_{-+{\bar{p}}q}{}^q}
\def \ythreexx {F_{-1{\bar{p}}q}{}^q}
\def \yfourxx {F_{- \bar{1} {\bar{p}}q}{}^q}
\def \yfivexx {F_{+1{\bar{p}}q}{}^q}
\def \ysixxx {F_{+ \bar{1} {\bar{p}}q}{}^q}
\def \ysevenxx {F_{-+1 q_1 q_2} \epsilon^{q_1 q_2}{}_{\bar{p}}}
\def \yeightxx {F_{-+\bar{1} q_1 q_2} \epsilon^{q_1 q_2}{}_{\bar{p}}}
\def \yninexx {F_{-1 \bar{1} q_1 q_2} \epsilon^{q_1 q_2}{}_{\bar{p}}}
\def \ytenxx {F_{+ 1 \bar{1} q_1 q_2} \epsilon^{q_1 q_2}{}_{\bar{p}}}

\def \conexx {\Omega_{q_1 , q_2 \bar{1}}  \epsilon^{q_1 q_2}{}_{\bar{p}}}
\def \ctwoxx {\Omega_{q,}{}^q{}_{\bar{p}}}
\def \cthreexx {\Omega_{q_1 , q_2 -}  \epsilon^{q_1 q_2}{}_{\bar{p}}}
\def \cfourxx {\Omega_{q_1 , q_2 +}  \epsilon^{q_1 q_2}{}_{\bar{p}}}
\def \cfivexx {\Omega_{q_1 , q_2 1}  \epsilon^{q_1 q_2}{}_{\bar{p}}}

\def \donexx {\Omega_{1 , q_1 q_2}  \epsilon^{q_1 q_2}{}_{\bar{p}}}
\def \dtwoxx {\Omega_{1,\bar{1} {\bar{p}}}}
\def \dthreexx {\Omega_{1,{\bar{p}}-}}
\def \dfourxx {\Omega_{1,{\bar{p}}+}}
\def \dfivexx {\Omega_{1,{\bar{p}}1}}

\def \aonexx {\Omega_{{\bar{p}},q}{}^q}
\def \atwoxx {\Omega_{{\bar{p}}, 1 \bar{1}}}
\def \athreexx {\Omega_{{\bar{p}},-+}}
\def \afourxx {\Omega_{{\bar{p}},\bar{1}-}}
\def \afivexx {\Omega_{{\bar{p}},1-}}
\def \asixxx {\Omega_{{\bar{p}},+1}}
\def \asevenxx {\Omega_{{\bar{p}},+\bar{1}}}

\def \bonexx {\Omega_{\bar{1}, q_1 q_2}  \epsilon^{q_1 q_2}{}_{\bar{p}}}
\def \btwoxx {\Omega_{\bar{1},{\bar{p}}-}}
\def \bthreexx {\Omega_{\bar{1},{\bar{p}}+}}
\def \bfourxx {\Omega_{\bar{1},{\bar{p}} 1}}
\def \bfivexx {\Omega_{\bar{1},\bar{1}{\bar{p}}}}

\def \eonexx {\Omega_{-, q_1 q_2}  \epsilon^{q_1 q_2}{}_{\bar{p}}}
\def \etwoxx {\Omega_{-,\bar{1} {\bar{p}}}}
\def \ethreexx {\Omega_{-,{\bar{p}}-}}
\def \efourxx {\Omega_{-,+{\bar{p}}}}
\def \efivexx {\Omega_{-,{\bar{p}}1}}

\def \tonexx {\Omega_{+, q_1 q_2}  \epsilon^{q_1 q_2}{}_{\bar{p}}}
\def \ttwoxx {\Omega_{+,\bar{1}{\bar{p}}}}
\def \tthreexx {\Omega_{+,{\bar{p}}-}}
\def \tfourxx {\Omega_{+,+{\bar{p}}}}
\def \tfivexx {\Omega_{+,1{\bar{p}}}}

\def \dfxx {D_{\bar{p}} f}
\def \dgxx {D_{\bar{p}} g}

\def \vone {G_{-+1}}
\def \vtwo {G_{-+ \bar{1}}}
\def \vthree {G_{1p}{}^p}
\def \vfour {G_{\bar{1} p}{}^p}
\def \vfive {G_{- 1 \bar{1}}}
\def \vsix {G_{+ 1 \bar{1}}}
\def \vseven {G_{234}}
\def \veight {G_{\bar{2} \bar{3} \bar{4}}}
\def \vnine {G_{+p}{}^p}
\def \vten {G_{-p}{}^p}

\def \zone {F_{-+1p}{}^p}
\def \zonec {F_{-+\bar{1}p}{}^p}
\def \ztwo {F_{-1\bar{1}p}{}^p}
\def \zthree {F_{+1 \bar{1}p}{}^p}
\def \zfour {F_{-1234}}
\def \zfourc {F_{-\bar{1} \bar{2} \bar{3} \bar{4}}}
\def \zfive {F_{-+234}}
\def \zfivec {F_{-+ \bar{2} \bar{3} \bar{4}}}
\def \zsix {F_{+1 \bar{2} \bar{3} \bar{4}}}
\def \zsixc {F_{+ \bar{1} 234}}

\def \hone {\Omega_{p,}{}^p{}_{\bar{1}}}
\def \honec {\Omega_{{\bar{p}},}{}^{{\bar{p}}}{}_1}
\def \htwo {\Omega_{p,}{}^p{}_+}
\def \htwoc  {\Omega_{{\bar{p}},}{}^{{\bar{p}}}{}_+}
\def \hthree {\Omega_{p}{}^p{}_-}
\def \hthreec {\Omega_{{\bar{p}},}{}^{{\bar{p}}}{}_-}
\def \hfour {\Omega_{p_1 , p_2 p_3} \epsilon^{p_1 p_2 p_3}}
\def \hfourc {\Omega_{{\bar{p}}_1 , {\bar{p}}_2 {\bar{p}}_3}
\epsilon^{{\bar{p}}_1 {\bar{p}}_2 {\bar{p}}_3}}
\def \hfive {\Omega_{p,}{}^p{}_{1}}
\def \hfivec {\Omega_{{\bar{p}},}{}^{{\bar{p}}}{}_{\bar{1}}}

\def \mone {\Omega_{1,p}{}^p}
\def \monec {\Omega_{\bar{1},p}{}^p}
\def \mtwo {\Omega_{1,1 \bar{1}}}
\def \mtwoc {\Omega_{\bar{1},1 \bar{1}}}
\def \mthree {\Omega_{1,1-}}
\def \mthreec {\Omega_{\bar{1}, \bar{1}-}}
\def \mfour {\Omega_{1,-+}}
\def \mfourc {\Omega_{\bar{1},-+}}
\def \mfive {\Omega_{1, \bar{1}-}}
\def \mfivec {\Omega_{\bar{1},1-}}
\def \msix {\Omega_{1,1+}}
\def \msixc {\Omega_{\bar{1}, \bar{1}+}}
\def \mseven {\Omega_{1, \bar{1}+}}
\def \msevenc {\Omega_{\bar{1},1+}}

\def \sone {\Omega_{-,p}{}^p}
\def \stwo {\Omega_{-,1 \bar{1}}}
\def \sthree {\Omega_{-,-+}}
\def \sfour {\Omega_{-,1-}}
\def \sfourc {\Omega_{-,\bar{1}-}}
\def \sfive {\Omega_{-,+1}}
\def \sfivec {\Omega_{-,+\bar{1}}}

\def \qone {\Omega_{+,p}{}^p}
\def \qtwo {\Omega_{+,1\bar{1}}}
\def \qthree {\Omega_{+,-+}}
\def \qfour {\Omega_{+,1-}}
\def \qfourc {\Omega_{+,\bar{1}-}}
\def \qfive {\Omega_{+,+1}}
\def \qfivec {\Omega_{+,+\bar{1}}}

\def \DFone {D_+ f}
\def \DFtwo {D_- f}
\def \DFthree {D_1 f}
\def \DFfour {D_{\bar{1}} f}

\def \DGone {D_+ g}
\def \DGtwo {D_- g}
\def \DGthree {D_1 g}
\def \DGfour {D_{\bar{1}} g}

\def \Pone {P_+}
\def \Ptwo {P_-}
\def \Pthree {P_1}
\def \Pfour {P_{\bar{1}}}

\def\be{\begin{equation}}
\def\ee{\end{equation}}

\def \bi {\bibitem}
\def \la{\label}

\begin{document}
\date{November 2002}
\begin{titlepage}
\begin{center}

\vspace{2.0cm}
{\Large \bf The $G_2$  spinorial  geometry of supersymmetric IIB backgrounds}\\[.2cm]

\vspace{1.5cm}
 {\large  U. Gran$^1$, J. Gutowski$^2$ and  G. Papadopoulos$^1$}

 \vspace{0.5cm}
${}^1$ Department of Mathematics\\
King's College London\\
Strand\\
London WC2R 2LS, UK\\
\vspace{0.5cm}
${}^2$ DAMTP, Centre for Mathematical Sciences\\
University of Cambridge\\
Wilberforce Road, Cambridge, CB3 0WA, UK
\end{center}

\vskip 1.5 cm
\begin{abstract}

We solve the Killing spinor equations of supersymmetric IIB backgrounds which admit
one supersymmetry and the Killing spinor has stability subgroup $G_2$ in $Spin(9,1)\times U(1)$.
We find that such  backgrounds admit a time-like Killing vector field  and the geometric
structure of the spacetime reduces from $Spin(9,1)\times U(1)$ to $G_2$. We determine the type of $G_2$
structure that the spacetime admits by computing the covariant derivatives of the
spacetime forms associated with the Killing spinor bilinears.

We also solve the Killing spinor equations of backgrounds with two supersymmetries and
$Spin(7)\ltimes \bR^8$-invariant spinors,  and four supersymmetries with $SU(4)\ltimes \bR^8$- and
with $G_2$-invariant spinors.
We show that the Killing spinor equations factorize in two sets, one involving the geometry
and the five-form flux, and the other the three-form flux and the scalars. In the $Spin(7)\ltimes \bR^8$
and $SU(4)\ltimes \bR^8$ cases, the spacetime admits a parallel null vector field and so the spacetime
metric can be locally described  in terms of Penrose coordinates adapted to the associated rotation
free, null, geodesic congruence. The transverse space of the congruence is a $Spin(7)$ and a $SU(4)$ holonomy
manifold, respectively. In the $G_2$ case, all the fluxes vanish and the spacetime  is the product of a
three-dimensional Minkowski space with a
holonomy $G_2$ manifold.
\end{abstract}
\end{titlepage}
\newpage
\setcounter{page}{1}
\renewcommand{\thefootnote}{\arabic{footnote}}
\setcounter{footnote}{0}

\setcounter{section}{0}
\setcounter{subsection}{0}
\newsection{Introduction}

In the last few years there has been some progress towards a systematic understanding of
supersymmetric solutions of IIB supergravity.
The maximal supersymmetric solutions of IIB supergravity have
been classified in \cite{jfgpa}. It has been found that they are
locally isometric to Minkowski space, $AdS_5\times S^5$ \cite{schwarz} and the maximal
supersymmetric plane wave \cite{georgea}.
More recently, it has been found that there are three types of supersymmetric
IIB backgrounds with one supersymmetry characterized by the stability subgroup
of the Killing spinor in $Spin(9,1)\times U(1)$ \cite{gju}. These stability subgroups are
$Spin(7)\ltimes\bR^8$,
$SU(4)\ltimes \bR^8$ and $G_2$. The Killing spinor equations of IIB supergravity
have been solved for the $Spin(7)\ltimes\bR^8$- and
$SU(4)\ltimes \bR^8$-invariant Killing spinors  \cite{gju} using a method proposed in \cite{uggp}.
Some progress has also been made  using the G-structures method  \cite{jones}.

The main aim of this paper is to solve the Killing spinor equations
of IIB supergravity for supersymmetric backgrounds that admit a $G_2$-invariant Killing spinor.
This result together with those in \cite{gju} complete the task of solving
the Killing spinor equations for  IIB backgrounds with one supersymmetry.
To achieve this, we follow the steps suggested in \cite{uggp}. First, we  describe the spinors in terms
of forms, and we  use the
 gauge symmetry of the  Killing spinor equations
 to bring the Killing spinor into a canonical or normal form.
Then the Killing spinor equations are  solved by utilizing a basis in the space of spinors
given in \cite{gju}, see also appendix A.
It has been shown in \cite{gju} that the $Spin(9,1)\times U(1)$ gauge group of the
IIB Killing spinor equations can be used to bring the $G_2$-invariant Killing spinor
in the canonical form
\bea
\epsilon &=& f(1+e_{1234})- i g (e_{15}+e_{2345})~,
\la{caform}
\eea
 where $f, g$ are real functions of the spacetime\footnote{This spinor can be simplified somewhat
 using  the spinor transformation $e^{b \Gamma_{05}}$ which for an appropriate choice of $b$ one can set $f^2=g^2$. However
 this choice  does not lead to a substantial simplification
 of the computation.}.

 The bosonic fields of IIB supergravity are the spacetime metric, two real scalars,
 the axion $\sigma$
and the dilaton $\phi$, which are combined into a complex one-form field strength $P$,
two three-form field strengths $G_1$ and $G_2$ which are combined to a complex three-form
field strength $G$, and a self-dual
five-form field strength $F$.
 Substituting the spinor (\ref{caform}) into the Killing spinor equations of IIB supergravity
 and expanding the result in the spinor basis given in appendix A, we derive
 a linear system for the fluxes, geometry and spacetime derivatives of the functions
 $f,g$ which determine the Killing spinor. It turns out that it is convenient to solve first this
 linear system for the complex  field strengths $G$ and $P$.
 The remaining equations are conditions on the self-dual five-form field
 strength $F$, the geometry as represented by the Levi-Civita connection $\Omega$ and
 the spacetime derivatives of $f,g$. These equations are also solved to express
 $F$ in terms of the geometry. After this is achieved, there are some remaining equations
 which contain only the Levi-Civita connection $\Omega$ and the spacetime derivatives of $f,g$.
 These equations are interpreted as the restrictions on the geometry of the spacetime imposed
 by the Killing spinor equations and describe the geometry of the spacetime of a background
 that admits a $G_2$-invariant Killing spinor.

We  investigate the geometric  conditions that arise from the Killing spinor equations
of IIB backgrounds with  Killing spinor (\ref{caform}).  In particular, we compute
the spacetime form bi-linears of the Killing spinor $\epsilon$ and $\tilde\epsilon=C*\epsilon$, where $C$ is the
charge conjugation matrix.
We find that the spacetime admits a timelike one-form, and  two spacelike one-forms appropriately twisted with $L^2$, where
$L$ is the line bundle\footnote{If the spacetime admits a $Spin(9,1)_c$ structure, $L$ may not be well-defined but $L^2$
is.} of the axion-dilaton system of IIB supergravity.   The spacetime also admits a $G_2$-invariant three-form $\phi$ and so a
 (geometric) $G_2$-structure.
We identify the type of the $G_2$-structure of the spacetime  by computing the
covariant derivative of the spacetime form  spinor bilinears mentioned above and we organize
the geometric conditions in terms of $G_2$ representations. It turns out that the geometry is characterized by five
conditions, three of which are conditions on the three one-forms. The other two are restrictions on the $G_2$-invariant
 three-form $\phi$. One of the conditions implies that the timelike one-form  is associated to
 a {\it  Killing vector field} $X$. In addition one of the conditions on $\phi$ expresses the singlet of $d\phi$ in the decomposition
 of $\Lambda^4(\bR^7)=\Lambda_{\bf1}^4(\bR^7)\oplus \Lambda_{\bf7}^4(\bR^7)\oplus \Lambda^4_{\bf 27}(\bR^7)$ under $G_2$
 in terms of derivatives of the one-forms and so resembles the condition of a {\it weak or nearly parallel} $G_2$-structure. However,
 this $G_2$-structure arises from a reduction of $Spin(9,1)\times U(1)$ to $G_2$ rather than the `standard'
 reduction of $Spin(7)$ to $G_2$
 which has been investigated in \cite{grayb}, see also \cite{ivanov, salamonb}. The other condition on $\phi$
 implies that the seven-dimensional part of ${\cal L}_X\phi$, in the decomposition
$\Lambda^3(\bR^7)=\Lambda_{\bf1}^3(\bR^7)\oplus \Lambda_{\bf7}^3(\bR^7)\oplus \Lambda^3_{\bf 27}(\bR^7)$ under $G_2$,
  vanishes. The remaining conditions impose
 restrictions on the commutators of the vector fields associated with the one-forms.

Next we investigate backgrounds with $N$ Killing spinors, $N>1$. In particular, we focus on supersymmetric backgrounds
for which the Killing spinors are invariant under some subgroup $H\subset Spin(9,1)$. In this
case the Killing spinors can be written as
\bea
\epsilon_r=\sum_{i=1}^k f_{ri} \eta_i~,
\eea
where $f_{ri}$ are functions on the spacetime and $\eta_i$ is a basis in the space of $H$-invariant spinors.
We analyze the Killing spinor equations
and we demonstrate that they simplify whenever the background admits the maximal number of $H$-invariant Killing spinors, $k=N$.
This is because the matrix of functions $f$ is invertible and this can be used to arrange such that only the
derivative term of the supercovariant derivative depends on $f$. We term these backgrounds as maximally supersymmetric
$H$-backgrounds.
In IIB supergravity, we find that there is an additional
simplifying feature. For any $H\subset Spin(9,1)$, there is a basis in the space of $H$-invariant spinors whose elements  come
in complex conjugate pairs. This can be used
to show that the Killing spinor equations of IIB supergravity \,{\it factorize}\, for all maximally supersymmetric $H$-backgrounds.
In particular, the terms in the Killing spinor equations involving the complex fluxes $P$ and $G$ separate from those that contain the fluxes $F$
and the geometry.

As an application of the above results to maximally supersymmetric $H$-backgrounds,
we solve the Killing spinor equations of IIB backgrounds with two supersymmetries
and $Spin(7)\ltimes\bR^8$-invariant spinors, and with four supersymmetries and $SU(4)\ltimes \bR^8$- and $G_2$-invariant
spinors. We find that in both the $Spin(7)\ltimes\bR^8$ and $SU(4)\ltimes \bR^8$ cases, the spacetime admits
a {\it parallel}, {\it null}, vector field, $X$, and the {\it holonomy of the Levi-Civita connection} is reduced either to
$Spin(7)\ltimes\bR^8$ or to $SU(4)\ltimes \bR^8$, respectively. The spacetime metric in both cases can be written
 as
\bea
ds^2= 2 dv\big(du+\a(y, v)+\b_I(y, v) dy^I\big)+ \gamma_{IJ}(y, v) dy^I dy^J
\eea
by adapting  Penrose  coordinates \cite{penrose} along the rotation free, null, geodesic congruence generated by $X=\frac{\partial}{\partial u}$.
 The transverse space $B$
of the null congruence, $u,v={\rm constant}$~
spanned by the coordinates $y^I$, is either a $Spin(7)$ or Calabi-Yau eight-dimensional manifolds, respectively.
We also investigate the geometry of IIB backgrounds with four supersymmetries and $G_2$-invariant spinors. We find that the
spacetime is the product of the three-dimensional Minkowski and a $G_2$ manifold $B$, i.e.
\bea
ds^2=ds^2(\bR^{1,2})+ ds^2(B)~.
\eea
In addition, the Killing spinor equations imply that  all the fluxes $P,G,F$ vanish.

There are at most two $Spin(7)\ltimes\bR^8$-invariant spinors in the complex Weyl representation
of $Spin(9,1)$. So  there are IIB backgrounds with either  one or two $Spin(7)\ltimes\bR^8$-invariant
Killing spinors. The geometry of  backgrounds with one supersymmetry has been determined in \cite{gju}.
In this paper, we have also solved the Killing spinor equations for backgrounds with two supersymmetries.
Thus we have completed the task of classifying the geometries of IIB supersymmetric backgrounds with
$Spin(7)\ltimes\bR^8$-invariant Killing spinors.

This paper is organized as follows: In section two, we  describe the Killing spinor equations of IIB supergravity,
give the conditions on the geometry for backgrounds with one $G_2$-invariant Killing spinor and determine
the $G_2$-structure that such backgrounds admit. The linear system and its solution for backgrounds with
one $G_2$-invariant Killing spinor is presented in appendix B. In section three, we investigate the properties
of the Killing spinor equations for backgrounds with extended supersymmetry and demonstrate a factorization
of the Killing spinor equations of IIB supergravity for certain backgrounds. In section four,
we determine the geometry of backgrounds with two $Spin(7)\ltimes\bR^8$-invariant Killing spinors, and  in appendix
C, we present the associated linear system and its solution. In section five, we investigate the geometry of
 backgrounds with four $SU(4)\ltimes\bR^8$-invariant Killing spinors, and  in appendix
D, we give the associated linear system and its solution. In section six, we determine the geometry of backgrounds
with four $G_2$-invariant Killing spinors, and  in appendix
E, we present the associated linear system and its solution. In section seven, we give our conclusions. In appendix A,
we summarize our spinor conventions and compute some spacetime form bilinears of $G_2$-invariant spinors.


\newsection{ Backgrounds with $G_2$-invariant Killing spinors }

\subsection{Killing spinor equations and a linear system}

As we have mentioned in the introduction, the bosonic fields of IIB supergravity are the spacetime metric $g$, a complex
one-form field strength, a complex three-form
field strength $G$, and a self-dual
five-form field strength $F$, $F_{M_1\dots M_5}=- {1\over 5!}
\epsilon_{M_1\dots M_5}{}^{N_1\dots N_5}
F_{N_1\dots N_5}$, where $\epsilon_{01\dots9}=1$.  The precise definition of the field strengths $P, G$ and $F$
is given in \cite{schwarz} and we shall not repeat the formulae here.
The Killing spinor equations of IIB supergravity are the
vanishing conditions of the gravitino and the supersymmetric partners of the scalars supersymmetry transformations
\cite{west,  schwarz} evaluated on the
bosonic fields of the theory\footnote{We use a mostly plus
convention for the metric.
To relate this  to the conventions of  \cite{schwarz}, one
takes  $\Gamma^A\rightarrow
i\Gamma^A$ and every time a index is lowered there is also
an additional minus sign, see also \cite{gju}.}. These turn into equations for the supersymmetry  parameter $\epsilon$
which we take  to be
a commuting spinor.
The  Killing spinor
equations of IIB supergravity can be interpreted as a parallel transport equation for the
supercovariant connection ${\cal D}$
\be
{\cal D}_M\epsilon=\tilde\nabla_M \epsilon+{i\over 48} \Gamma^{N_1\dots N_4 } \epsilon
 F_{N_1\dots N_4 M}
 -{1\over 96} (\Gamma_{M}{}^{N_1N_2N_3}
G_{N_1N_2N_3}-9 \Gamma^{N_1N_2} G_{MN_1N_2}) (C\epsilon)^*=0
\la{kseqna}
\ee
and an algebraic equation
\be
A\epsilon=P_M \Gamma^M (C\epsilon)^*+ {1\over 24} G_{N_1N_2N_3} \Gamma^{N_1N_2N_3} \epsilon=0~,
\la{kseqnb}
\ee
where
$$
\tilde\nabla_M=D_M+{1\over4} \Omega_{M,AB} \Gamma^{AB}~,~~~~~~D_M=\partial_M-{i\over2}Q_M
$$
is the
spin connection, $\nabla_M=\partial_M+{1\over4} \Omega_{M,AB} \Gamma^{AB}$,
twisted with $U(1)$ connection $Q_M$, $Q_M^*=Q_M$, $\epsilon$
is a (complex) Weyl spinor,
$\Gamma^{0\dots 9}\epsilon=-\epsilon$, and  $C$ is a charge conjugation
 matrix\footnote{In the basis of gamma matrices
chosen in \cite{schwarz}, $C=1$, and so it has been neglected,
see however \cite{becker}.}. (For our spinor conventions see appendix A.)
For a superspace formulation of IIB supergravity
see \cite{howe}.

The gauge group of the supercovariant derivative ${\cal D}$ is
$Spin(9,1)\times U(1)$. Under such local transformations the fluxes
rotate up to a local Lorentz rotation and an appropriate $U(1)$
gauge transformation of the connection $Q$. As we have mentioned in
the introduction, the Killing spinor can be written up to a
$Spin(9,1)$ gauge transformation as  (\ref{caform}). This spinor is
$G_2$-invariant and exhibits a $SU(3)\subset G_2$ manifest
invariance. This and the choice of the basis in the space of spinors
given in (\ref{hbasisa}) make it convenient to decompose the fluxes
and geometry in $SU(3)$ representations. For this, we split the
spacetime frame indices $A=(+,-,1, \bar 1, p, \bar p)$, $p=1,2,3$.
We then substitute the Killing spinor $\epsilon$ (\ref{caform}) into
the Killing spinor equations (\ref{kseqna}) and  (\ref{kseqnb}) and
perform the gamma matrix algebra such that these equations are
expressed in the basis (\ref{hbasisa}). Then we set every component
in this basis to zero. In this way, we derive a linear system for
the fluxes, spacetime geometry and spacetime derivatives of the
functions $f,g$ which determine the Killing spinor $\epsilon$. This
computation is described in more detail in appendix B.

Next we solve the linear system. As we have mentioned in the introduction, we first solve for the
complex fluxes $P$ and $G$. It turns out that not all components of these fluxes are specified
by the Killing spinor equations. After choosing the independent components the linear system is solved
for the remaining components of $P$ and $G$ in terms of $F$, $\Omega$ and the
derivatives of $f,g$ in (\ref{caform}). Having done this, the rest of the equations of the linear
system become conditions for the components of $F$, the geometry represented by the Levi-Civita connection
$\Omega$, the $U(1)$ (real) connection $Q$ and the spacetime derivatives of the functions $f,g$.
Since $F$, $Q$, $\Omega$, $f$ and $g$ are real, one has to also consider the complex
conjugates of the equations for $F$, $Q$, $\Omega$, $f$ and $g$. It turns out that these equations
do not determine all components of $F$. After choosing the independent ones, the linear system is solved
for the remaining components of $F$ in terms of the geometry $\Omega$, $Q$ and
the derivatives of $f,g$. Some equations remain. These involve the geometry $\Omega$, $Q$ and the spacetime
derivatives of $f,g$. We interpret these equations as conditions on the geometry of spacetime of a background
that admits a $G_2$-invariant Killing spinor.  The expressions for the fluxes mentioned above
are given in appendix B. The computation described above is lengthy but routine and it is convenient to organize
it in $SU(3)$ representations. In what follows, we shall investigate the conditions on the geometry in more detail.

\subsection{Geometry}

\subsubsection{Geometric conditions}

It is convenient to arrange the conditions on the geometry of the spacetime of backgrounds
with a $G_2$-invariant Killing spinor in terms of $SU(3)$
representations. The geometric conditions that transform under the  trivial
representation of $SU(3)$ are
\bea
g^2 \Omega_{\bar1,-\bar1}-f^2 \Omega_{\bar1,+\bar1}&=&0
\la{scalarcon1}
\eea
\bea
2 \partial_-\log f+\Omega_{-,-+}=0
\la{scalarcon2}
\eea
\bea
2f\partial_1f+ f^2\Omega_{1,-+}+g^2\Omega_{-,-1}-f^2\Omega_{-,+1}&=&0
\la{scalarcon3}
\eea
\bea
2 g\partial_1 g- g^2 \Omega_{1,-+}- g^2 \Omega_{+,-1}+ f^2 \Omega_{+,+1}&=&0
\la{scalarcon4}
\eea
\bea
2f\partial_+ f-2g\partial_-g+ g^2 \Omega_{-,-+}+ f^2 \Omega_{+,-+}&=&0
\la{scalarcon5}
\eea
\bea
2\partial_+\log g+\Omega_{+,-+}&=&0
\la{scalarcon6}
\eea
\bea
g^2\Omega_{\bar1,-1}- f^2 \Omega_{\bar1,+1}+ g^2\Omega_{-,1\bar1}- f^2 \Omega_{+,1\bar1}&=&0
\la{scalarcon7}
\eea
\bea
g^2 \Omega_{\bar r,-}{}^{\bar r}-f^2 \Omega_{\bar r,+}{}^{\bar r}+ g^2 \Omega_{-,r}{}^r-f^2 \Omega_{+,r}{}^r&=&0
\la{scalarcon8}
\eea
\bea
\Omega_{-,+\bar 1}-\Omega_{-,+1}-\Omega_{+,-\bar 1}+\Omega_{+,-1}&=&0
\la{scalarcon9}
\eea
\bea
g^2 Q_--f^2 Q_++2 f g \Omega_{-,+1}-2fg \Omega_{+,-1}&=&0
\la{scalarcon10}
\eea
\bea
\Omega_{\bar1,r}{}^r- \Omega_{1, r}{}^r+ \epsilon^{\bar r_1\bar r_2\bar r_3} \Omega_{\bar r_1,\bar r_2\bar r_3}
+\epsilon^{ r_1 r_2 r_3} \Omega_{ r_1, r_2 r_3}
\cr
+  \Omega_{\bar r, \bar 1}{}^{\bar r}- \Omega_{ r, \bar 1}{}^{r}- \Omega_{\bar r,  1}{}^{\bar r}
+\Omega_{ r,  1}{}^{ r}- {g^2\over 2f^2} \Omega_{-,-\bar 1}- {g^2\over 2f^2} \Omega_{-,- 1}
\cr
+{f^2\over2 g^2} \Omega_{+,+\bar 1}
+ {f^2\over2 g^2}  \Omega_{+,+1}-  \Omega_{+,- 1}+\Omega_{-,+1}&=&0~.
\la{scalarcon11}
\eea

The geometric conditions that transform under the fundamental representation of $SU(3)$  are
\bea
2 f\partial_pf+f^2 \Omega_{p,-+}+ g^2 \Omega_{-,-p}- f^2 \Omega_{+,-p}&=&0
\la{vectorcon1}
\eea
\bea
2 g\partial_pg-g^2 \Omega_{p,-+}- g^2 \Omega_{+,-p}+ f^2 \Omega_{+,+p}&=&0
\la{vectorcon2}
\eea
\bea
g^2\Omega_{1,-p}-f^2\Omega_{1,+p}+ g^2 \Omega_{p,-1}-f^2 \Omega_{p,+1}&=&0
\la{vectorcon3}
\eea
\bea
g^2 \Omega_{p,-\bar 1}- f^2 \Omega_{p,+\bar 1}+ g^2 \Omega_{p,-1}
- f^2 \Omega_{p,+1}- g^2 \Omega_{-,p\bar1}
\cr
-g^2 \Omega_{-,p 1}
+ f^2 \Omega_{+, p\bar 1}+ f^2 \Omega_{+,p1}&=&0
\la{vectorcon4}
\eea
\bea
g^2 \Omega_{\bar 1, -p}-f^2 \Omega_{\bar 1,+p}
-g^2 \Omega_{p,-1}+ f^2 \Omega_{p,+1}+g^2\Omega_{-,p\bar 1}
\cr
+g^2 \Omega_{-,p1}- f^2 \Omega_{+,p\bar 1}-f^2 \Omega_{+,p1}&=&0
\la{vectorcon5}
\eea
\bea
\Omega_{-,+p}-\Omega_{+,-p}&=&0
\la{vectorcon6}
\eea
\bea
g^2 \epsilon^{\bar r_1\bar r_2}{}_p \Omega_{\bar r_1,-\bar r_2}-f^2
\epsilon^{\bar r_1\bar r_2}{}_p \Omega_{\bar r_1,+\bar r_2}+2g^2 \Omega_{p,-1}-2f^2 \Omega_{p,+1}
\cr
-g^2 \epsilon^{\bar r_1\bar r_2}{}_p \Omega_{-,\bar r_1\bar r_2}+
f^2 \epsilon^{\bar r_1\bar r_2}{}_p \Omega_{+,\bar r_1\bar r_2}+2 g^2 \Omega_{-,1p}-2 f^2 \Omega_{+,1p}&=&0
\la{vectorcon7}
\eea

The conditions that transforms under the fundamental representation and its conjugate of $SU(3)$ are
\bea
-g^2 \Omega_{\bar q,-p}+ f^2 \Omega_{\bar q,+p}
- g^2 \Omega_{p,-\bar q}+  f^2 \Omega_{p,+\bar q}=0~,
\la{pbqcon}
\eea
and the conditions that transform under the symmetric product of the fundamental representation are
\bea
g^2 \Omega_{\bar p,-\bar q}-f^2 \Omega_{\bar p,+\bar q}+ g^2 \Omega_{\bar q,-\bar p}- f^2 \Omega_{\bar q,+\bar p}=0~,
\la{pqcon}
\eea
The above conditions on the geometry can be simplified by going to the gauge  $f=g$. In this gauge, the conditions
(\ref{scalarcon2}), (\ref{scalarcon5}) and (\ref{scalarcon6}), imply that
\bea
\Omega_{-,-+}=0~.
\eea
Consequently, $\partial_- f=0$. Some additional simplification also occurs because the dependence
of the conditions on $f$ can be eliminated apart from the terms containing spacetime derivatives. In what follows,
we shall examine the geometric conditions without choosing the gauge $f=g$.

The conditions on the geometry (\ref{scalarcon1})-(\ref{pqcon}) have been described in terms of $SU(3)\subset G_2$
representations. However, it  should be possible to  re-expressed them in terms  of $G_2$ representations. This is
because as one may expect  the existence of a $G_2$-invariant spinor reduces the
structure group of spacetime from $Spin(9,1)\times U(1)$ to $G_2$. To make this manifest, we shall examine the
spacetime form  bilinears that can be constructed from the Killing spinor $\epsilon$ of the background.

\subsection{Spacetime forms from spinor bilinears}

The Killing spinors of IIB supergravity are  complex Weyl spinors. Because the charge conjugation matrix $C$ acts
non-trivially on such spinors, one can construct for any complex Weyl spinor $\epsilon$ another spinor
$\tilde\epsilon=C*\epsilon$. If $\epsilon$ is a Killing spinor, $\tilde \epsilon$ may not be Killing but it
is defined on the spacetime, see also \cite{gju}. It turns out that $\tilde\epsilon$ is needed to understand the geometric conditions
which arise from the Killing spinor equations of IIB supergravity.
After some inconsequential rescaling with a numerical factor,  the Killing spinor can  be written as
\be
\epsilon={1\over\sqrt{2}}[f (1+e_{1234})-i g (e_{15}+e_{2345})]~,
\ee
and
 \bea
 \tilde\epsilon=C*\epsilon={1\over\sqrt{2}}
[f (1+e_{1234})+i g (e_{15}+e_{2345})]~.
\eea
It is then straightforward from the results in appendix \ref{spinorforms}
to find that
the one-forms are
\bea
\kappa(\epsilon, \epsilon)&=&-f^2 (-e^0+e^5)-g^2 (e^0+e^5)+2ifg e^1
\cr
\kappa(\epsilon, \tilde\epsilon)&=&-f^2 (-e^0+e^5)+ g^2 (e^0+e^5)
\cr
\kappa(\tilde\epsilon, \tilde\epsilon)&=&-f^2 (-e^0+e^5)-g^2 (e^0+e^5)-2ifg e^1
\la{oneforms}
\eea
the three-form is
\bea
\xi(\tilde\epsilon, \epsilon)&=&-2i fg [{\rm Re}\, \hat\chi
-e^6\wedge \hat\omega-e^0\wedge e^1\wedge
e^5    ]
\la{threeform}
\eea
and the five-forms are
\bea
\tau(\epsilon, \epsilon)&=& -f^2 (-e^0+e^5)\wedge[{\rm Re}\, \chi-{1\over2}\omega\wedge \omega]
\cr
&+&g^2 (e^0+e^5)\wedge[ e^1\wedge {\rm Re}\,\hat\chi
+{1\over2} \hat\omega\wedge\hat\omega-\hat\omega
\wedge e^1\wedge e^6+ e^6\wedge {\rm Im}\,\hat\chi]
\cr
&+&2ifg [e^0\wedge e^5\wedge {\rm Re}\,\hat\chi- e^1\wedge e^6\wedge
{\rm Im}\,\hat\chi
\cr
&+&{1\over2} e^1\wedge \hat\omega\wedge \hat\omega
+ \hat\omega\wedge e^6\wedge e^0\wedge e^5]~,
\eea
\bea
\tau(\tilde\epsilon, \epsilon)&=& -f^2 (-e^0+e^5)\wedge[{\rm Re}\, \chi-{1\over2}\omega\wedge \omega]
\cr
&-&g^2 (e^0+e^5)\wedge[ e^1\wedge {\rm Re}\,\hat\chi
+{1\over2} \hat\omega\wedge\hat\omega-\hat\omega
\wedge e^1\wedge e^6+ e^6\wedge {\rm Im}\,\hat\chi]~,
\eea
and $\tau(\tilde\epsilon, \tilde\epsilon)$ can be derived from $\tau(\epsilon, \epsilon)$ after setting $g\rightarrow -g$,
where $\hat\chi$ and $\hat\omega$ are given in appendix \ref{spinorforms}.

The spacetime form bilinears of the pairs $(\epsilon, \epsilon)$ and $(\tilde\epsilon, \tilde\epsilon)$
transform under the $U(1)$ subgroup of the $Spin(9,1)\times U(1)$ gauge group of IIB supergravity. So, there are
line bundle valued forms on the spacetime. In fact, they carry equal but opposite charge. As a consequence the form
$\kappa(\epsilon,\epsilon)\wedge \kappa(\tilde\epsilon, \tilde\epsilon)\wedge \kappa(\epsilon, \tilde\epsilon)$ is a well-defined three
form on the spacetime.  These will be used
later to investigate the conditions on the geometry of spacetime.

\subsection{Analysis of the geometric conditions}

To analyze the geometric conditions, we introduce a frame $e^+, e^-, e^{\a}, e^{\bar\a}$ such that
the metric is written as
\bea
ds^2= 2(e^+ e^-+ \delta_{\a\bar\b} e^{\a} e^{\bar\b})~.
\eea
It is convenient to what follows to suitably normalize the various forms that are associated with
the spinor bilinears. In particular we shall focus on the three one-forms (\ref{oneforms})
and the three-form (\ref{threeform}) which are used below to analyze the geometric conditions.
First observe that the one-form $\kappa(\epsilon, \tilde\epsilon)$
is timelike and denote with $X$ the associated vector\footnote{We have rescaled
the one-form and changed the overall sign for convenience.}
$\kappa=-{1\over\sqrt {2} }\kappa(\epsilon, \tilde\epsilon)$, i.e.
\bea
X= f^2 e_+- g^2 e_-~,
\eea
where $e_A$ is the coframe of $e^A$, $e^A{}_M e^M{}_B=\delta^A{}_B$.
Then $g(X,X)=-4 f^2 g^2$. Next we define
\bea
\hat\kappa={1\over 2i\sqrt {-g(X,X)}} (\kappa(\epsilon, \epsilon)-\kappa(\tilde\epsilon, \tilde\epsilon))= e^1
\eea
and
\bea
\kappa'=
-{1\over2\sqrt {2} }
\big(\kappa(\epsilon, \epsilon)+\kappa(\tilde\epsilon, \tilde\epsilon)\big)=f^2 e^-+g^2 e^+~.
\eea
it is also convenient to define the null one forms
 $\lambda=g^2 e^+$ and $\mu=f^2  e^-$. As we have mentioned,
$\kappa$ is timelike, while $\hat\kappa$ and $\kappa'$ are spacelike.

Observe that the three form $\xi(\epsilon, \tilde \epsilon)$ is the sum of two terms. The first term can be recognized
as the standard  $G_2$-invariant three-form $\phi$. The second term is proportional to $\kappa\wedge \hat\kappa\wedge \kappa'$.
Because of this, we can separate the two parts and
after an appropriate normalization with the length of $X$, we write
\bea
\phi={\rm Re} \hat\chi- e^6\wedge \hat\omega~.
\la{gtthreeform}
\eea
Since the spacetime admits a no-where vanishing $G_2$-invariant three-form, it admits a $G_2$-structure.

The above geometric data impose some topological conditions on the spacetime.
Suppose that the spacetime $M$ is smooth, it admits a  $G_2$-structure and the line bundle $L$ associated
with the IIB scalar fields is trivial.
This implies that $f,g\not=0$ and that $\hat\kappa, \kappa'$ are well-defined one-forms on $M$. The
forms $\kappa, \hat\kappa, \kappa'$ do not vanish. As a result, the tangent bundle $TM$ decomposes as $TM=I^3\oplus F$, where
$I^3$ is a trivial bundle of rank three. In addition the vector bundle $\Lambda^3F$ admits
a no-where vanishing section $\phi$. If $L$ is not trivial, there are again topological restrictions like for example
that the bundle of three-forms of the spacetime, $\Lambda^3(M)$, admits for example two no-where vanishing sections $\phi$ and
$\kappa\wedge \hat\kappa\wedge \kappa'$.

Next we turn to the geometric conditions implied by the Killing spinor equations. Since the spacetime $M$
admits a $G_2$-structure, one expects that the conditions on the geometry can be expressed as vanishing
conditions on the covariant derivative with respect to the Levi-Civita connection
 of the invariant forms $ \hat\kappa, \lambda, \mu$
and $\phi$. This is an adaptation of the results of Gray-Hervella for manifolds with a $U(n)$-structure \cite{gray}
 to this case. However here there is an additional ingredient due to the fact that the spinors $\epsilon, \tilde \epsilon$
 are twisted with respect to the scalar field $U(1)$ connection $Q_M$ of the line bundle $L$. As a result, the
 one-forms are also twisted. In particular $\kappa(\epsilon, \epsilon), \kappa(\tilde\epsilon,\tilde \epsilon)$
 are twisted with $L^2$ and its dual, respectively, but $\kappa$ and $\phi$
 are not. This implies that the covariant derivative of the form should be taken with the $U(1)$ twisted
 connection $\tilde\nabla$ instead of the Levi-Civita  where appropriate. In particular observe that
 \bea
 \tilde \nabla_M \kappa(\epsilon, \epsilon)=\nabla_M \kappa(\epsilon, \epsilon)- i Q_M \kappa(\epsilon, \epsilon)
 \cr
 \tilde \nabla_M \kappa(\tilde\epsilon,\tilde \epsilon)=\nabla_M \kappa(\tilde\epsilon, \tilde\epsilon)+i
  Q_M \kappa(\tilde\epsilon, \tilde\epsilon)~.
  \eea
These equations can be used to compute the (twisted) covariant derivatives of $\hat\kappa, \lambda$ and $\mu$.

 As we have mentioned, we have  expressed the geometric conditions in terms of $SU(3)$ representations.
 Since $SU(3)\subset G_2$ and the spacetime admits a $G_2$-structure,
 one expects that the various geometric conditions will  combine and organize themselves in $G_2$ representations.
 Indeed, we shall demonstrate that this is the case and in this way, we  provide
 an additional check on our results.
First, the  six conditions  (\ref{scalarcon1})-(\ref{scalarcon6}) ,  (\ref{vectorcon1}), (\ref{vectorcon2})
 (\ref{vectorcon3}),
(\ref{pbqcon}), (\ref{pqcon}),
and
\bea
g^2 \Omega_{\bar 1, -1}-f^2 \Omega_{\bar 1,+ 1}+g^2 \Omega_{1, -\bar 1}-f^2 \Omega_{1,+ \bar 1}&=&0
\cr
g^2 \big(\Omega_{p,-\bar 1}+\Omega_{\bar 1, -p}\big)- f^2 \big( \Omega_{\bar 1,+p}+\Omega_{p,+\bar 1}\big)&=&0
\eea
which are derived from the real part of (\ref{scalarcon7}) and the sum of (\ref{vectorcon4}) and (\ref{vectorcon5}), respectively,
imply that the vector field $X$ associated with the one-form $\kappa$ is Killing, i.e.
\bea
\nabla_A\kappa_B+\nabla_B\kappa_A=0~.
\la{killing}
\eea
Next observe that
\bea
X^B(\tilde\nabla_A \hat \kappa_B-\tilde\nabla_B \hat \kappa_A)=0~,~~~A=\bar 1, p
\la{commute}
\eea
implies  (\ref{scalarcon7}) and  (\ref{vectorcon4}).
Next consider
\bea
X^B\nabla^C\phi_{BCA}+{1\over12} X^B \nabla_B\phi_{CDF} {}^*\phi^{CDF}{}_A=0~,~~~~A=\bar 1, p
\la{exphi}
\eea
where ${}^*\phi$ is the  dual of $\phi$ with respect to the $G_2$-invariant  (volume) form $d{\rm vol}= e^2 \wedge e^3
\wedge e^4\wedge e^6\wedge e^7\wedge e^8\wedge e^9$. This condition implies  (\ref{scalarcon8})
and  (\ref{vectorcon7}). Similarly,  (\ref{scalarcon11}) can be written as
\bea
{1\over12} {}^*\phi^{ABCD} \nabla_A \phi_{BCD}+ {g^2\over f^2} \tilde\nabla_-\hat\kappa_--{f^2\over g^2}
\tilde\nabla_+\hat\kappa_+ =0
\la{phiphi}
\eea
The remaining equations can be written as
\bea
 f^2  \tilde \nabla_+\lambda_A-  g^2 \tilde\nabla_-\mu_A =0~,~~~~A=\bar 1, p~.
 \la{lammu}
 \eea

 To conclude the geometric content of the Killing spinor equations of IIB supergravity for backgrounds
 with one $G_2$-invariant Killing spinor is described by the equations (\ref{killing}), (\ref{commute}), (\ref{exphi}),
 (\ref{phiphi}) and (\ref{lammu}).

 The geometric interpretation of (\ref{killing}) is straightforward. In particular, one can adapt coordinates
 with respect to $X$,  $X=\frac{\partial}{\partial t}$, and write the metric as
 \bea
 ds^2=-4 f^2 g^2 (dt+ \b)^2+ ds^2_{(9)}
 \eea
 where $\b$ is a one-form in the nine remaining directions independent of $t$ and $ds^2_{(9)}$ is the metric
 in the space of orbits of $X$ again independent of $t$.
 Using (\ref{killing}) and the orthogonality of $\kappa$ and $\hat\kappa$, we can write the condition (\ref{commute})  as
 \bea
 {\cal L}_X \hat\kappa_A=0~,~~~ A=\bar 1, p~.
\eea
However since $\hat\kappa$ is twisted with respect to $L^2$, the Lie derivative ${\cal L}_X$ has an
additional term involving the connection $Q$. Similarly, the condition (\ref{exphi}) can be rewritten
as
\bea
{\cal L}_X \phi_{ABC} ~{}^*\phi^{ABC}{}_D=0~,~~~D=\bar 1, p~.
\la{weakcon}
\eea
Clearly, this condition implies that the seven-dimensional part of  ${\cal L}_X\phi$, in the decomposition of three forms
 $\Lambda^3(\bR^7)=\Lambda_{\bf1}^3(\bR^7)\oplus \Lambda_{\bf7}^3(\bR^7)\oplus \Lambda_{\bf 27}^3(\bR^7)$ in terms of $G_2$ representations,
 vanishes. The Lie derivative of $\phi$ along $X$ may be interpreted as an infinitesimal deformation of $\phi$. Then (\ref{weakcon})
 implies that this deformation is a deformation along the moduli of a seven-dimensional submanifold since it
 vanishes along some diffeomorphism
 directions.
The condition (\ref{phiphi}) implies that the singlet of $d\phi$, in the decomposition of four forms
 $\Lambda^4(\bR^7)=\Lambda_{\bf1}^4(\bR^7)\oplus \Lambda_{\bf7}^4(\bR^7)\oplus \Lambda_{\bf 27}^4(\bR^7)$ in terms of $G_2$ representations,
 is expressed in terms of derivatives of the one-form $\hat\kappa$. The condition (\ref{phiphi})  resembles
 the  {\it weak or nearly parallel}
 $G_2$-structure that one finds  in the standard reduction of a $Spin(7)$-structure to $G_2$ \cite{grayb}. Of course here,
  the reduction that takes place is from $Spin(9,1)\times U(1)$ to $G_2$ and there are far more  $G_2$ classes
  than those investigated for the standard case.

\newsection{Backgrounds with extended supersymmetry}

\subsection{Invariant spinors and extended supersymmetry}

A class of supersymmetric backgrounds are those for which  the Killing spinors are invariant under a subgroup $H$ of the gauge
group $G$ of the Killing spinor equations of a supergravity theory.
Typically $H$ includes some of the Berger holonomy groups like $SU(n)$, $G_2$ and $Spin(7)$.
 Let $\Delta^H$ be the space of $H$-invariant spinors  i.e. $H \eta=\eta$ for  every $\eta\in \Delta^H$. $\Delta^H$ is
a subspace of some representation $\Delta$ of the gauge group $G$, and $\Delta$  is the space of spinors of the
supergravity theory. Next define the group $\hat H$ as the subgroup of $G$ which preserves the subspace $\Delta^H$ but
not necessarily the individual spinors in $\Delta^H$, i.e. $\hat H \Delta^H\subset \Delta^H$. It turns out that $H$
is a normal subgroup of $\hat H$ and  $\hat G=\hat H/H$ is the factor group.
Clearly $\hat G$ preserves the subspace $\Delta^H$,  $\hat G \Delta^H\subset \Delta^H$.

Let $\eta_i$, $i=1,\dots, k$, be a basis
in the space of $H$-invariant spinors, $\Delta^H$.
The Killing spinors of a background with $N$ supersymmetries and $H$-invariant
Killing spinors\footnote{If $N<k$, the Killing spinors may exhibit a larger symmetry than $H$ which in some cases
is an extension of $H$ with a discrete group \cite{amus, mcinnes}.} can be written as
\bea
\epsilon_r=\sum_{i=1}^kf_{ri} \eta_i~,~~~~r=1,\dots,N~,
\eea
where $f_{ri}$ are functions of the spacetime and $r=1,\dots, N$,  $N\leq k$. For backgrounds with exactly $N$ supersymmetries
 the rank of the matrix $f=(f_{ri})$
of coefficients is $N$. If a background admits $N=k$ supersymmetries, then $f=(f_{ri})$ is invertible. We term  such backgrounds
as maximally  supersymmetric with $H$-invariant spinors or maximally supersymmetric $H$-backgrounds for short.

The Killing spinor equations of maximally supersymmetric $H$-backgrounds exhibit some simplification. To see this,
suppose that the Killing spinor equations of a supergravity theory are a parallel transport equation of some supercovariant connection
${\cal D}$ and a set of algebraic equations collectively denoted by $A$. In such case, the Killing spinor
equations  can be written as
\bea
{\cal D}_M \epsilon_r= \sum_{i=1}^N(D_M f_{ri} \eta_i+ f_{ri}{\cal D}_M \eta_i)=0~,
\cr
A \epsilon_r=\sum_{i=1}^N f_{ri} A\eta_i=0~.
\eea
Since $f=(f_{ri})$ is invertible, these equations imply that
\bea
\sum_{j=1}^N(f^{-1}D_M f)_{ij} \eta_j+ {\cal D}_M \eta_i=0
\cr
A\eta_i=0~.
\la{magicf}
\eea
The simplification consists of eliminating the dependence of the ${\cal D}_M \eta_i$ term in the first Killing spinor
equation and of the second Killing spinor equation on the functions $f_{ri}$.
Observe that this simplification is not possible for backgrounds with $H$-invariant spinors which are not
 maximally supersymmetric, i.e. $N<k$.
This is because the matrix $f=f_{ri}$ cannot be inverted and so some dependence on the functions $f_{ri}$ may remain.
However some simplification can still be made by an appropriate choice of the basis $\eta_i$.

Suppose that we apply the method of \cite{uggp} to solve the  parallel transport equation in (\ref{magicf}). In such a case
apart from a set of vanishing conditions for the fluxes and geometry, one finds that there is a parallel transport equation for the
functions $f$ of the type
\bea
(f^{-1}\partial_M f)_{ij}+ (C_M)_{ij}=0~,
\la{prpr}
\eea
where $C$ is a connection constructed from the Levi-Civita connection of the spacetime and the fluxes of the
supergravity theory.  The connection $C$ can be thought of as the restriction of the supercovariant connection
of the supergravity theory on the subbundle of the Killing spinors.
It is well-known for such parallel transport equations that a necessary and sufficient
condition for the existence of a solution $f$ is that the holonomy of $C$ to be the identity, ${\rm hol}(C)=1$. This in particular
implies that the curvature $K=dC-C\wedge C$ should vanish, $K=0$.
For backgrounds with one supersymmetry this is a `mild' condition. However, it becomes rather
restrictive for backgrounds with extended supersymmetry.
Assuming that ${\rm hol}(C)=1$, the solution of (\ref{prpr}) can be written as
\bea
f(x)= f_0~ P{\rm exp}(-\int_\gamma C)~,
\eea
where $f_0$ is a constant matrix and the integral is over a path $\gamma$ from a fixed point in spacetime to $x$.
In fact one can choose $f_0$ to be the identity matrix using the property of (\ref{prpr})
to be invariant under the transformation $f\rightarrow g f$, where $g$ is an invertible constant $N\times N$ matrix.

It remains to determine whether there is a gauge such that the matrix $f$ can be taken to be the identity matrix, $f=1$,
and so Killing spinors of a maximally supersymmetric $H$-background can be chosen such that $\epsilon_i=\eta_i$.
It is well-known that this is always the case for Riemannian backgrounds with parallel spinors but it is not apparent
in the context of supergravity because the holonomy and  the gauge group of the supercovariant connection can be different.
To give a sufficient condition for the existence of the gauge $f=1$, let $s$ be the Lie algebra that $C$ takes values in.
Typically $s$ is a subalgebra of the Lie algebra of the holonomy group of the supercovariant connection.
It is clear that one can choose the gauge $f=1$, if this gauge can be reached with a gauge transformation in $G$ which preserves
$\Delta^H$, i.e.
with a $\hat G$ gauge transformation of the supercovariant derivative. This implies that a sufficient condition
for choosing the gauge $f=1$ is that $s$ is a subalgebra of the Lie algebra $\hat g$ of $\hat G$.

\subsection{Maximally supersymmetric $H$-backgrounds in IIB supergravity}

The  considerations of the previous section have a straightforward application in IIB supergravity. In this case
$G=Spin(9,1)\times U(1)$. We shall
consider three cases of backgrounds with $H$-invariant spinors. These are  $Spin(7)\ltimes\bR^8$,
$SU(4)\ltimes\bR^8$ and $G_2$. In addition, we shall show that the terms which contain the $F$ and $G$ fluxes
in the Killing spinor equation of the supercovariant connection
of any IIB maximally supersymmetric $H$-background  factorize, $H\subset Spin(9,1)$.

There are two $Spin(7)\ltimes\bR^8$
invariant spinors in the (complex) Weyl representation of $Spin(9,1)$. So there may exist backgrounds
with $Spin(7)\ltimes\bR^8$-invariant Killing spinors with either one or two supersymmetries.
The Killing spinor equations
of backgrounds with one $Spin(7)\ltimes\bR^8$-invariant Killing spinors have been solved in  \cite{gju}.
So it remains to investigate the case with two $Spin(7)\ltimes\bR^8$-invariant Killing spinors.
These are the maximally supersymmetric
$Spin(7)\ltimes\bR^8$-backgrounds. One can choose the basis in the space of
$Spin(7)\ltimes\bR^8$-invariant spinors to be
\bea
\eta_1=(1+e_{1234})~,~~~~\eta_2=i \eta_1=i (1+e_{1234})~.
\eea
Substituting these into (\ref{magicf}), we find after a straightforward computation using $C*\eta_1=\eta_1$
and $C*\eta_2=-\eta_2$
that
\bea
{1\over2}[(f^{-1}D_M f)_{11}+(f^{-1}D_M f)_{22}+ i(f^{-1}D_M f)_{12} -i (f^{-1}D_M f)_{21}]\eta_1
\cr
+{1\over4}\Omega_{M,AB} \Gamma^{AB} \eta_1+{i\over 48} \Gamma^{N_1\dots N_4 } \eta_1
 F_{N_1\dots N_4 M}=0
 \cr
{1\over2}[(f^{-1}D_M f)_{11}-(f^{-1}D_M f)_{22}+ i(f^{-1}D_M f)_{12} +i (f^{-1}D_M f)_{21}]\eta_1
\cr
-{1\over 96} (\Gamma_{M}{}^{N_1N_2N_3}
G_{N_1N_2N_3}-9 \Gamma^{N_1N_2} G_{MN_1N_2}) \eta_1=0
\cr
P_M \Gamma^M \eta_1=0
\cr
 G_{N_1N_2N_3} \Gamma^{N_1N_2N_3} \eta_1=0~.
 \la{maxspins}
\eea
The second condition can be further simplified using the last. In particular acting with a gamma matrix on the last equation
in (\ref{maxspins})
and using the resulting expression after skew-symmetrization, we find that the second condition can be written as
\bea
[(f^{-1}D_M f)_{11}-(f^{-1}D_M f)_{22}+ i(f^{-1}D_M f)_{12} +i (f^{-1}D_M f)_{21}]\eta_1
\cr
+{1\over 4} \Gamma^{N_1N_2} G_{MN_1N_2}\eta_1=0~.
\eea
This condition is a parallel transport type of equation for the spinor $\eta_1$.
It is surprising that parts of the gravitino Killing spinor equation which contain the $F$ and $G$ fluxes factorize.
This is a profound simplification in the Killing spinor equations because the main difficulty
in the cases studied in \cite{gju} and in earlier sections is the mixing of these two terms in the Killing spinor equations.

There are four $SU(4)\ltimes\bR^8$-invariant spinors. Therefore, there are IIB backgrounds with one, two, three and four
$SU(4)\ltimes\bR^8$-invariant Killing spinors. The Killing spinor equations of backgrounds with one $SU(4)\ltimes\bR^8$-invariant Killing spinor
have been solved in \cite{gju}. Here, we shall investigate the Killing spinor equations of maximally supersymmetric
$SU(4)\ltimes\bR^8$-backgrounds. A basis in the space of $SU(4)\ltimes\bR^8$-invariant spinors is
\bea
\eta_1=1+e_{1234}~,~~~\eta_2&=&i(1-e_{1234})~,~~~
\cr
\eta_3=i\eta_1=i(1+e_{1234})~,~~~\eta_4&=&i\eta_2=-(1-e_{1234})~.
\eea
Substituting these into (\ref{magicf}) and using $C*\eta_1=\eta_1$, $C*\eta_2=\eta_2$, $C*\eta_3=-\eta_3$ and $C*\eta_4=-\eta_4$,
we find after some straightforward computation that
\bea
{1\over2}[(f^{-1}D_M f)_{11}+ i(f^{-1}D_M f)_{13}
+(f^{-1}D_M f)_{33}-i(f^{-1}D_M f)_{31}]\eta_1
\cr
+{1\over2}[(f^{-1}D_M f)_{12}+i (f^{-1}D_M f)_{14}
+(f^{-1}D_M f)_{34}-i(f^{-1}D_M f)_{32}]\eta_2
\cr
+{1\over4}\Omega_{M,AB} \Gamma^{AB} \eta_1+{i\over 48} \Gamma^{N_1\dots N_4 } \eta_1
 F_{N_1\dots N_4 M}=0
 \cr
{1\over2}[(f^{-1}D_M f)_{21}+ i(f^{-1}D_M f)_{23}
+(f^{-1}D_M f)_{43}-i(f^{-1}D_M f)_{41}]\eta_1
\cr
+{1\over2}[(f^{-1}D_M f)_{22}+i (f^{-1}D_M f)_{24}
+(f^{-1}D_M f)_{44}-i(f^{-1}D_M f)_{42}]\eta_2
\cr
+{1\over4}\Omega_{M,AB} \Gamma^{AB} \eta_2+{i\over 48} \Gamma^{N_1\dots N_4 } \eta_2
 F_{N_1\dots N_4 M}=0
 \cr
[(f^{-1}D_M f)_{11}+ i(f^{-1}D_M f)_{13}
-(f^{-1}D_M f)_{33}+i(f^{-1}D_M f)_{31}]\eta_1
\cr
+[(f^{-1}D_M f)_{12}+i (f^{-1}D_M f)_{14}
-(f^{-1}D_M f)_{34}+i(f^{-1}D_M f)_{32}]\eta_2
\cr
+{1\over 4}  \Gamma^{N_1N_2} G_{MN_1N_2} \eta_1=0
\cr
[(f^{-1}D_M f)_{21}+ i(f^{-1}D_M f)_{23}
-(f^{-1}D_M f)_{43}+i(f^{-1}D_M f)_{41}]\eta_1
\cr
+[(f^{-1}D_M f)_{22}+i (f^{-1}D_M f)_{24}
-(f^{-1}D_M f)_{44}+i(f^{-1}D_M f)_{42}]\eta_2
\cr
+{1\over 4}  \Gamma^{N_1N_2} G_{MN_1N_2} \eta_2=0
\cr
P_M \Gamma^M \eta_1=0~,~~~P_M \Gamma^M \eta_2=0~,
\cr
 G_{N_1N_2N_3} \Gamma^{N_1N_2N_3} \eta_1=0~,~~~G_{N_1N_2N_3} \Gamma^{N_1N_2N_3} \eta_2=0~.
 \la{maxsuf}
\eea
Again the parts of the gravitino Killing spinor equation which contain the $F$ and $G$ fluxes
of the maximally supersymmetric $SU(4)\ltimes \bR^8$ backgrounds factorize.
We shall show that this is a generic property of maximally supersymmetric $H$-backgrounds in IIB supergravity.

There are four $G_2$-invariant spinors in the (complex) Weyl representation of $Spin(,1)$. Therefore, there are
IIB backgrounds with one, two, three and four
$G_2$-invariant Killing spinors. The Killing spinor equations of backgrounds with one $G_2$-invariant Killing spinor
have been solved in section two and the appendices. Here, we shall investigate the Killing spinor
equations of maximally supersymmetric
$G_2$-backgrounds. A basis in the space of $G_2$-invariant spinors is
\bea
\eta_1=1+e_{1234}~,~~~\eta_2&=&e_{15}+e_{2345}~,~~~
\cr
\eta_3=i\eta_1=i(1+e_{1234})~,~~~\eta_4&=&i\eta_2=i(e_{15}+e_{2345})~.
\eea
Substituting these into (\ref{magicf}) and using $C*\eta_1=\eta_1$, $C*\eta_2=\eta_2$, $C*\eta_3=-\eta_3$ and $C*\eta_4=-\eta_4$,
we find after some straightforward computation the same Killing spinor equations as for the maximally supersymmetric
$SU(4)\ltimes\bR^8$-backgrounds (\ref{maxsuf}) but now $\eta_1=1+e_{1234}$ and $\eta_2=e_{15}+e_{2345}$.

It remains to show that the $F$ and $G$ terms of the Killing spinor equation associated with the
supercovariant derivative factorize for any IIB maximally supersymmetric $H$-background, $H\subset Spin(9,1)$. It is clear
from the special cases we have investigated above that to show this, it is sufficient to show that there
is a basis in $\Delta^H$ of the type $(\eta_i, i \eta_i, i=1, \dots, k)$, where $\eta_i$ are Majorana spinors.
Let $\eta\in \Delta^H$. Then $C*\eta$ is also in $\Delta^H$ because $C*$ commutes with the elements of
$Spin(9,1)$ and so with the elements $H\subset Spin(9,1)$. Since $(C*)^2=1$,
 a basis can be chosen in $\Delta^H$ for which the basis elements have eigenvalues either $+1$ or $-1$. If we denote
 with $(\eta_i, i=1, \dots, k)$ the basis elements in $\Delta^H$ with eigenvalue $+1$, then those with eigenvalue $-1$ are
 given by $i \eta_i$, where $\eta_i$ are Majorana spinors. Using this, one can show the factorization of the
 Killing spinor equation associated with the supercovariant derivative using arguments similar to those we have demonstrated
 above for the three special cases.

\newsection{Maximally supersymmetric $Spin(7)\ltimes\bR^8$-backgrounds}

\subsection{The conditions on the geometry}

The conditions on the geometry of spacetime for maximally supersymmetric $Spin(7)\ltimes \bR^8$ backgrounds
derived in appendix C can be listed as
\bea
(f^{-1}\partial_A f)_{11}-(f^{-1}\partial_A f)_{22}&=& 0~,~~~A=-,+, \a, \bar\a
\cr
 (f^{-1}\partial_A f)_{12}+ (f^{-1}\partial_Af)_{21}&=&0~,~~~
A=-,+, \a, \bar\a
\cr
(f^{-1}\partial_+ f)_{11}+(f^{-1}\partial_+ f)_{22}+\Omega_{+,-+}&=&0~,
\cr
(f^{-1}\partial_+ f)_{12}-(f^{-1}\partial_+ f)_{21}-Q_+&=&0~,
\cr
(f^{-1}\partial_- f)_{11}+(f^{-1}\partial_- f)_{22}+\Omega_{-,-+}&=&0~,
\cr
(f^{-1}\partial_- f)_{12}-(f^{-1}\partial_- f)_{21}- Q_-&&
\cr
+{1\over2} F_{-\g}{}^\g{}_\d{}^\d+{1\over6}
F_{-\a_1\a_2\a_3\a_4} \epsilon^{\a_1\a_2\a_3\a_4}&=&0~,
\cr
(f^{-1}\partial_\a f)_{11}+(f^{-1}\partial_\a f)_{22}+\Omega_{\a,-+}&=&0
\cr
(f^{-1}\partial_\a f)_{12}-(f^{-1}\partial_\a f)_{21}-Q_\a&=&0~,
\la{parc}
\eea
and
\bea
\Omega_{+,+\a}=0~,~~~\Omega_{+,\a}{}^\a&=&0~,~~~
\cr
\Omega_{+,\bar\a\bar\b}-{1\over2} \epsilon_{\bar\a\bar\b}{}^{\g\d} \Omega_{+,\g\d}&=&0~,
\cr
\Omega_{-,+\a}=0~,~~~~\Omega_{-,\a}{}^\a&=&0~,
\cr
\Omega_{-,\a\b}-{1\over2} \epsilon_{\a\b}{}^{\bar\g\bar\d} \Omega_{-,\bar\g\bar\d}&=&0~,
\cr
\Omega_{\bar\a,+\b}=0~,~~~\Omega_{\a,+\b}&=&0~,
\cr
\Omega_{\bar\a, \b_1\b_2}-{1\over2} \epsilon_{\b_1\b_2}{}^{\bar\g_1\bar\g_2} \Omega_{\bar\a, \bar\g_1\bar\g_2}&=&0~,
\cr
\Omega_{\a,\b}{}^\b&=&0~.
\la{geospins}
\eea
Most of the above conditions have the interpretation of a parallel transport equation.
The conditions for the fluxes will be summarized at the end of the section.

\subsection{The geometry of spacetime}

To investigate the geometry of maximally supersymmetric $Spin(7)\ltimes\bR^8$-backgrounds, let us focus
on the equations for the functions $f$  in (\ref{parc}). It is straightforward to see that these can be rewritten
as
\bea
\partial_A f+f C=0~,
\la{fpartrans}
\eea
where $C={1\over2}\Omega_{A,-+}\tau_0 +{1\over2}\hat Q_A \tau_1$,  $\hat Q_A= Q_A$ for $A=+, \a, \bar\a$ and
\bea
\hat Q_-=Q_--{1\over2} F_{-\g}{}^\g{}_\d{}^\d-{1\over6}
F_{-\a_1\a_2\a_3\a_4} \epsilon^{\a_1\a_2\a_3\a_4}~,
\eea
$\tau_0$ is the identity $2\times2$ matrix and $\tau_1$ is the $2\times 2$ skew-symmetric matrix with $(\tau_1)_{12}=-1$.
Therefore, the connection $C$ takes values in the abelian Lie algebra $s=\bR\oplus u(1)$.
As we have mentioned in section three, the parallel transport equation  (\ref{fpartrans})  has a solution iff the holonomy of the
connection $C$ is the identity. In particular the curvature of $K$ should vanish and this
gives
\bea
\Omega_{A,-+}=\partial_Aa
\cr
\hat Q_A=\partial_A b~,
\eea
for some functions $a,b$ of spacetime.  Solving the parallel transport equation to express $f$ in terms of $a, b$, one finds that
\bea
f(x)=f_0 e^{-{1\over2} [a(x)\tau_0+ b(x) \tau_1]} = e^{-{1\over2} a(x)}  f_0 [\cos({1\over2} b(x))\tau_0- \sin({1\over2} b(x))\tau_1]~,
\eea
where $f_0$ is a constant $2\times 2$ real matrix. As we have explained in section three, one can use the
invariance $f\rightarrow gf$, $g$ constant
$2\times 2$ matrix,  to set $f_0=1$. In such case,
the Killing spinors can be written\footnote{If one does not use $f_0=1$ , then  $\epsilon_1$ and $\epsilon_2$
in (\ref{kspinspins})
are in addition multiplied by the constant matrix $f_0$.} as
\bea
\epsilon_1&=& e^{-{1\over2} a(x)} e^{{i\over2}b(x)} \eta_1
\cr
\epsilon_2&=& i e^{-{1\over2} a(x)} e^{{i\over2} b(x)} \eta_1~.
\la{kspinspins}
\eea
It turns out that in this case, there is a gauge for which the functions $f=1$.
 In particular,  the exponential factor  can be
gauged away with the gauge transformation $e^{-{1\over2}a(x)\Gamma^{+-}}$ which is in the $Spin(9,1)$ gauge symmetry
of the supercovariant connection and the phase can also be gauged away with the $U(1)$ gauge transformation
$e^{{i\over2}b(x)}$. The result is that the Killing spinors are constant and are given in this gauge by
$\epsilon_1=\eta_1$ and $\epsilon_2=i \eta_1$.

In the gauge where the  the Killing spinors
are constant, it becomes apparent that {\it the holonomy of the  Levi-Civita connection} of  backgrounds
with maximal $Spin(7)\ltimes \bR^8$ supersymmetry  {\it is contained } in $Spin(7)\ltimes \bR^8$.
This follows from the conditions in (\ref{parc})  of   the Killing spinor equations
which in this gauge imply that  $\Omega_{A,-+}=0$.
In addition the conditions in  (\ref{geospins}) imply that the non-vanishing components of the
Levi-Civita connection  are $\Omega_{A, \a\b}$, $\Omega_{A, \a\bar\b}$ with
$\Omega_{A,\d}{}^\d=0$ and $\Omega_{A, \a \b}={1\over2} \epsilon_{\a\b}{}^{\bar\g\bar\d} \Omega_{A,\bar\g\bar\d}$,
 and $\Omega_{A,-\a}$. Clearly the Levi-Civita connection takes values in Lie algebra of $Spin(7)\ltimes \bR^8$.

 The spacetime form spinor bilinears are a one-form $\kappa=e^-$ and a five-form $\tau$. From these, after an
 appropriate normalization,
 one can construct the familiar $Spin(7)$-invariant form
 \bea
 \psi= i_{\kappa} \tau= {\rm Re}\chi-{1\over 2}\omega\wedge \omega~,
 \eea
 where $\chi$ and $\omega$ are given in appendix A.
To investigate the properties of the one-form $\kappa$, it is convenient to work in the gauge
where the Killing spinors are constant. Of course the geometry of the spacetime does not depend
on the specific gauge we use. It turns out that
 the  null vector field
$X=e_+$ associated with $\kappa$ is parallel with respect to the Levi-Civita connection $\nabla$. In particular this implies
that $X$ is Killing, self-parallel and that the associated null geodesic congruence is rotation free. Because of this,
we can introduce Penrose coordinates \cite{penrose}, see also \cite{blau}, and write the metric of the spacetime as
\bea
ds^2= 2 dv(du+ \alpha  dv+ \b_I dy^I)+ \gamma_{IJ} dy^I dy^J~,
\la{penmetr}
\eea
where $X=e_+={\partial\over \partial u}$ and $v$ is the Hamilton-Jacobi function of the null geodesics \cite{patricot}.
Since in addition $X$ is Killing, then $\alpha=\alpha(v, y)$, $\beta_I=\beta_I(v, y)$
and $\gamma_{IJ}=\gamma_{IJ}(v,y)$.
In this coordinate system, it is natural to introduce the frame
\bea
e^-=dv~,~~~e^+=du+ \alpha  dv+ \b_I dy^I~,~~~e^\a=e^\a{}_I dy^I~,~~~e^{\bar\a}=e^{\bar\a}{}_I dy^I~.
\eea
The associated coframe is
\bea
e_-={\partial\over\partial v}-\a {\partial\over\partial u}~,~~e_+={\partial\over\partial u}~,~~
~e_\a= e^I{}_\a {\partial\over\partial y^I}-\b_\a {\partial\over\partial u}~,~~~
e_{\bar\a}= e^I{}_{\bar\a} {\partial\over\partial y^I}-\b_{\bar\a} {\partial\over\partial u}~,
\eea
where $e^I{}_\a e^\b{}_I=\delta^\b{}_\a$ and $\b_\a=\b_I e^I{}_\a$, and similarly for $e^I{}_{\bar\a}$ and
$\b_{\bar\a}$.
The torsion free condition for this frame implies that
\bea
\Omega_{+,-\a}=0~,~~~\Omega_{+,-\bar\a}=0~,~~~\Omega_{+,\a\b}=0~,~~~\Omega_{+,\a\bar\b}=0~.
\eea
The rest of the components of the connection can be computed from the frame in the usual way.

The spacetime is the Lorentzian extension of an one-parameter family of manifolds with holonomy $Spin(7)$ equipped
with a line bundle. To see
this observe that the eight-dimensional manifolds $B$ given by $u, v={\rm const}$ admit a metric with holonomy $Spin(7)$.
The component
$\alpha  dv+ \b_I dy^I$ of the metric can be thought of as the connection of a line bundle over the family and $u$  as the
coordinate along the fibre direction.  The component
$\Omega_{v,IJ}$ of the connection restricts to a two-form on $B$ which takes values in $spin(7)$.
In this family one can compute the deformation of the $Spin(7)$-invariant form $\psi$ to find
that all the components of $\nabla\psi$ vanish apart from
\bea
\nabla_A\psi_{B_1B_2B_3-}=\Omega_{A,-}{}^B\,\psi_{B_1B_2B_3B}~,~~~~
A=-, \a, \bar\a~,~~~B_1,B_2,B_3,B=\a, \bar\a~.
\eea
Conversely, one can solve all the geometric conditions arising from the Killing spinor equations
by (i) considering a spacetime with   a  metric as in (\ref{penmetr}) which in addition has
 the property  that the submanifold $B$ given by  $u, v={\rm const}$ has
holonomy $Spin(7)$, and (ii) requiring that $\Omega_{-,AB}$, $A,B=\a, \bar\a$ takes values in $spin(7)$.
The latter condition gives a restriction on the dependence of the metric of $B$ on the deformation
parameter $v$. In particular, we have\footnote{Our form notation is $\omega={1\over k!} \omega_{i_1\dots i_k} dx^{i_1}\wedge \dots \wedge
dx^{i_k}$ and $(d\omega)_{i_1\dots i_{k+1}}= (k+1) \partial_{[i_1}\omega_{i_2\dots i_{k+1}]}$. }

\bea
{1\over2}(d\beta)_{AB}+ \partial_v e_{I[A} e^I{}_{B]}-{1\over 2} \psi_{AB}{}^{CD} \big (
{1\over2}(d\beta)_{CD}+\partial_v e_{I[C} e^I{}_{D]}\big)=0~,~~~A,B=\a,\bar\a~.
\eea
A solution of this is to take the frame $e^\a$ to be independent of $v$ and $d\beta$ to take values in
$spin(7)$.

\subsection{Fluxes}

It remains to give a geometric interpretation of the conditions on the fluxes, see appendix C. As we have already
mentioned  the Killing spinor equations imply that all the components of $G$ vanish apart from $G_{-AB}$ which takes
values in $spin(7)$.  The only non-vanishing component of $P$ is $P_-$. The conditions implied
by the Killing spinor equations on the flux $F$ have been given in terms of $SU(4)\subset Spin(7)$ representations
in appendix C. However it is expected that they should be re-expressed in terms of $Spin(7)\ltimes \bR^8$ representations.
This is because, as we have seen in the previous section, the (geometric) structure group of
the spacetime of maximal supersymmetric $Spin(7)\ltimes \bR^8$-backgrounds
 reduces from $Spin(9,1)\ltimes \bR^8$ to $Spin(7)\ltimes \bR^8$. In particular one find that in addition
to the self-duality condition of $F$
\bea
F_{-+ABC}&=&0~,
\cr
F_{+AC_1C_2C_3} \psi^{C_1C_2C_3}{}_B&=&0
\cr
F^{\bf 7}_{-A_1A_2A_3A_4}&=&0~,~~~A=\a,\bar\a
\eea
 and
\bea
- Q_-+{1\over24}
F_{-ABCD} \psi^{ABCD}&=&0~,
\eea
where $F^{\bf 7}_{-A_1A_2A_3A_4}$ denotes the seven-dimensional irreducible representation of $spin(7)$ in the
decomposition of $\Lambda^4(\bR^8)=\Lambda^4_{\bf 1}(\bR^8)\oplus\Lambda^4_{\bf 7}(\bR^8)\oplus
 \Lambda^4_{\bf 27}(\bR^8)\oplus \Lambda^4_{\bf 35}(\bR^8)$.

\newsection{Maximally supersymmetric $SU(4)\ltimes\bR^8$-backgrounds}

\subsection{Conditions on the geometry}

The conditions on the geometry of maximally supersymmetric $SU(4)\ltimes\bR^8$-backgrounds have
been derived in appendix D and can be summarized as follows:
\bea
(f^{-1} \partial_A f)_{11} = (f^{-1} \partial_A f)_{22} = (f^{-1} \partial_A f)_{33} = (f^{-1} \partial_A f)_{44}~,&&
\cr
(f^{-1} \partial_A f)_{23} = (f^{-1} \partial_A f)_{32} = (f^{-1} \partial_A f)_{14} = (f^{-1} \partial_A f)_{41} &=&0~,
\cr
(f^{-1} \partial_A f)_{12} = - (f^{-1} \partial_A f)_{21} = (f^{-1} \partial_A f)_{34} = -(f^{-1} \partial_A f)_{43}~,&&
\cr
(f^{-1} \partial_A f)_{13} = - (f^{-1} \partial_A f)_{31} = (f^{-1} \partial_A f)_{24} = - (f^{-1} \partial_A f)_{42}~,&&
\cr
(f^{-1} \partial_- f)_{11} = (f^{-1} \partial_- f)_{22} = (f^{-1} \partial_- f)_{33} = (f^{-1} \partial_- f)_{44}~,&&
\cr
(f^{-1} \partial_A f)_{11} + {1 \over 2} \Omega_{A,-+} &=&0~,
\cr
2 (f^{-1} \partial_A f)_{13}-Q_A &=& 0
\cr
2i (f^{-1} \partial_A f)_{21}-\Omega_{A, \beta}{}^\beta &=&0~,
\cr
(f^{-1} \partial_- f)_{11}+{1 \over 2} \Omega_{-,-+} &=&0~,
\cr
(f^{-1} \partial_- f)_{13} + (f^{-1} \partial_- f)_{24} +{1 \over 2} F_{- \gamma}{}^\gamma{}_\beta{}^\beta-Q_- &=&0~,
\cr
i ((f^{-1} \partial_- f)_{21}+(f^{-1} \partial_- f)_{43})-\Omega_{-,\beta}{}^\beta &=& 0~,
\cr
(f^{-1} \partial_- f)_{13} - (f^{-1} \partial_- f)_{24} + {1 \over 12}(F_{- \alpha_1 \alpha_2
\alpha_3 \alpha_4} \epsilon^{\alpha_1 \alpha_2
\alpha_3 \alpha_4}
\qquad&&
\cr
+ F_{- {\bar{\alpha}}_1 {\bar{\alpha}}_2 {\bar{\alpha}}_3 {\bar{\alpha}}_4}
\epsilon^{{\bar{\alpha}}_1 {\bar{\alpha}}_2 {\bar{\alpha}}_3 {\bar{\alpha}}_4})&=&0~,
\cr
(f^{-1} \partial_- f)_{14}-(f^{-1} \partial_- f)_{32}  -{i \over 12}
(F_{- \alpha_1 \alpha_2
\alpha_3 \alpha_4} \epsilon^{\alpha_1 \alpha_2
\alpha_3 \alpha_4}
\qquad&&
\cr
- F_{- {\bar{\alpha}}_1 {\bar{\alpha}}_2 {\bar{\alpha}}_3 {\bar{\alpha}}_4}
\epsilon^{{\bar{\alpha}}_1 {\bar{\alpha}}_2 {\bar{\alpha}}_3 {\bar{\alpha}}_4})&=&0~,
\cr
 (f^{-1}\partial_- f)_{21}- (f^{-1}\partial_- f)_{43}
+i (f^{-1}\partial_- f)_{23} +i (f^{-1}\partial_- f)_{41}+{i\over2}G_{- \beta}{}^\beta&=&0~,
\cr
(f^{-1} \partial_- f)_{ij}+(f^{-1} \partial_- f)_{ji} &=&0~,
\la{parcsu}
\eea
where in the last equation $i \neq j$ and $i,j=1,2,3,4$,  $A=+,\a, \bar\a$, and
\bea
\Omega_{\alpha_1, \alpha_2 \alpha_3} =0~,~~~
\Omega_{{\bar{\alpha}} , \beta_1 \beta_2} =0~,~~~
\Omega_{\alpha, + {\bar{\beta}}} =0~,~~~\Omega_{+,+\a}&=&0~,~~~
\cr
\Omega_{-,\alpha_1 \alpha_2} =0~,~~~
\Omega_{-,+\alpha} =0~,~~~
\Omega_{+, \beta_1 \beta_2}=0~,~~~
\Omega_{\alpha_1 , + \alpha_2} &=&0~.
\la{geosuf}
\eea
The above condition appear rather involved. However, most of them have the interpretation of a parallel transport
equation.

\subsection{The geometry of spacetime}

As in the case of $Spin(7)\ltimes \bR^8$-invariant Killing spinors, the conditions (\ref{parcsu}) can be written
as a parallel transport equation
\bea
f^{-1}\partial_A f+C_A=0~,
\eea
where $C_A={1\over2}\hat\Omega_A^0 t_0+ {1\over2}\hat\Omega_A^r I_r+ {1\over2}\tilde \Omega_A^r J_r$,
for some choice of self-dual $I_r$ and anti-self dual $J_r$ $4\times4$ matrices such that $I_r^2=J_r^2=-1$, $I_3=I_1 I_2$,
$J_3=J_1 J_2$ and $I_r J_s=J_s I_r$ and $t_0$ is the identity $4\times4$ matrix. In addition, we have
\bea
\hat\Omega_A^0&=& \Omega_{A,-+}
\cr
\hat\Omega^1_A&=& i \Omega_{A, \b}{}^\b
\cr
\hat\Omega_-^2&=&{1\over 12} (F_{-\a_1\a_2\a_3\a_4} \epsilon^{\a_1\a_2\a_3\a_4}+F_{-\bar\a_1\bar\a_2\bar\a_3\bar\a_4}
 \epsilon^{\bar\a_1\bar\a_2\bar\a_3\bar\a_4})
 \cr
 \hat\Omega_-^3&=&{i\over 12} (F_{-\a_1\a_2\a_3\a_4} \epsilon^{\a_1\a_2\a_3\a_4}-F_{-\bar\a_1\bar\a_2\bar\a_3\bar\a_4}
 \epsilon^{\bar\a_1\bar\a_2\bar\a_3\bar\a_4})
 \cr
 \tilde\Omega_A^1&=&- \hat Q_A
 \cr
 \tilde\Omega^2_-&=&-{i\over4}(G_{-\b}{}^\b+{}^*G_{-\b}{}^\b)
 \cr
 \tilde\Omega^3_-&=&{1\over4} (G_{-\b}{}^\b-{}^*G_{-\b}{}^\b)
 \eea
the remaining components vanishing, and
\bea
\hat Q_A=Q_A~,~~~A=+, \a, \bar\a~,~~~~\hat Q_-=Q_--{1\over2} F_{-\b}{}^\b{}_\g{}^\g~.
\eea
The connection $C$ takes values
in the Lie algebra $s=so(4)\oplus \bR=su(2)\oplus su(2)\oplus \bR$. A necessary and sufficient condition
for the parallel transport equation
to have a solution is that the holonomy of $C$ should be the identity. This in particular implies that
the curvature $K=dC-C\wedge C$ should vanish. However unlike the $Spin(7)\ltimes \bR^8$ case, $s=so(4)\oplus \bR$ is not
a subalgebra of $spin(9,1)\oplus u(1)$. This can be seen by identifying the generator of the $U(1)$ rotations in
$so(4)\oplus \bR$, which is $I_1$, and observing that it does not commute with all the generators of $so(4)\oplus \bR$.

The vanishing components $\hat K^0=0, \hat K^1=0$ and $\tilde K^1=0$ of the curvature $K=0$ imply
that
\bea
\partial_{[A} \Omega_{B],-+}=0
\cr
\partial_{[A} \Omega_{B],\b}{}^\b=0
\cr
\partial_{[A} \hat Q_{B]}=0~,
\eea
which in turn can be solved as
\bea
\Omega_{A,-+}&=&\partial_A a
\cr
\hat Q_A&=&\partial_A b
\cr
i \Omega_{A,\b}{}^\b&=&\partial_A c~,
\eea
for some real  functions $a,b,c$.
It turns out that one can find commuting gauge $Spin(9,1)\times U(1)$ transformations to set the above components
of the $C$ connection to zero. The gauge transformations for setting $\Omega_{A,-+}=0$ and $\hat Q_A=0$ are as in the
$Spin(7)\ltimes \bR^8$ case. The transformation for setting $\Omega_{A,\b}{}^\b=0$ is
$e^{-{1\over2} c \Gamma^{16}}$. This can be verified by an explicit computation. In particular, this implies
that the Cartan subalgebra of $\bR\oplus so(4)$ generated by $\{\tau_0, I_1, J_1\}$ lies in $\hat g\subset spin(9,1)\oplus u(1)$.
 After choosing the gauge $\Omega_{A,-+}=\hat Q_A=\Omega_{A,\b}{}^\b=0$, the remaining conditions
of $K=0$ imply that
\bea
\partial_A\hat\Omega_-^r=\partial_A\tilde\Omega_-^r=0~,~~~r=2,3~,~~~A=\a, \bar\a, +~.
\la{flatcon}
\eea
We shall return to these equations later.

The spacetime form Killing spinor bilinears are generated by an one-form $\kappa=e^-$, a two form $\omega$ and the standard holomorphic $SU(4)$
invariant form $\chi$. It turns out that the null vector field $X=e_+$ associated with $\kappa$ is parallel. This implies that
it is Killing, self-parallel and the associated geodesic congruence is rotation free.
So the spacetime metric can be written locally in Penrose coordinates as in the $Spin(7)\ltimes \bR^8$ case. In particular, we have
\bea
ds^2=2 dv(du+\a dv+\b_I dy^I)+\gamma_{IJ} dy^I dy^J~,
\la{pensu}
\eea
where $X={\partial\over\partial u}$, $v$ is the Hamilton-Jacobi function of the null geodesic congruence
and  $\a=\a(v, y)$, $\b_I=\b_I(v, y)$ and $\gamma_{IJ}=\gamma_{IJ}(v,y)$.
We introduce the frame
\bea
e^-=dv~,~~~e^+=du+\a dv+\b_I dy^I~,~~~ e^\a=e^\a{}_I dy^I~,~~~e^{\bar\a}=e^{\bar\a}{}_I dy^I~.
\la{fpensu}
\eea
The torsion free condition of the Levi-Civita connection in this frame reveals the additional conditions
\bea
\Omega_{+,-\a}=0~,~~~\Omega_{+,-\bar\a}=0~,~~~\Omega_{+, \a\bar\b}=0~.
\eea

Next we compute the covariant derivatives of the two-form $\omega$ and the holomorphic (4,0) form $\chi$ to find
that they vanish apart from the components
\bea
&&\nabla_A \omega_{B-}=\Omega_{A,-}{}^C \omega_{BC}~,
\cr
&&\nabla_A \chi_{B_1B_2B_3-}=\Omega_{A,-}{}^C \chi_{B_1B_2B_3C}~,~~~A=-, \a, \bar\a~,~~~
B, B_1, B_2, B_3=\a, \bar\a ~.
\eea
This in particular reveals that the submanifold of the spacetime given
by $v, u={\rm const}$ is an eight-dimensional Calabi-Yau manifold.
The spacetime is  a one-parameter family of Calabi-Yau manifolds equipped with a line bundle with fibre coordinate $u$ and connection
$\a dv+\b_I dy^I$.
Conversely, given an one-parameter family of Calabi-Yau manifolds equipped with a line bundle, one can write a metric
as in (\ref{pensu}). The Killing spinor equations impose some additional conditions on the way that the Calabi-Yau metrics
depend on the deformation parameter $v$. These conditions arise from the requirement that $\Omega_{-, AB}$ takes values
in $su(4)$ and can be written as
\bea
{1\over2}(d\beta)_{\a\g}+\partial_v e_{I[\a} e^I{}_{\g]}=0~,~~~(d\beta)_{\a\bar\g} \delta^{\a\bar\g}
-e^I{}_{\a} \partial_v e^{\a}{}_{I}+e^I{}_{\bar\a} \partial_v e^{\bar \a}{}_{I}=0~.
\eea
One solution of these conditions is to take the metric $\gamma$ to be independent of $v$ and $d\beta$ to take values in $su(4)$.

The conditions on fluxes of the background required by supersymmetry are already written in $SU(4)$ representations. So we shall
not repeat the formulae.
It remains to write the conditions (\ref{flatcon}) in the frame (\ref{fpensu}). Indeed these can be written
as
\bea
\partial_I F_{-\a_1\a_2\a_3\a_4} \epsilon^{\a_1\a_2\a_3\a_4}=0~,~~~\partial_I G_{-\b}{}^\b=0~,
\cr
\partial_uF_{-\a_1\a_2\a_3\a_4} \epsilon^{\a_1\a_2\a_3\a_4}=0~,~~~\partial_uG_{-\b}{}^\b=0~,
\la{flusuf}
\eea
i.e. they depend only on the coordinate $v$.

\subsection{Fluxes}

Apart from the conditions on the fluxes that we have derived in (\ref{flusuf}), the Killing spinor equations imply
that all components of $P_A$ vanish apart from $P_-$ which remains unconstrained. Similarly all the components of
$G$ vanish apart from $G_{-\a\bar\b}$. The trace of this component is constrained as in (\ref{flusuf}).
  The F-fluxes satisfy the conditions
\bea
F_{+ \alpha {\bar{\beta}}_1 {\bar{\beta}}_2 {\bar{\beta}}_3} =0~,~~~
F_{-+ \beta_1 \beta_2 \beta_3} &=&0~,
\cr
F_{- \beta_1 \beta_2 \gamma}{}^\gamma =0~,~~~
F_{+\alpha {\bar{\beta}} \gamma}{}^\gamma &=&0
\cr
F_{-+ \alpha  {\bar{\beta}}_1 {\bar{\beta}}_2 } =0~,~~~F_{-\a}{}^\a{}_\b{}^\b&=&2 Q_-~,
\eea
in addition to (\ref{flusuf}) and to the self-duality condition. These conditions  can
be easily derived from  (\ref{xindep1}).
In addition that last condition follows from the gauge $\hat Q_A=0$.

\newsection{Maximally supersymmetric $G_2$-backgrounds}

\subsection{Conditions on the geometry and fluxes}

The solution of the linear system for maximally supersymmetric $G_2$-backgrounds has been given in appendix E.
The conditions on the geometry of such backgrounds are
\bea
(f^{-1} \partial_M f)_{11} = (f^{-1} \partial_M f)_{33}=-(f^{-1} \partial_M f)_{22}=-(f^{-1} \partial_M f)_{44}~,&&
\cr
(f^{-1} \partial_M f)_{13}=-(f^{-1} \partial_M f)_{31}=(f^{-1} \partial_M f)_{24}=-(f^{-1} \partial_M f)_{42}~,&&
\cr
(f^{-1} \partial_M f)_{23}=(f^{-1} \partial_M f)_{32}=(f^{-1} \partial_M f)_{14}=(f^{-1} \partial_M f)_{41}&=&0~,
\cr
(f^{-1} \partial_M f)_{12}=(f^{-1} \partial_M f)_{34}=\Omega_{M,+\bar{1}}~,&&
\cr
(f^{-1} \partial_M f)_{21}=(f^{-1} \partial_M f)_{43}=\Omega_{M,1-}~,&&
\cr
(f^{-1} \partial_M f)_{11} +{1 \over 2} \Omega_{M,-+}&=&0~,
\cr
2 (f^{-1} \partial_M f)_{13}-Q_M &=& 0~,
\la{parcg2}
\eea
 and
\bea
\Omega_{M,1q}-{1 \over 2} \Omega_{M, {\bar{r}}_1 {\bar{r}}_2} \epsilon^{{\bar{r}}_1 {\bar{r}}_2}{}_q &=&0
\cr
\Omega_{M,\bar{1}q}+{1 \over 2} \Omega_{M, {\bar{r}}_1 {\bar{r}}_2} \epsilon^{{\bar{r}}_1 {\bar{r}}_2}{}_q &=&0
\cr
\Omega_{M,+q} &=&0
\cr
\Omega_{M,-q} &=&0
\cr
\Omega_{M,q}{}^q &=&0
\cr
\Omega_{M,1 \bar{1}} &=&0
\cr
\Omega_{M,-1}-\Omega_{M,-\bar{1}} &=&0
\cr
\Omega_{M,+1}-\Omega_{M,+\bar{1}} &=&0
\eea
for $M=+,-,1,\bar{1},p,\bar{p}$.
Moreover, it turns out that the fluxes $P,G$ and $F$ vanish.

\subsection{The geometry of spacetime}

As in the previous cases, we focus on the conditions involving the the functions $f$. It is straightforward to see that these equations
can be written as a parallel transport equation
\bea
 f^{-1}\partial_A f+ C_A=0
\la{paralgt}
\eea
 of a connection $C$, where
 \bea
 C_A={1\over2}\Omega_{A,-+} t_0+{1\over2} Q_A t_1+  \Omega_{A,+1} t_2+ \Omega_{A,-1} t_3~.
 \eea
 The connection $C$ is real
 and the $4\times 4$ matrices $t_0, t_1, t_2$ and $t_3$ are easily computed from the conditions of the geometry listed in the previous
 section. The commutators of these matrices are
 \bea
 [t_1, t_0]=[t_1, t_2]=[t_1, t_3]=0~,
 \cr
 [t_0, t_2]=2 t_2~,~~~[t_0, t_3]=-2 t_3~,~~~[t_2, t_3]=-t_0~.
 \eea
Therefore $C$ is  an $s=\bR\oplus sl(2, \bR)$ connection. As we have already mentioned, the
existence of a solution for  (\ref{paralgt}) requires that the
curvature $K=dC-C\wedge C$ vanishes. In addition,  one can choose a gauge using $Spin(9,1)\times U(1)$ transformations
 such that  $f=1$. To see this, first observe that the part of the solution
 along the central generator $t_1$ can be gauged away using a $U(1)$ gauge transformation
 as in the $Spin(7)\ltimes \bR^8$ and $SU(4)\ltimes \bR^8$ investigated previously.
 The remaining $SL(2,\bR)$  part of the solution can be gauged away with a $SL(2,\bR)\subset Spin(9,1)$ transformation.
 The Lie algebra  $sl(2,\bR)\subset spin(9,1)$ is spanned by the elements $\Gamma^{05}, \Gamma^{01}, \Gamma^{15}$.
 Therefore in this case $s\subset \hat g$.

 In the gauge that $f=1$, the geometric conditions are very simple. All the components of the Levi-Civita connection
 vanish apart from the traceless part of $\Omega_{M, p\bar q}$, $\Omega_{M,1q}$, $\Omega_{M, {\bar{r}}_1 {\bar{r}}_2}$
 and $\Omega_{M,\bar{1}q}$. In addition, the last three   satisfy the conditions
 \bea
 \Omega_{M,1q}-{1 \over 2} \Omega_{M, {\bar{r}}_1 {\bar{r}}_2} \epsilon^{{\bar{r}}_1 {\bar{r}}_2}{}_q &=&0
\cr
\Omega_{M,\bar{1}q}+\Omega_{M,1q}&=&0~.
\la{fincon}
\eea
To interpret these conditions, it is helpful to find the spacetime forms constructed as Killing spinor
bilinears. In the gauge we have chosen above, the Killing spinors are $\eta_1, \eta_2, i\eta_1, i\eta_2$. Therefore
the linearly independent spacetime form bilinears are generated by $\eta_1$ and $\eta_2$.  A straightforward computation using the
results of appendix  \ref{spinorforms} reveals that the spacetime forms are generated by three one-forms
$\kappa= e^-$, $\hat\kappa$ and $\kappa'=e^+$, and  the $G_2$-invariant three-form $\phi$ given in (\ref{gtthreeform}).
The second condition in (\ref{fincon}) together with $\Omega_{A,1\bar1}=0$, $\Omega_{M, +1}=0$ and $\Omega_{M,-1}=0$
imply that $\Omega_{A, 1B}+\Omega_{A, \bar 1 B}=0$, i.e. the connection vanishes along the $e^1$ direction.
The supersymmetry conditions also imply that $\Omega_{A,+B}=\Omega_{A,-B}=0$. From these conditions one concludes
that the one-forms $e^+, e^-, e^1$ are parallel. In addition the first condition in (\ref{fincon})
implies that the three-form $\phi$ is parallel with respect to the Levi-Civita connection.
These imply that the spacetime is the
product $M=\bR^{1,2}\times B$, where $\bR^{1,2}$ is the three-dimensional Minkowski space and $B$ is a manifold
with holonomy $G_2$ as  may have been expected because the fluxes vanish.

\newsection{Concluding Remarks}

There are three classes of  IIB supergravity backgrounds with one supersymmetry characterized by the
stability subgroup of the Killing spinor in $Spin(9,1)\times U(1)$. These are $Spin(7)\ltimes \bR^8$,
$SU(4)\ltimes \bR^8$ and $G_2$. The Killing spinor equations for the first two have been solved in \cite{gju}.
In this paper, we have solved the Killing spinor equations of IIB supergravity for backgrounds with one
$G_2$-invariant Killing spinor. We have found that such backgrounds admit a timelike Killing vector field,
two spacelike one-forms twisted with respect to line bundle of IIB scalar fields,
and a $G_2$-invariant three-form. As expected the spacetime admits a $G_2$-structure which we have specified
by computing the covariant derivative of the above forms.

We have also investigated backgrounds with extended supersymmetry. In particular we focused on backgrounds
that admit Killing spinors which are invariant under a subgroup $H$ of $Spin(9,1)\times U(1)$. We have
shown that in IIB supergravity the Killing spinor equations of backgrounds with a maximal number of $H$-invariant spinors
factorize, i.e.  the $P,G$  and $F$ fluxes are separated in the Killing spinor equations. The resulting
equations are straightforward to solve. In particular, we have solved the Killing spinor equations
of backgrounds with two supersymmetries and $Spin(7)\ltimes \bR^8$-invariant spinors, and
of backgrounds with four  supersymmetries and $SU(4)\ltimes \bR^8$- and $G_2$-invariant spinors.
In both the $Spin(7)\ltimes \bR^8$ and $SU(4)\ltimes \bR^8$ cases, the spacetime metric can be written
in Penrose coordinates and the space transverse to the null geodesic congruence is either a $Spin(7)$
or a $G_2$ manifold, respectively. In addition the {\it holonomy of the Levi-Civita connection} in both
cases is contained in $Spin(7)\ltimes \bR^8$ and in $SU(4)\ltimes \bR^8$, respectively.
 In the $G_2$ case, the spacetime is the product
$\bR^{1,2}\times B$, where $B$ is a $G_2$ manifold.

It appears that in many cases of interest, i.e. the cases for which the spinors have a non-trivial
stability subgroup in $Spin(9,1)\times U(1)$, the IIB Killing spinor equations are tractable.
In particular, one can easily solve the Killing spinor equations for any  background
that exhibits maximal supersymmetry with $H$-invariant spinors for some $H\subset Spin(9,1)$. These include backgrounds
with eight and sixteen supersymmetries. Another class of backgrounds are those that admit $N$
$H$-invariant Killing spinors but $N$ is smaller than the number of $H$-invariant spinors
in  the (complex) Weyl representation of $Spin(9,1)$. Many examples of such backgrounds are already known in the literature,
e.g. \cite{klebanov, volkov, nunez, aatgp, minasian, cvetic, warner}. To illustrate this further ,
there are four $SU(4)\ltimes \bR^8$-invariant spinors in the Weyl representation of $Spin(9,1)$.
So far the Killing spinor equations have been
solved for one such Killing spinor and for all four Killing spinor. However, two more cases remain
to be tackled, one with two and the other three Killing spinors.
For such backgrounds,
the Killing spinor equations do not factorize. So it is expected that these backgrounds will exhibit
a more involved structure. Nevertheless, the  machinery is in place  to investigate
 all  supersymmetric backgrounds.

\section*{Acknowledgements}

The work of U.G.~is funded by the
Swedish Research Council. This work has been partially supported by the
PPARC grant PPA/G/O/2002/00475. J.G. thanks EPSRC for support.

\setcounter{section}{0}

\appendix{Spinors}

\subsection{Spinors from forms}

It has been known for sometime that spinors can be described in terms of forms, see
e.g.\cite{wang, lawson, harvey}. In particular, the description of $Spin(9,1)$ spinors
in terms of forms used in the context of IIB supergravity can be found in  \cite{gju}.
Here for completeness, we shall briefly summarize some of the key formulae.
For a more detail account see \cite{gju}.

Let  $U=\bR<e_1,\dots,e_5>$ be a vector space spanned by $e_1,\dots,e_5$ orthonormal vectors.
The space of Dirac $Spin(9,1)$ spinors is
$\Delta_c=\Lambda^*(U\otimes \bC)$ . This decomposes into
two complex chiral representations
according to the degree of the form $\Delta_c^+=\Lambda^{{\rm even}}(U\otimes \bC)$
and  $\Delta_c^-=\Lambda^{{\rm odd}}(U\otimes \bC)$.
These are the complex Weyl representations
of $Spin(9,1)$.
The gamma matrices are represented on $\Delta_c$ as
\bea
\Gamma_0\eta&=& -e_5\wedge\eta +e_5\lc\eta~,~~~~
\Gamma_5\eta= e_5\wedge\eta+e_5\lc \eta
\cr
\Gamma_i\eta&=& e_i\wedge \eta+ e_i\lc \eta~,~~~~~~i=1,\dots,4
\cr
\Gamma_{5+i}\eta&=& i e_i\wedge\eta-ie_i\lc\eta~,
\eea
where $\lc$ is the adjoint of $\wedge$ with respect to the auxiliary inner
product $<,>$.
The gamma matrices have been chosen such that
  $\{\Gamma_i; i=1,\dots, 9\}$  are Hermitian
and $\Gamma_0$ is anti-Hermitian with respect to the (auxiliary)
inner product
\be
<z^a e_a, w^b e_b>=\sum_{a=1}^{5}  (z^a)^* w^a~,~~~~
\ee
on $U\otimes \bC$  and then extended to $\Delta_c$,
where $(z^a)^*$ is the standard
complex conjugate\footnote{In \cite{uggp} we denote
the standard complex
of $\eta$ with $\bar\eta$ instead of $\eta^*$ that we use here.} of $z^a$.
The above gamma matrices
satisfy the Clifford algebra relations
$\Gamma_A\Gamma_B+\Gamma_B \Gamma_A=2 \eta_{AB}$ with respect to the
Lorentzian inner product as expected.
The Dirac inner product on the space of spinors $\Delta_c$ is
$D(\eta,\theta)=<\Gamma_0\eta, \theta>$.

The  $Pin (9,1)$ (Majorana) invariant inner product is
\be
B(\eta,\theta)= <B(\eta^*), \theta>~,~~~~~~~~
\ee
where  $B=\Gamma_{06789}$. Observe that $B(\eta, \theta)=-B(\theta,\eta)$.
It is well-known that $Spin(9,1)$ admits two inequivalent
Majorana-Weyl representations.
The Majorana condition on the complex Weyl
representations is imposed  by setting  the
Dirac conjugate spinor  to be equal to the Majorana conjugate one. Equivalently,
one can impose the reality condition using an anti-linear map
 which commutes with the generators
of $Spin(9,1)$ and squares to one. It turns out that it is convenient to chose as a
reality condition
\be
\eta=-\Gamma_0 B(\eta^*)~,
\ee
or equivalently
\be
\eta^*=\Gamma_{6789}\eta~.
\la{rcon}
\ee
The map $C=\Gamma_{6789}$ is also called charge conjugation matrix. The reality condition
can also be expressed as
$\eta=C *\eta= C(\eta^*)$.

The spacetime forms associated with pair of spinors $\eta,\theta$ are
\be
\alpha(\eta, \theta)={1\over k!} B(\eta,\Gamma_{A_1\dots A_k} \theta)
e^{A_1}\wedge\dots\wedge e^{A_k}~,~~~~~~~k=0,\dots, 9~.
\la{forms}
\ee
If both spinors are of the same chirality, then
it is sufficient to compute the forms up to degree $k\leq 5$.
This is because the forms with degrees  $k\geq 6$
are related to those with degrees $k\leq 5$ with a
Hodge duality operation. The forms of middle dimension
are either self-dual or anti-self-dual.

To solve the Killing spinor equations, it is convenient to use an oscillator
basis  in the space of spinors $\Delta_c$.  For this  write
\be
\Gamma_{\bar\a}= {1\over \sqrt {2}}(\Gamma_\a+i \Gamma_{\a+5})~,~~~~~~~~~
\Gamma_\pm={1\over \sqrt{2}} (\Gamma_5\pm\Gamma_0)
~,~~~~~~~~~\Gamma_{\a}= {1\over \sqrt {2}}(\Gamma_\a-i \Gamma_{\a+5})~.
\la{hbasis}
\ee
Observe that the Clifford algebra relations in the above basis are
$\Gamma_A\Gamma_B+\Gamma_B\Gamma_A=2 g_{AB}$,   where the non-vanishing
components of the metric are
$g_{\a\bar\b}=\delta_{\a\bar\b}, g_{+-}=1$. In addition, we define
$\Gamma^B=g^{BA} \Gamma_A$.
The $1$ spinor is a Clifford  vacuum, $\Gamma_{\bar\a}\,1=\Gamma_+\, 1=0$
and  the representation $\Delta_c$
can be constructed by acting on $1$ with the creation operators
$\Gamma^{\bar\a}, \Gamma^+$
or equivalently any spinor can be written as
\be
\eta= \sum_{k=0}^5 {1\over k!}~ \phi_{\bar a_1\dots \bar a_k}~
 \Gamma^{\bar a_1\dots\bar a_k} 1~,~~~~\bar a=\bar\a, +~,
 \la{hbasisa}
\ee
i.e. $\Gamma^{\bar a_1\dots\bar a_k} 1$, for $k=0,\dots,5$, is a basis in the
space of (Dirac) spinors.

\subsection{Spacetime form bilinears and $G_2$-invariant spinors} \la{spinorforms}

The $G_2$-invariant spinors are linear combinations of $1$, $e_{1234}$, $e_{15}$
and $e_{2345}$ spinors. Here we  compute
the spacetime forms associated with all the pairs of the these spinors.
In particular, we have the one-forms
\bea
\kappa(e_{1234},1)&=& e^0-e^5~,
\cr
\kappa(1, e_{2345})&=&-e^1-i e^6~,~~~\kappa(e_{1234}, e_{15})=-e^1+ie^6~,
\cr
\kappa(e_{2345}, e_{15})&=& e^0+e^5~,
\eea
the three-forms
\bea
\xi(e_{1234}, 1)&=&i (e^0-e^5)\wedge \omega~,
\cr
\xi(1, e_{15})&=& \hat\chi~,
\cr
\xi(1, e_{2345})&=&i (e^1+ie^6)\wedge \hat\omega+(e^1+ie^6)\wedge e^0\wedge e^5~,
\cr
\xi(e_{1234}, e_{15})&=&-i (e^1-ie^6)\wedge \hat\omega+(e^1-ie^6)\wedge e^0\wedge e^5~,
\cr
\xi(e_{1234}, e_{2345})&=&\hat\chi^*
\cr
\xi(e_{2345}, e_{15})&=&i (e^0+ e^5)\wedge \hat\omega- i (e^0+ e^5)\wedge e^1\wedge e^6~,
\eea
and the five-forms
\bea
\tau(1,1)&=&(e^0-e^5)\wedge \chi~,
\cr
\tau(e_{1234}, e_{1234})&=& (e^0-e^5)\wedge \chi^*~,
\cr
\tau(e_{1234}, 1)&=&- {1\over2} (e^0-e^5)\omega\wedge \omega~,
\cr
\tau(1, e_{15})&=&-\hat\chi
\wedge [e^0\wedge e^5+ i e^1\wedge e^6]~,
\cr
\tau(1, e_{2345})&=& (e^1+i e^6)\wedge [{1\over2} \hat\omega\wedge \hat\omega
-i \hat\omega\wedge e^0\wedge e^5]~,
\cr
\tau(e_{1234}, e_{15})&=& (e^1-ie^6)\wedge [{1\over2}\hat\omega\wedge \hat\omega
+i \hat\omega\wedge e^0\wedge e^5]~,
\cr
\tau(e_{1234}, e_{2345})&=&\hat\chi^*
\wedge [- e^0\wedge e^5+ i e^1\wedge e^6]~,
\cr
\tau(e_{2345}, e_{15})&=& (e^0+e^5)\wedge [-{1\over2}\hat\omega\wedge \hat\omega
+\hat\omega\wedge e^1\wedge e^6]~,
\cr
\tau(e_{15}, e_{15})&=& (e^0+e^5)\wedge (-e^1+i e^6)\wedge \hat\chi~,
\cr
\tau(e_{2345}, e_{2345})&=& -(e^0+e^5)\wedge (e^1+i e^6)\wedge
\hat\chi^*~,
\eea
where
\bea
\omega&=& e^1\wedge e^6+ e^2\wedge e^7+e^3\wedge e^8+e^4\wedge e^9~,
\cr
\chi&=&(e^1+i e^6)\wedge
(e^2+i e^7)\wedge (e^3+i e^8)\wedge (e^4+i e^9)~,
\cr
\hat\omega&=&e^2\wedge e^7+e^3\wedge e^8+e^4\wedge e^9~,
\cr
\hat\chi&=&(e^2+ie^7)\wedge (e^3+i e^8)\wedge (e^4+i e^9)~.
\eea
Observe that $\kappa(\theta, \eta)=\kappa(\eta, \theta)$, $\xi(\theta, \eta)=-\xi(\eta, \theta)$
and $\tau(\theta, \eta)=\tau(\eta, \theta)$.
We use the above expressions to compute the spacetime form bi-linears
of Killing  spinors.

\newsection{Backgrounds with a $G_2$-invariant Killing spinor}
\subsection{Preliminaries}

In this appendix, we give the details of the derivation of the conditions on the geometry and fluxes
imposed by the Killing spinor equations for backgrounds with one $G_2$-invariant Killing spinor.
The conditions we derive on the geometry have already been summarized and investigated in section two.
The calculation is performed by decomposing the geometry and fluxes in $SU(3)\subset G_2$ representations.
Because of the self-duality condition of the five-form flux $F$,
no all components of $F$ are independent. To determine the independent
components of $F$, we write the
self-duality condition in $SU(3)$ representations.
In particular, we find the following relations (up to complex conjugation):
\bea
\label{eqn:dualrela}
F_{1 \bar{1} p_1 p_2 p_3} &=& F_{-+p_1 p_2 p_3}~,
\cr
F_{1 \bar{1} p_1 p_2 \bar{q}} &=& -2 g_{\bar{q} [p_1} F_{p_2] -+r}{}^r -F_{-+p_1 p_2 \bar{q}}~,
\cr
F_{1 p_1 p_2 p_3 \bar{q}} &=&{1 \over 2} \epsilon_{p_1 p_2 p_3} \epsilon_{\bar{q}}{}^{r_1 r_2} F_{-+1r_1 r_2}~,
\cr
F_{1 p_1 p_2 \bar{q}_1 \bar{q}_2} &=&(g_{p_2 \bar{q}_1}g_{p_1 \bar{q}_2}-g_{p_1 \bar{q}_1}g_{p_2 \bar{q}_2})
F_{-+1r}{}^r
\cr
&+&g_{p_2 \bar{q}_1} F_{-+1\bar{q}_2 p_1}+g_{p_1 \bar{q}_2}F_{-+1\bar{q}_1 p_2}
-g_{p_1 \bar{q}_1} F_{-+1\bar{q}_2 p_2}-g_{p_2 \bar{q}_2}F_{-+1\bar{q}_1 p_1}~,
\cr
F_{1 p \bar{q}_1 \bar{q}_2 \bar{q}_3} &=& {1 \over 2} \epsilon_{\bar{q}_1 \bar{q}_2 \bar{q}_3}
\epsilon_p{}^{\bar{r}_1 \bar{r}_2} F_{-+1 \bar{r}_1 \bar{r}_2}~,
\cr
F_{p_1 p_2 p_3 \bar{q}_1 \bar{q}_2} &=& - \epsilon_{p_1 p_2 p_3}
\epsilon_{\bar{q}_1 \bar{q}_2}{}^\ell F_{-+1 \bar{1} \ell}~,
\eea
and
\bea
\label{eqn:dualrelb}
F_{+p_1 p_2 p_3 \bar{q}} &=& -{1 \over 2} \epsilon_{p_1 p_2 p_3} \epsilon_{\bar{q}}{}^{r_1 r_2} F_{+1 \bar{1} r_1 r_2}~,
\cr
F_{+p_1 p_2 \bar{q}_1 \bar{q}_2} &=& (g_{p_1 \bar{q}_1} g_{p_2 \bar{q}_2}-g_{p_2 \bar{q}_1} g_{p_1 \bar{q}_2})
F_{+1 \bar{1} r}{}^r
\cr
&+& g_{p_2 \bar{q}_1} F_{+1 \bar{1} p_1 \bar{q}_2} + g_{p_1 \bar{q}_2} F_{+1 \bar{1} p_2 \bar{q}_1}
-g_{p_1 \bar{q}_1} F_{+1 \bar{1} p_2 \bar{q}_2} - g_{p_2 \bar{q}_2} F_{+1 \bar{1} p_1 \bar{q}_1}~,
\cr
F_{+1p_1 p_2 p_3}&=&0~,
\cr
F_{+1 p r}{}^r &=&0~,
\cr
F_{+1p \bar{q}_1 \bar{q}_2} &=&  g_{p [\bar{q}_1} F_{\bar{q}_2 ]+1r}{}^r~,
\eea
and
\bea
\label{eqn:dualrelc}
F_{-p_1 p_2 p_3 \bar{q}} &=& {1 \over 2} \epsilon_{p_1 p_2 p_3} \epsilon_{\bar{q}}{}^{r_1 r_2} F_{-1 \bar{1} r_1 r_2}~,
\cr
F_{-p_1 p_2 \bar{q}_1 \bar{q}_2} &=& -(g_{p_1 \bar{q}_1} g_{p_2 \bar{q}_2}-g_{p_2 \bar{q}_1} g_{p_1 \bar{q}_2})
F_{-1 \bar{1} r}{}^r
\cr
&-& g_{p_2 \bar{q}_1} F_{-1 \bar{1} p_1 \bar{q}_2} - g_{p_1 \bar{q}_2} F_{-1 \bar{1} p_2 \bar{q}_1}
+g_{p_1 \bar{q}_1} F_{-1 \bar{1} p_2 \bar{q}_2} + g_{p_2 \bar{q}_2} F_{-1 \bar{1} p_1 \bar{q}_1}~,
\cr
F_{-1 p_1 p_2 \bar{q}} &=& - g_{\bar{q} [p_1}F_{p_2] -1r}{}^r~,
\cr
F_{-1 \bar{q} r}{}^r &=&0~,
\cr
F_{-1 \bar{q}_1 \bar{q}_2 \bar{q}_3} &=&0~.
\eea

The components $F_{+1 \bar{q}_1 \bar{q}_2 \bar{q}_3}$ and $F_{-1 p_1 p_2 p_3}$
(and their complex conjugates)
are not constrained by the duality. Our strategy is to simplify the linear system obtained from the
Killing spinor equations by first re-writing terms of the type $F_{1 \bar{1} p_1 p_2 p_3}$,
$F_{1 \bar{1} p_1 p_2 \bar{q}}$, $F_{1 p_1 p_2 p_3 \bar{q}}$, $F_{1 p_1 p_2 \bar{q}_1 \bar{q}_2}$,
$F_{1 p \bar{q}_1 \bar{q}_2 \bar{q}_3}$ and $F_{p_1 p_2 p_3 \bar{q}_1 \bar{q}_2}$, and their
complex conjugates, which appear on the
left-hand-side of ({\ref{eqn:dualrela}}) in terms of $F_{-+p_1 p_2 p_3}$, $F_{-+p_1 p_2 \bar{q}}$,
$F_{-+1 p_1 p_2}$, $F_{-+1 p \bar{q}}$ and $F_{-+1 \bar{q}_1 \bar{q}_2}$ and their complex
conjugates.
Similarly, we use the equations ({\ref{eqn:dualrelb}}) and ({\ref{eqn:dualrelc}})
to substitute for the fluxes that appear in the left-hand-side into the linear system.

 \subsection{The linear system}

Setting $\epsilon= f (1+e_{1234}) +(ig/\sqrt{2}) \Gamma^+ (e_1 +
e_{234})$ into the Killing spinor equation (\ref{kseqnb}), where $f$ and $g$ are real functions,
and using the basis in the space of spinor given in appendix A,
 we obtain from the algebraic Killing spinor equations the following linear system

\bea
 f( P_{\bar{p}} +{1 \over 4} G_{-+\bar{p}}
+{1 \over 4} G_{\bar{p}q}{}^q+{1 \over 4} G_{\bar{p}1\bar 1} -{1
\over 4} \epsilon_{\bar{p}}{}^{qr} G_{qr1})+\frac{i g}{2}(G_{\bar p
1-}-\frac{1}{2}G_{qr-}\ep^{qr}{}_{\bar p}) =0~,
\la{y4y}
\eea
\bea
 f( P_{\bar{1}} +{1 \over 4} G_{-+\bar{1}}
+{1 \over 4} G_{\bar{1}q}{}^q +{1 \over 2} G_{234})+ig(-P_- +{1\over
4}G_{-p}{}^p-{1\over 4}G_{-1\bar 1})=0~,
\la{y5y}
\eea
\bea
f( P_p +{1 \over 4}  G_{-+p} -{1 \over 4} G_{p q}{}^q-{1 \over 4}
G_{p 1\bar 1} -{1 \over 4}  \epsilon_p{}^{{\bar{q}} {\bar{r}}}
G_{{\bar{q}} {\bar{r}}\bar{1}})+{i g\over 2}(G_{p\bar 1
-}-\frac{1}{2}G_{\bar q\bar r -}\ep^{\bar q\bar r}{}_{p} )  =0~,
\la{y6y}
\eea
\bea
f( P_1 +{1 \over 4}  G_{-+1} -{1 \over 4} G_{1 p}{}^p +{1 \over 2}
G_{{\bar{2}} {\bar{3}}
{\bar{4}}})-ig(P_-+\frac{1}{4}G_{-p}{}^p-\frac{1}{4}G_{-1\bar
1})=0~,
\la{y7y}
\eea
\bea
f( P_+ +{1 \over 4}  G_{+ p}{}^p+{1 \over 4}  G_{+ 1\bar
1})+ig(P_1-{1\over 4}G_{1p}{}^p- {1\over
4}G_{1+-}+\frac{1}{2}G_{234}) =0~,
\la{y8y}
\eea
\bea
f( P_+ -{1 \over 4}  G_{+ p}{}^p-{1 \over 4}  G_{+ 1\bar
1})+ig(P_{\bar 1}+\frac{1}{4}G_{\bar 1 p}{}^p-\frac{1}{4}G_{\bar 1
+-} +\frac{1}{2}G_{\bar 2\bar 3\bar 4}) =0~,
\la{y9y}
\eea
\be
f( G_{+{\bar{p}} {\bar{q}}}+ \epsilon_{{\bar{p}} {\bar{q}}}{}^{r}
G_{+ r 1})-igG_{\bar p\bar q
1}+2ig(P_r+\frac{1}{4}G_{rs}{}^s-\frac{1}{4}G_{r1\bar
1}-\frac{1}{4}G_{r+-})\ep^r{}_{\bar p \bar q}=0~,
\la{y10y}
\ee
and
\be
f( G_{+{\bar{1}} {\bar{p}}}-{1 \over 2} \epsilon_{ {\bar{p}}}{}^{q
r} G_{+ q r})+2ig(-P_{\bar p}+\frac{1}{4}G_{\bar p
q}{}^q-\frac{1}{4}G_{\bar p 1\bar 1}+\frac{1}{4}G_{\bar p
+-}+\frac{1}{4}G_{qr\bar 1}\ep^{qr}{}_{\bar p})=0~.
\la{y11y}
\ee

Next we turn into the Killing spinor equation associated with the
supercovariant derivative (\ref{kseqna}). In particular the linear equations associated
with
${\cal D}_p\epsilon=0$  are

\bea
\label{x12x}
D_pf+f({1\over2}\Omega_{p,q}{}^q+{1\over2}\Omega_{p,1\bar1}+
{1\over2} \Omega_{p,-+} +i F_{p-+q}{}^q +i
F_{p-+1\bar1}
\cr
+{1\over4} G_{pq}{}^q +{1\over4} G_{p1\bar1}+{1\over4} G_{-+p})
+ig(\Om_{p,1-}+iF_{p1-q}{}^q -\frac{1}{2}G_{p1-}) =0~,
\eea

\bea
\label{x13x}
 f (\Omega_{p,\bar q_1\bar q_2}-2i g_{p [{\bar{q}}_1} F_{{\bar{q}_2}]-+r}{}^r
-2i g_{p [{\bar{q}}_1} F_{{\bar{q}_2}]-+1 \bar{1}}
+2i F_{p -+\bar q_1\bar q_2}
 + {1\over2} G_{p\bar q_1\bar q_2}
-{1\over4} g_{p[\bar q_1} G_{\bar q_2]r}{}^r
\cr
 -{1\over4} g_{p[\bar
q_1} G_{\bar q_2]1\bar1} -{1\over4} g_{p[\bar q_1} G_{\bar q_2]-+} -
\Omega_{p,1 r} \epsilon^{r}{}_{\bar q_1\bar q_2}
 -{1\over4} G_{p 1 r} \epsilon^{r}{}_{\bar q_1\bar q_2})
 \cr
 +ig(2iF_{p \bar q_1 \bar q_21-}+\frac{1}{2}g_{p[\bar q_1}G_{\bar q_2]1-}
 +(\Om_{p,r-}
 -\frac{1}{4}G_{pr-})\ep^r{}_{\bar q_1\bar q_2})=0~,
 \eea

\bea
\label{x14x}
 f( \Omega_{p,\bar1\bar q}+iF_{-+ \bar{1}r}{}^r g_{p \bar{q}}
  +2i F_{p -+\bar1\bar q}+
 {1\over2} G_{p\bar1\bar q} +{1\over8}
g_{p\bar q} G_{\bar1r}{}^r +{1\over8} g_{p\bar q} G_{\bar1-+}
\cr
- {1\over2} \Omega_{p,r_1 r_2}\epsilon^{r_1 r_2}{}_{\bar q}-{1\over4} g_{p\bar
q} G_{234})
+ig(-\Om_{p,\bar q-} -iF_{-r}{}^r{}_{1 \bar{1}} g_{p \bar{q}}
+2iF_{p\bar q 1\bar1-}
\cr
+\frac{1}{2}G_{p\bar q-}-\frac{1}{8}g_{p\bar
q}G_{-r}{}^r
+\frac{1}{8}g_{p\bar q}G_{1\bar1-})=0~,
 \eea

 \bea
 \label{x15x}
D_pf+f (  -{1\over2} \Omega_{p,q}{}^q-{1\over2}
\Omega_{p,1\bar1}+{1\over2} \Omega_{p,-+}
\cr
-{1\over8} G_{pq}{}^q-{1\over8} G_{p1\bar1}+{1\over8} G_{-+p}
+i F_{-+ \bar{1} \bar{q}_1 \bar{q}_2} \epsilon^{\bar{q}_1 \bar{q}_2}{}_p
 +{1\over8}
 \epsilon_{p}{}^{\bar q_1\bar q_2}
  G_{\bar1\bar q_1\bar q_2})
\cr
+ig(-i F_{- 1 \bar{1} {\bar{q}}_1 {\bar{q}}_2} \epsilon^{{\bar{q}}_1 {\bar{q}}_2}{}_p
-\frac{1}{8} G_{\bar q_1\bar
q_2-}\ep_{p}{}^{\bar q_1\bar
q_2}+\Om_{p,\bar1-}-\frac{1}{4}G_{p\bar1-})=0~,
\eea

\bea
 \label{x16x}
f({1\over2} \Omega_{p,+\bar q}+{i\over2} g_{p \bar{q}}F_{+1 \bar{1} r}{}^r
+i F_{p+\bar q 1\bar1} +{1\over16}g_{p\bar q}
G_{+r}{}^r+{1\over16}g_{p\bar q} G_{+1\bar1}- {1\over4} G_{+p\bar
q})
\cr
+ig(\frac{1}{2}\Om_{p,\bar
q1}+iF_{p\bar
q+-1}+{i \over 2} F_{-+1r}{}^r g_{p \bar{q}}-\frac{1}{4}G_{p\bar q1}
\cr
+\frac{1}{16}g_{p\bar q}G_{1r}{}^r+\frac{1}{16}g_{p\bar
q}G_{1+-}-\frac{1}{4}\Om_{p,r_1r_2}\ep^{r_1r_2}{}_{\bar
q}+\frac{1}{8}g_{p\bar q}G_{234})=0~,
\eea

\bea
 \label{x17x}
f({1\over2} \Omega_{p,+\bar1}+{i\over2} F_{p+\bar1q}{}^q  -
{1\over4} G_{+p\bar1} )+\frac{i}{2}D_{p}g+\frac{ig}{2}(-\frac{1}{2}
\Om_{p,1\bar1}+\frac{1}{2}\Om_{p,+-}+\frac{1}{2}\Om_{p,q}{}^q
\cr
-iF_{-+pr}{}^r
-iF_{p+-1\bar1}-\frac{1}{4}G_{pq}{}^q+\frac{1}{4}G_{p1\bar1}-\frac{1}{4}G_{p+-})
 =0~,
\eea

\bea
 \label{x18x}
f({i \over 2} \epsilon_p{}^{\bar{q}_1 \bar{q}_2} F_{+1 \bar{1} \bar{q}_1 \bar{q}_2}
+{1\over16} \ep_{p}{}^{\bar q_1 \bar
q_2} G_{\bar q_1\bar q_2+} + {1\over2} \Omega_{p,+1}
- {1\over 8}  G_{+p1})+\frac{i}{2}D_{p}g
\cr
+ig({i \over 2}\epsilon_p{}^{\bar{q}_1 \bar{q}_2}F_{-+1 \bar{q}_1 \bar{q}_2}
+{1\over16} \ep_{p}{}^{\bar q_1 \bar q_2} G_{\bar q_1\bar q_2 1}
+\frac{1}{2}(-\frac{1}{2}\Om_{p,r}{}^r+\frac{1}{2}\Om_{p,1\bar1}
\cr
+\frac{1}{2}\Om_{p,+-}
+\frac{1}{8}G_{pq}{}^q-\frac{1}{8}G_{p1\bar1}-\frac{1}{8}G_{p+-}))=0~,
\eea

and

\bea
 \label{x19x}
f(i F_{p+\bar1\bar q_1\bar q_2} +{1\over4} g_{p[\bar q_1} G_{\bar
q_2]\bar1+} - {1\over2} \Omega_{p,+r}  \epsilon^r{}_{\bar q_1\bar
q_2}
+ {1\over 8} G_{+pr} \epsilon^r{}_{\bar q_1\bar
q_2})
\cr
+ig(\frac{1}{2}\Om_{p,\bar q_1\bar q_2} -iF_{-+p \bar{q}_1 \bar{q}_2}
-i F_{-+r}{}^r{}_{[\bar{q}_1}g_{\bar{q}_2]p}
+i F_{-+ 1 \bar{1} [\bar{q}_1}g_{\bar{q}_2]p}
-\frac{1}{4}G_{p\bar
q_1\bar q_2}
\cr
+\frac{1}{8}g_{p [\bar q_1}G_{\bar q_2]r}{}^r-\frac{1}{8}g_{p [\bar
q_1}G_{\bar q_2]1\bar1}+\frac{1}{8}g_{p [\bar q_1}G_{\bar q_2]+-}
-(\frac{1}{2}\Om_{p,r\bar1} -\frac{1}{8}G_{pr\bar1}
)\ep^r{}_{\bar q_1\bar q_2} )=0~.
\eea

The linear equations associated with ${\cal D}_1\epsilon=0$ are

\bea
 \label{x20x}
D_1f+f( {1\over2}\Omega_{1,p}{}^p+{1\over2}\Omega_{1,1\bar1}+
{1\over2} \Omega_{1,-+} +i F_{1-+p}{}^p
\cr
+{1\over4} G_{1p}{}^p +{1\over4} G_{-+1})
+ig(\Om_{1,1-}-2iF_{1234-})=0~,
\eea

\bea
 \label{x21x}
 f (\Omega_{1,\bar q_1\bar q_2}
  +2i F_{1 -+\bar q_1\bar q_2} +
 {1\over2} G_{1\bar q_1\bar q_2}
-  \Omega_{1,1 r}
 \epsilon^{r}{}_{\bar q_1\bar q_2})
 \cr
 +ig(\Om_{1,r-}-iF_{1r-s}{}^s
  -\frac{1}{2}G_{1r-})\ep^r{}_{\bar
q_1\bar q_2}=0~,
 \eea

\bea
 \label{x22x}
 f (\Omega_{1,\bar1\bar q}-i F_{-+\bar{q}r}{}^r
  +i F_{1 -+\bar1\bar q}
-{1\over8} G_{\bar qr}{}^r+{3\over8} G_{\bar q1\bar1} -{1\over8}
G_{\bar q-+}
\cr
- ({1\over2} \Omega_{1,r_1 r_2}
 +{1\over8} G_{1 r_1 r_2} )
 \epsilon^{r_1 r_2}{}_{\bar q})
\cr
+ig(-\Om_{1,\bar q-}+\frac{1}{4}G_{1\bar
q-}-(iF_{1r_1r_2\bar1-}+\frac{1}{8}G_{r_1
r_2-})\ep^{r_1r_2}{}_{\bar q})=0~,
 \eea

 \bea
  \label{x23x}
D_1f+f (  -{1\over2} \Omega_{1,p}{}^p -{1\over2}
\Omega_{1,1\bar1}+{1\over2} \Omega_{1,-+}  -{1\over8}
G_{1p}{}^p+{1\over8} G_{-+1}
\cr
-2iF_{-+\bar{2} \bar{3} \bar{4}} -{1\over4}
G_{\bar2\bar3\bar4})+ig(\Om_{1,\bar1-}
-iF_{1\bar1-q}{}^q-\frac{1}{8}G_{-p}{}^p-\frac{3}{8}G_{1\bar1-})=0~,
\eea

\bea
 \label{x24x}
f({1\over2} \Omega_{1,+\bar p}+{i\over2} F_{1+\bar pq}{}^q -
{1\over4} G_{+1\bar p})
\cr
+ig(\frac{1}{2}\Om_{1,\bar
p1}
-(\frac{1}{4}\Om_{1,r_1r_2}
+\frac{i}{2}F_{1r_1r_2+-}-\frac{1}{8}G_{1r_1r_2}
)\ep^{r_1r_2}{}_{\bar p})=0~,
\eea

\bea
 \label{x25x}
f({1\over2} \Omega_{1,+\bar1}+{i\over2} F_{1+\bar1q}{}^q
+{1\over16}G_{+q}{}^q-{3\over16} G_{+1\bar1}
\cr
+\frac{i}{2}D_{1}g+\frac{ig}{2}(-\frac{1}{2}\Om_{1,1\bar1}+\frac{1}{2}\Om_{1,+-}+\frac{1}{2}\Om_{1,q}{}^q
\cr
-\frac{1}{8}G_{1q}{}^q-\frac{1}{8}G_{1+-}-2iF_{-+234}-\frac{1}{4}G_{234})
=0~,
\eea

\bea
 \label{x26x}
f(i F_{1+\bar2\bar3\bar4}  + {1\over2} \Omega_{1,+1}
)+\frac{i}{2}D_{1}g+\frac{ig}{2}(-\frac{1}{2}\Om_{1,r}{}^r+\frac{1}{2}\Om_{1,1\bar1}
\cr
+\frac{1}{2}\Om_{1,+-} -iF_{1+-q}{}^q
+\frac{1}{4}G_{1q}{}^q-\frac{1}{4}G_{1+-})=0~,
\eea

and

\bea
 \label{x27x}
f(i F_{1+\bar1\bar q_1\bar q_2} +{1\over8}G_{\bar q_1\bar q_2+} -
{1\over2} \Omega_{1,+r} \epsilon^r{}_{\bar
q_1\bar q_2} +{1\over 8}  G_{+1r} \epsilon^r{}_{\bar q_1\bar
q_2})
\cr
+ig(\frac{1}{2}\Om_{1,\bar q_1\bar q_2} -\frac{1}{8}G_{1\bar
q_1\bar q_2}-(\frac{1}{2}\Om_{1,r\bar1} +\frac{i}{2}F_{-+rs}{}^s
\cr
+\frac{i}{2}F_{1r\bar1+-}-\frac{3}{16}G_{1r\bar1}+\frac{1}{16}G_{rs}{}^s-\frac{1}{16}G_{r+-}
)\ep^r{}_{\bar q_1\bar q_2})=0~.
\eea

The linear equations associated with ${\cal D}_{\bar p}\epsilon=0$ are

\bea
 \label{x28x}
D_{\bar p}f+f({1\over2} \Omega_{\bar p,q}{}^q+{1\over2} \Omega_{\bar
p,1\bar1}+{1\over2} \Omega_{\bar p, -+}
\cr
 +{1\over8} (G_{\bar pq}{}^q+G_{\bar
p1\bar1}+ G_{\bar p-+}) +iF_{-+1q_1 q_2} \epsilon^{q_1 q_2}{}_{\bar{p}}
 +{1\over8} \epsilon_{\bar p}{}^{q_1 q_2} G_{1 q_1 q_2})
\cr
+ig(\Om_{\bar p,1-}-\frac{1}{4}G_{\bar
p1-}+i F_{-1 \bar{1} q_1 q_2} \epsilon^{q_1 q_2}{}_{\bar{p}}-\frac{1}{8}\ep_{\bar p}{}^{q_1
q_2}G_{q_1q_2-})=0~,
\eea

\bea
 \label{x29x}
f(\Omega_{\bar p, \bar q_1\bar q_2}
+{1\over4} G_{\bar 2\bar 3\bar 4}\ep_{\bar p\bar q_1\bar q_2} -(\Omega_{\bar
p,1r}
\cr
-i F_{-+1 s}{}^s g_{r \bar{p}}+2i F_{\bar p-+1r})
\epsilon^{r}{}_{\bar q_1\bar q_2}
 -(-{1\over8} g_{\bar pr}
G_{1s}{}^s +{1\over2}G_{\bar p1r}+{1\over8} g_{\bar pr} G_{1-+})
\epsilon^{r}{}_{\bar q_1\bar q_2})
\cr
+ig(\Om_{\bar
p,r-}+i g_{r \bar{p}}F_{-s}{}^s{}_{1 \bar{1}}+2iF_{\bar pr1\bar1
-}-\frac{1}{8}g_{\bar p r}G_{-s}{}^s
-\frac{1}{2}G_{\bar
pr-}+\frac{1}{8}g_{\bar p r}G_{1\bar1-})\ep^r{}_{\bar q_1\bar
q_2}=0~,
\eea

\bea
 \label{x30x}
f(\Omega_{\bar p, \bar1\bar q}+{1\over4} G_{\bar p\bar1\bar q}
-({1\over2}\Omega_{\bar p,r_1 r_2}
-i F_{-+s}{}^s{}_{[r_1} g_{r_2] \bar{p}} -i F_{-+1 \bar{1} [r_1}g_{r_2] {\bar{p}}}
+i F_{\bar
p-+r_1 r_2}) \epsilon^{r_1 r_2}{}_{\bar q}
\cr
-({1\over8} g_{\bar pr_1} G_{r_2s}{}^s +{1\over8} g_{\bar pr_1}
G_{r_21\bar1} +{1\over4}G_{\bar pr_1r_2}-{1\over8} g_{\bar p r_1}
G_{r_2-+}) \epsilon^{r_1r_2}{}_{\bar q})
\cr
+ig(-\Om_{\bar p,\bar q-}+\frac{1}{4}G_{\bar p\bar q-}
-(iF_{\bar
pr_1r_2\bar1-}+\frac{1}{4}g_{\bar p r_1}G_{r_2
\bar1-})\ep^{r_1r_2}{}_{\bar q})=0~,
\eea

\bea
 \label{x31x}
D_{\bar p}f+f(-{1\over2} \Omega_{\bar p,q}{}^q-{1\over2}
\Omega_{\bar p,1\bar1}+{1\over2} \Omega_{\bar p,-+} -i F_{\bar p-+q}{}^q
\cr
-i F_{\bar p-+1\bar1} -{1\over4} G_{\bar pq}{}^q-{1\over4}
G_{\bar p1\bar1}+{1\over4} G_{\bar p-+})+ig(\Om_{\bar
p,\bar1-}-iF_{\bar p\bar1-q}{}^q-\frac{1}{2}G_{\bar
p\bar1-})=0~,
\eea

\bea
 \label{x32x}
f({1\over2} \Omega_{\bar p,+\bar q} - {1\over8} G_{+\bar p\bar
q}
-{i\over2} F_{\bar p+1 r_1 r_2} \epsilon^{r_1 r_2}{}_{\bar q}
-{1\over8}  G_{1r+} \epsilon^{r}{}_{\bar p\bar
q})+ig(\frac{1}{2}\Om_{\bar p,\bar q1}-\frac{1}{8}G_{\bar p\bar
q1}
\cr
-(\frac{1}{4}\Om_{\bar p,r_1r_2} +{i \over 2}F_{-+s}{}^s{}_{[r_1}g_{r_2] \bar{p}}
-{i \over 2} F_{-+1 \bar{1} [r_1}g_{r_2] \bar{p}}
-\frac{i}{2} F_{-+ \bar{p} r_1 r_2}
-\frac{1}{8}G_{\bar pr_1r_2}-\frac{1}{16}g_{\bar p
r_1}G_{r_2s}{}^s
\cr
+\frac{1}{16}g_{\bar p r_1}G_{r_21\bar1}+\frac{1}{16}g_{\bar p
r_1}G_{r_2+-} )\ep^{r_1r_2}{}_{\bar q})=0~,
\eea

\bea
 \label{x33x}
f({1\over2} \Omega_{\bar p,+\bar1}-{1\over8} G_{+\bar p\bar1}
-{i \over 2}\epsilon_{\bar{p}}{}^{q_1 q_2} F_{+1 \bar{1}q_1 q_2}
+{1\over16}  G_{q_1q_2+} \epsilon^{q_1q_2}{}_{\bar p})
\cr
+\frac{i}{2}D_{\bar p}g+\frac{ig}{2}(-\frac{1}{2}\Om_{\bar
p,1\bar1}+\frac{1}{2}\Om_{\bar p,+-}+\frac{1}{2}\Om_{\bar p,q}{}^q
\cr
-\frac{1}{8}G_{\bar pq}{}^q+\frac{1}{8}G_{\bar
p1\bar1}-\frac{1}{8}G_{\bar p+-}
+i \epsilon_{\bar{p}}{}^{q_1 q_2}F_{-+\bar{1}q_1 q_2}
+\frac{1}{8}\ep_{\bar p}{}^{q_1q_2}G_{q_1q_2\bar1})=0~,
\eea

\bea
 \label{x34x}
f({1\over2} \Omega_{\bar p,+1}
-{i\over2} F_{\bar p+1q}{}^q +{1\over4} G_{\bar p+1})
\cr
+\frac{i}{2}D_{\bar p}g+\frac{ig}{2}(-\frac{1}{2}\Om_{\bar
p,r}{}^r+\frac{1}{2}\Om_{\bar p,1\bar1} +\frac{1}{2}\Om_{\bar p,+-}
+iF_{-+ \bar{p} r}{}^r
\cr
+iF_{\bar
p+-1\bar1} +\frac{1}{4}G_{\bar pq}{}^q-\frac{1}{4}G_{\bar
p1\bar1}-\frac{1}{4}G_{\bar p+-})=0~,
\eea

and

\bea
 \label{x35x}
f(-{1\over2} \Omega_{\bar p,+r}
-{i\over2}g_{r \bar{p}} F_{+1 \bar{1} q}{}^q
+iF_{\bar p+r1\bar1}
\cr
+{1\over16} g_{\bar pr} G_{+s}{}^s +{1\over16} g_{\bar pr}
G_{+1\bar1}-{1\over4} G_{\bar p+r}) \epsilon^r{}_{\bar q_1\bar q_2}
+ig(\frac{1}{2}\Om_{\bar p,\bar q_1\bar q_2}
\cr
-\frac{1}{8}G_{\bar 2\bar
3\bar 4}\ep_{\bar p\bar q_1\bar q_2} -(\frac{1}{2}\Om_{\bar p,r\bar1} +iF_{\bar
pr\bar1+-}-{i \over 2}F_{-+\bar{1} s}{}^s g_{r \bar{p}}
\cr
-\frac{1}{4}G_{\bar pr\bar1} -\frac{1}{16}g_{\bar p
r}G_{\bar1s}{}^s+\frac{1}{16}g_{\bar p r}G_{\bar1+-})\ep^r{}_{\bar
q_1\bar q_2})=0~.
\eea

The linear equations associated with ${\cal D}_{\bar 1}\epsilon=0$ are

\bea
 \label{x36x}
D_{\bar1}f+f({1\over2} \Omega_{\bar1,p}{}^p+{1\over2}
\Omega_{\bar1,1\bar1}+{1\over2} \Omega_{\bar1, -+}
+{1\over8} (G_{\bar1p}{}^p+ G_{\bar1-+}) -2iF_{-+234}
-{1\over4}  G_{234})
\cr
+ig(\Om_{\bar1,1-}+iF_{\bar11-q}{}^q
+\frac{1}{8}G_{-q}{}^q-\frac{3}{8}G_{\bar11-})=0~,
\eea

\bea
 \label{x37x}
f(\Omega_{\bar1, \bar q_1\bar q_2}+{1\over4} G_{\bar1\bar q_1\bar q_2}
-2({1\over2}\Omega_{\bar1,1r} +{i \over 2}F_{-+rs}{}^s
\cr
+{i\over2} F_{\bar1-+1r}) \epsilon^{r}{}_{\bar q_1\bar q_2}
-({1\over8} G_{rs}{}^s -{3\over8} G_{r1\bar1}-{1\over8}G_{r-+}) \epsilon^{r}{}_{\bar q_1\bar q_2})
\cr
+ig(2iF_{\bar1 \bar q_1 \bar q_2 1-}+\frac{1}{4}G_{\bar q_1 \bar
q_2-}+(\Om_{\bar1,r-}
-\frac{1}{4}G_{\bar1r-})\ep^r{}_{\bar q_1\bar q_2})=0~,
\eea

\bea
 \label{x38x}
f(\Omega_{\bar1, \bar1\bar p}-({1\over2}\Omega_{\bar1,q_1 q_2}
+i F_{\bar1-+q_1 q_2}
+{1\over4}G_{\bar1 q_1 q_2})\epsilon^{q_1 q_2}{}_{\bar p})
\cr
+ig(-\Om_{\bar1,\bar p-}-iF_{\bar1\bar
p-r}{}^r+\frac{1}{2}G_{\bar1\bar p-})=0~,
\eea

\bea
 \label{x39x}
D_{\bar1}f+f(-{1\over2} \Omega_{\bar1,p}{}^p-{1\over2}
\Omega_{\bar1,1\bar1}+{1\over2} \Omega_{\bar1,-+} -i F_{\bar1-+p}{}^p
-{1\over4} G_{\bar1p}{}^p+{1\over4}
G_{\bar1-+})
\cr
+ig(-2iF_{\bar1\bar2\bar3\bar4-}+\Om_{\bar1
,\bar1-})=0~,
\eea

\bea
 \label{x40x}
f({1\over2} \Omega_{\bar1,+\bar p} -{1\over8} G_{+\bar1\bar p} -({i\over2}
F_{\bar1+1q_1q_2}+{1\over16}  G_{q_1q_2+}) \epsilon^{q_1q_2}{}_{\bar
p})
\cr
+ig(\frac{1}{2}\Om_{\bar1,\bar p1}
-{i \over 2} F_{-+ \bar{p}r}{}^r
+\frac{i}{2}F_{\bar1\bar p+-1}-\frac{3}{16}G_{\bar p1\bar1}
-\frac{1}{16}G_{\bar pr}{}^r-\frac{1}{16}G_{\bar
p+-}
\cr
-(\frac{1}{4}\Om_{\bar1,r_1r_2}-\frac{1}{16}G_{\bar1r_1r_2}
)\ep^{r_1r_2}{}_{\bar p})=0~,
\eea

\bea
 \label{x41x}
f({1\over2} \Omega_{\bar1,+\bar1}+i F_{\bar1+234})+\frac{i}{2}
D_{\bar 1}g+\frac{ig}{2}(-\frac{1}{2}\Om_{\bar1
,1\bar1}+\frac{1}{2}\Om_{\bar1,+-}+\frac{1}{2}\Om_{\bar1,q}{}^q
\cr
+iF_{\bar1+-q}{}^q
-\frac{1}{4}G_{\bar1q}{}^q-\frac{1}{4}G_{\bar1+-})=0~,
\eea

\bea
 \label{x42x}
f({1\over2} \Omega_{\bar1,+1}
-{i\over2} F_{\bar1+1p}{}^p -{1\over16}  G_{+p}{}^p+{3\over16}
G_{+1\bar1})
\cr
+\frac{i}{2}D_{\bar1}g+ig(-iF_{-+\bar{2} \bar{3} \bar{4}}-\frac{1}{8}G_{\bar2\bar3\bar4
}+\frac{1}{2}(-\frac{1}{2}\Om_{\bar1,r}{}^r+\frac{1}{2}\Om_{\bar1,1\bar1}
\cr
+\frac{1}{2}\Om_{\bar1,+-}
+\frac{1}{8}G_{\bar1q}{}^q-\frac{1}{8}G_{\bar1+-}))=0~,
\eea

and

\bea
 \label{x43x}
-f({1\over2} \Omega_{\bar1,+r} -{i\over2} F_{\bar1+rs}{}^s
+{1\over4} G_{\bar1+r})\epsilon^r{}_{\bar q_1\bar q_2}
\cr
+ig(\frac{1}{2}\Om_{\bar1,\bar q_1\bar q_2}
+iF_{\bar1\bar q_1\bar q_2+-}-\frac{1}{4}G_{\bar1\bar
q_1\bar q_2}-\frac{1}{2}\Om_{\bar1,r \bar{1}}\ep^r{}_{\bar q_1\bar q_2})=0~.
\eea

The linear equations associated with ${\cal D}_-\epsilon=0$ are

\bea
 \label{x44x}
D_-f+f({1\over2} \Omega_{-,p}{}^p+{1\over2}
\Omega_{-,1\bar1}+{1\over2} \Omega_{-,-+} +i F_{-p}{}^p{}_{1\bar1}
\cr
+{1\over4} G_{-p}{}^p +{1\over4} G_{-1\bar1} +2i  F_{-1234})
+ig\Om_{-,1-}=0~,
\eea

\bea
 \label{x45x}
f(\Omega_{-,\bar q_1\bar q_2} +2i F_{-\bar q_1\bar q_21\bar1} +{1\over2} G_{-\bar q_1\bar q_2}
\cr
 - (\Omega_{-,1r} -i F_{-1rs}{}^s) \epsilon^{r}{}_{\bar q_1\bar q_2}
-{1\over2}  G_{-1r} \epsilon^{r}{}_{\bar q_1\bar q_2})
+ig\Om_{-,r-}\ep^{r}{}_{\bar q_1\bar q_2}=0~,
\eea

\bea
 \label{x46x}
f(\Omega_{-,\bar1\bar p}+i F_{-\bar1\bar pq}{}^q +{1\over2}
G_{-\bar1\bar p} - ({1\over2} \Omega_{-,q_1 q_2} -iF_{-q_1 q_21\bar1}) \epsilon^{q_1 q_2}{}_{\bar
p}
\cr
-{1\over4}  G_{-q_1q_2} \epsilon^{q_1 q_2}{}_{\bar p})-ig\Om_{-,\bar
p-}=0~,
\eea

\bea
 \label{x47x}
D_-f+f(-{1\over2} \Omega_{-,p}{}^p-{1\over2}
\Omega_{-,1\bar1}+{1\over2}\Omega_{-,-+}+i F_{-p}{}^p{}_{1\bar1}
\cr
-{1\over4}  G_{-p}{}^p -{1\over4}  G_{-1\bar1}+2i
F_{-\bar1\bar2\bar3\bar4})+ig\Om_{-,\bar1-}=0~,
\eea

\bea
 \label{x48x}
f({1\over2} \Omega_{-,+\bar p}+{i\over2} F_{-+\bar pq}{}^q
+{i\over2} F_{-+\bar p1\bar1} -{1\over16} G_{\bar pq}{}^q-{1\over16}
G_{\bar p1\bar1}+{3\over16} G_{-+\bar p}
\cr
-{i\over2}  F_{-+1q_1q_2} \epsilon^{q_1q_2}{}_{\bar p} +{1\over 16}
G_{1q_1q_2} \epsilon^{q_1q_2}{}_{\bar p})+ig(\frac{1}{2}\Om_{-,\bar
p1}-\frac{1}{8}G_{-\bar p1}
\cr
-(\frac{1}{4}\Om_{-,r_1r_2}-\frac{1}{16}G_{-r_1r_2}
)\ep^{r_1r_2}{}_{\bar p})=0~,
\eea

\bea
 \label{x49x}
f({1\over2} \Omega_{-,+\bar1}+{i\over2} F_{-+\bar1q}{}^q
-{1\over16} G_{\bar1q}{}^q+{3\over16} G_{-+\bar1} +i  F_{-+234}
-{1\over 8}
G_{234})
\cr
+\frac{i}{2}D_{-}g+\frac{ig}{2}(-\frac{1}{2}\Om_{-,1\bar1}+\frac{1}{2}\Om_{-,+-}+\frac{1}{2}\Om_{-,q}{}^q
-\frac{1}{8}G_{-q}{}^q+\frac{1}{8}G_{- 1\bar1})=0~,
\eea

\bea
\label{x50x}
f(i F_{-+\bar2\bar3\bar4}  -{1\over8} G_{\bar2\bar3\bar4} +
{1\over2} \Omega_{-,+1}-{i\over2} F_{-+1r}{}^r
\cr
+ {1\over16} G_{1r}{}^r +{3\over16} G_{-+1}
)+\frac{i}{2}D_{-}g+ig(-\frac{1}{4}\Om_{-,r}{}^r+\frac{1}{4}\Om_{-,1\bar1}
\cr
+\frac{1}{4}\Om_{-,+-}
+\frac{1}{16}G_{-q}{}^q-\frac{1}{16}G_{-1\bar1})=0~,
\eea

and

\bea
 \label{x51x}
f(i F_{-+\bar{1}\bar q_1\bar q_2}  -{1\over8} G_{\bar{1} \bar q_1\bar q_2} -
({1\over2} \Omega_{-,+r}-{i\over2} F_{-+rs}{}^s-{i\over2}
F_{-+r1\bar1}) \epsilon^r{}_{\bar q_1\bar q_2}
\cr
- ({1\over16} G_{rs}{}^s +{1\over16} G_{r1\bar1}+{3\over16} G_{-+r})
\epsilon^r{}_{\bar q_1\bar q_2}) +ig(\frac{1}{2}\Om_{-,\bar q_1\bar
q_2}
\cr
-\frac{1}{8}G_{-\bar q_1\bar
q_2}-(\frac{1}{2}\Om_{-,r\bar1}-\frac{1}{8}G_{-r\bar1} )\ep^r{}_{\bar q_1\bar q_2})=0~.
\eea

The linear equations associated with ${\cal D}_+\epsilon=0$ are

\bea
 \label{x52x}
D_+f+f({1\over2} \Omega_{+,p}{}^p+{1\over2} \Omega_{+,1\bar1}
+{1\over2} \Omega_{+,-+}
+{1\over8} G_{+p}{}^p+{1\over8} G_{+1\bar1})
\cr
+ig(\Om_{+,1-}+iF_{+1-q}{}^q-\frac{1}{8}G_{1q}{}^q
+\frac{3}{8}G_{+-1}-2iF_{+234-}+\frac{1}{4}G_{234})=0~,
\eea

\bea
 \label{x53x}
f(\Omega_{+,\bar q_1\bar q_2}
 +{1\over4} G_{+\bar q_1\bar q_2}
-\Omega_{+,1r}\epsilon^{r}{}_{\bar q_1\bar q_2}
-{1\over4}  G_{+1r} \epsilon^{r}{}_{\bar q_1\bar q_2})
\cr
+ig(2iF_{+\bar
q_1 \bar q_2 1-}-\frac{1}{4}G_{\bar q_1 \bar q_2
1}+(\Om_{+,r-}-iF_{+r-s}{}^s
\cr
+iF_{+r1\bar1
-}-\frac{3}{8}G_{+r-}+\frac{1}{8}G_{rs}{}^s-\frac{1}{8}G_{r1\bar1})\ep^r{}_{\bar
q_1\bar q_2})=0~,
\eea

\bea
 \label{x54x}
f(\Omega_{+,\bar1\bar p}
 +{1\over4} G_{+\bar1\bar p}
-{1\over2} \Omega_{+,q_1 q_2} \epsilon^{q_1 q_2}{}_{\bar p}
-{1\over8}  G_{+q_1q_2} \epsilon^{q_1q_2}{}_{\bar p})
\cr
+ig(-\Om_{+,\bar
p-}-iF_{+\bar p-r}{}^r+iF_{+\bar p1\bar1-}+\frac{3}{8}G_{+\bar
p-}
\cr
+\frac{1}{8}G_{\bar pr}{}^r-\frac{1}{8}G_{\bar
p1\bar1}-(iF_{+r_1r_2\bar1-}-\frac{1}{8}G_{r_1
r_2\bar1})\ep^{r_1r_2}{}_{\bar q})=0~,
\eea

\bea
 \label{x55x}
D_+f+f(-{1\over2} \Omega_{+,p}{}^p-{1\over2}
\Omega_{+,1\bar1}+{1\over2} \Omega_{+,-+} -{1\over8}
G_{+p}{}^p
-{1\over8} G_{+1\bar1})
\cr
+ig(-2iF_{+\bar2\bar3\bar4-}+\frac{1}{4}
G_{\bar2\bar3\bar4}+\Om_{+,\bar1-}
-iF_{+\bar1-q}{}^q+\frac{1}{8}G_{\bar1 p}{}^p-\frac{3}{8}G_{+\bar1-})=0~,
 \eea

 \bea
  \label{x56x}
 \frac{f}{2}\Omega_{+,+\bar p}+ig(\frac{1}{2}\Om_{+,\bar
p1}+\frac{i}{2}F_{+\bar p1r}{}^r-\frac{1}{4}G_{+\bar p1}
\cr
-(\frac{1}{4}\Om_{+,r_1r_2}+\frac{i}{2}F_{+r_1r_21\bar1
}-\frac{1}{8}G_{+r_1r_2}
)\ep^{r_1r_2}{}_{\bar p})=0~,
 \eea

 \bea
  \label{x57x}
 \frac{f}{2}\Omega_{+,+\bar1}+\frac{i}{2}D_{+}g +\frac{ig}{2}(-\frac{1}{2}\Om_{+,1\bar1}
 +\frac{1}{2}\Om_{+,+-}+\frac{1}{2}\Om_{+,q}{}^q
\cr
-iF_{+ 1\bar1q}{}^q -\frac{1}{4}G_{+q}{}^q+\frac{1}{4}G_{+
1\bar1}+2iF_{+234\bar1})=0~,
 \eea

 \bea
  \label{x58x}
\frac{f}{2}\Omega_{+,+1}+\frac{i}{2}D_{+}g+ig(iF_{+\bar2\bar3\bar4
1}+\frac{1}{2}(-\frac{1}{2}\Om_{+,r}{}^r+\frac{1}{2}\Om_{+,1\bar1}
\cr
+\frac{1}{2}\Om_{+,+-}-iF_{+1\bar1q}{}^q
+\frac{1}{4}G_{+q}{}^q-\frac{1}{4}G_{+1\bar1}))=0~,
 \eea

and

\bea
 \label{x59x}
-\frac{f}{2}\Omega_{+,+r}\ep^{r}{}_{\bar q_1\bar
q_2}+ig(\frac{1}{2}\Om_{+,\bar q_1\bar q_2} -iF_{+\bar q_1\bar
q_21\bar1}-\frac{1}{4}G_{+\bar q_1\bar
q_2}
\cr
-(\frac{1}{2}\Om_{+,r\bar1} -\frac{i}{2}F_{+r\bar1
}{}_s{}^s-\frac{1}{4}G_{+r\bar1} )\ep^r{}_{\bar q_1\bar q_2})=0~.
 \eea

All the above linear equations  make the linear system associated with the
Killing spinor equations of IIB supergravity for backgrounds with one $G_2$-invariant Killing spinor.
The task is to solve this linear system
for the fluxes and to determine the conditions on the spacetime geometry.

\subsection{The solution to the linear system}

To proceed, we shall obtain equations from which the $G$-flux has been eliminated,
and then simplify these further to obtain constraints on the geometry.

In particular, take (\ref{x57x})+(\ref{x58x}) to obtain
\beq
\label{con1x}
{1 \over 2} f (\Omega_{+,+\bar{1}}+ \Omega_{+,+1})+iD_+ g +ig(-iF_{+1 \bar{1} q}{}^q
+i F_{+234 \bar{1}}+iF_{+ \bar{2} \bar{3} \bar{4}1}+{1 \over 2}\Omega_{+,+-}) =0~.
\eeq

{}From the imaginary part of this equation, we find

\beq
\label{con2x}
\partial_+ g +{1 \over 2}g \Omega_{+,+-}=0~.
\eeq

Similarly, take (\ref{x44x})+(\ref{x47x}) to obtain

\beq
\label{con3x}
2 D_- f +f (\Omega_{-,-+}+2iF_{-1 \bar{1} p}{}^p+2iF_{-1234}+2iF_{- \bar{1} \bar{2} \bar{3} \bar{4}})
+ig(\Omega_{-, \bar{1} -}+ \Omega_{-, 1 -})=0~.
\eeq

{}From the real part of this equation, we find

\beq
\label{con4x}
\partial_- f +{1 \over 2}f \Omega_{-,-+}=0~.
\eeq

Next, take $-ig$[(\ref{x49x})+(\ref{x50x})]$-{f \over 2}$[(\ref{x52x})+(\ref{x55x})] to obtain

\beq
\label{con5x}
g D_- g - f D_+ f + {g^2 \over 2} \Omega_{-,+-}-{f^2 \over 2} \Omega_{+,-+}
-{i \over 2}fg(\Omega_{-,+1}+\Omega_{-,+\bar{1}}+\Omega_{+,1-}+\Omega_{+,\bar{1}-})=0~.
\eeq

{}From the real part of this equation, we find

\beq
\label{con6x}
g \partial_- g -f \partial_+ f + {g^2 \over 2} \Omega_{-,+-}-{f^2 \over 2} \Omega_{+,-+} =0~.
\eeq

Consider $-2ig$[(\ref{x25x})+(\ref{x26x})+${1 \over 2}$(\ref{x52x})]+$2f$(\ref{x58x}) to obtain

\bea
\label{con7x}
2gD_1g +2g^2({1 \over 2} \Omega_{1,+-}+{1 \over 2}\Omega_{+1-}+iF_{-+1q}{}^q
-2iF_{-+234}) +f^2 \Omega_{+,+1}
\cr
-ig \partial_+ f +if \partial_+ g +ifg(\Omega_{1+\bar{1}}+\Omega_{1,+1}-\Omega_{+,p}{}^p
+\Omega_{+,+-})=0~.
\eea

Also take the complex conjugate of
$-2ig$[(\ref{x41x})+(\ref{x42x})+${1 \over 2}$(\ref{x55x})]+$2f$(\ref{x57x})
to obtain

\bea
\label{con8x}
2g {\bar{D}}_1 g + 2g^2 ({1 \over 2} \Omega_{1,+-}+{1 \over 2} \Omega_{+,1-} -iF_{-+1q}{}^q
+2iF_{-+234}) +f^2 \Omega_{+,+1}
\cr
-if \partial_+g +ig \partial_+ f
-ifg(\Omega_{1,+\bar{1}}+\Omega_{1,+1}-\Omega_{+,q}{}^q+\Omega_{+,+-})=0~.
\eea

Considering the sum of (\ref{con7x}) and (\ref{con8x}), we find

\beq
\label{con9x}
2g \partial_1 g -g^2 (\Omega_{1,-+}+\Omega_{+,-1})+f^2 \Omega_{+,+1}=0~.
\eeq

Take $f$[(\ref{x20x})+(\ref{x23x})-$2$(\ref{x50x})]+$ig$(\ref{x44x})
to find

\bea
\label{con10x}
2fD_1 f +f^2 (\Omega_{1,-+}-\Omega_{-,+1}+2iF_{-+1p}{}^p -4iF_{-+\bar{2} \bar{3} \bar{4}})
-g^2 \Omega_{-,1-}
\cr
-if \partial_- g +ig \partial_- f +ifg(\Omega_{1,1-}+\Omega_{1,\bar{1}-}+\Omega_{-,p}{}^p
+\Omega_{-,-+})=0~.
\eea

Next consider the complex conjugate of $f$[(\ref{x36x})+(\ref{x39x})-$2$(\ref{x49x})]+$ig$(\ref{x47x})
to find

\bea
\label{con11x}
2f{\bar{D}}_1 f +f^2 (\Omega_{1,-+}-\Omega_{-,+1}-2iF_{-+1p}{}^p +4iF_{-+\bar{2} \bar{3} \bar{4}})
-g^2 \Omega_{-,1-}
\cr
+if \partial_- g -ig \partial_- f -ifg(\Omega_{1,1-}+\Omega_{1,\bar{1}-}+\Omega_{-,p}{}^p
+\Omega_{-,-+})=0~.
\eea

Take the sum of (\ref{con10x}) and (\ref{con11x}) to obtain

\beq
\label{con12x}
2f \partial_1 f +f^2(\Omega_{1,-+}-\Omega_{-,+1})-g^2 \Omega_{-,1-}=0~.
\eeq

Next take the sum (\ref{x17x})+(\ref{x18x}),  consider the dual of (\ref{x53x})
and use this to eliminate the $G_{pq}{}^q$ dependence. Then take the
dual of (\ref{x59x}) to eliminate the remaining $G$-flux terms to obtain

\bea
\label{con13x}
2gD_p g +g^2 (\Omega_{p,+-}+\Omega_{+,p-}+2iF_{-+1 \bar{1} p}
-2i F_{-+pq}{}^q +2iF_{-+1\bar{q}_1 \bar{q}_2} \epsilon^{\bar{q}_1 \bar{q}_2}{}_p)
\cr
+f^2 \Omega_{+,+p}-ifg(\Omega_{p,+\bar{1}}+\Omega_{p+1}
-\Omega_{+1p}-\Omega_{+p \bar{1}}+\Omega_{+ \bar{q}_1 \bar{q}_2}\epsilon^{\bar{q}_1 \bar{q}_2}{}_p)=0~.
\eea

Also take the complex conjugate of $-2ig$[(\ref{x33x})+(\ref{x34x})-${1 \over 2}$(\ref{x54x})]+2f(\ref{x56x})
to get

\bea
\label{con14x}
2g{\bar{D}}_p g +g^2 (\Omega_{p,+-}+\Omega_{+,p-} -2i F_{-+1 {\bar{q}}_1 {\bar{q}}_2}
\epsilon^{{\bar{q}}_1 {\bar{q}}_2}{}_p
+2iF_{-+pq}{}^q -2iF_{-+1 \bar{1}p})
\cr
+f^2 \Omega_{+,+p}+ifg(\Omega_{p,+1}+\Omega_{p,+\bar{1}}-\Omega_{+,1p}-\Omega_{+,p\bar{1}}
+\Omega_{+,{\bar{q}}_1 {\bar{q}}_2}\epsilon^{{\bar{q}}_1 {\bar{q}}_2}{}_p)=0~.
\eea

Next taking the sum of ({\ref{con13x}}) and ({\ref{con14x}}), we find

\beq
\label{con15x}
2g \partial_p g +g^2(\Omega_{p,+-}+\Omega_{+,p-})+f^2 \Omega_{+,+p}=0~.
\eeq

Consider also the sum (\ref{x12x})+(\ref{x15x}). Use the dual of (\ref{x51x})
to eliminate the $G_{pq}{}^q$ dependence and then use the
dual of (\ref{x45x}) to remove the remaining $G$-flux terms. This leaves

\bea
\label{con16x}
2f D_p f + f^2(\Omega_{p,-+}-\Omega_{-,+p}+2i F_{-+\bar{1} {\bar{q}}_1 {\bar{q}}_2}\epsilon^{{\bar{q}}_1 {\bar{q}}_2}{}_p
+2iF_{-+pq}{}^q +2iF_{-+1 \bar{1}p})
\cr
-g^2 \Omega_{-,p-}+ifg(\Omega_{p,1-}+\Omega_{p,\bar{1}-}-\Omega_{-,p \bar{1}}-\Omega_{-,1p}
+ \Omega_{-, {\bar{q}}_1 {\bar{q}}_2} \epsilon^{{\bar{q}}_1 {\bar{q}}_2}{}_p)=0~.
\eea

Also take the complex conjugate of $f$[(\ref{x28x})+(\ref{x31x})-$2$(\ref{x48x})]-$ig$(\ref{x46x})
to get

\bea
\label{con17x}
2f {\bar{D}}_p f + f^2(\Omega_{p,-+}-\Omega_{-,+p}-2iF_{-+pq}{}^q-2iF_{-+p1 \bar{1}}
-2i F_{-+\bar{1} {\bar{q}}_1 {\bar{q}}_2} \epsilon^{{\bar{q}}_1 {\bar{q}}_2}{}_p)
\cr
-g^2 \Omega_{-,p-}-ifg(\Omega_{p, \bar{1}-}+\Omega_{p,1-}-\Omega_{-,p\bar{1}}-\Omega_{-,1p}
+ \Omega_{-, {\bar{q}}_1 {\bar{q}}_2} \epsilon^{{\bar{q}}_1 {\bar{q}}_2}{}_p)=0~,
\eea

Taking the sum of (\ref{con16x}) and (\ref{con17x}), we find

\beq
\label{con18x}
2f \partial_p f + f^2(\Omega_{p,-+}-\Omega_{-,+p}) -g^2 \Omega_{-,p-}=0~.
\eeq

Next consider the complex conjugate of $f$(\ref{x41x})-${ig \over 2}$(\ref{x39x})
to get

\bea
\label{con19x}
f^2 ({1 \over 2} \Omega_{1,+1}+iF_{+1 \bar{2} \bar{3} \bar{4}})+
g^2(-{1 \over 2} \Omega_{1,-1}+iF_{-1234})
\cr
-{i \over 2}(f \partial_1 g -g \partial_1 f) -{i \over 2}fg (-\Omega_{1,q}{}^q+\Omega_{1,+-})=0~.
\eea

Also take $f$(\ref{x26x})-${ig \over 2}$(\ref{x20x}) to get

\bea
\label{con20x}
f^2({1 \over 2} \Omega_{1,+1}-iF_{+1 \bar{2} \bar{3} \bar{4}})+g^2(-{1 \over 2} \Omega_{1,-1}-iF_{-1234})
\cr
+{i \over 2}(f \partial_1 g - g \partial_1 f) +{i \over 2} fg(-\Omega_{1,q}{}^q + \Omega_{1,+-})=0~.
\eea

Taking the sum of (\ref{con19x}) and (\ref{con20x}), we find

\beq
\label{con21x}
f^2 \Omega_{1,+1}-g^2 \Omega_{1,-1}=0~.
\eeq

Next take $ig$[(\ref{x23x})+(\ref{x36x})]-$2f$[(\ref{x25x})+(\ref{x42x})] to obtain

\bea
\label{con22x}
ig(\partial_1 f +\partial_{\bar{1}} f)-if(\partial_1 g + \partial_{\bar{1}} g)
-g^2 (\Omega_{1, \bar{1} -}+ \Omega_{\bar{1},1-}-2iF_{-1 \bar{1} q}{}^q)
\cr
-f^2 (\Omega_{1,+\bar{1}}+\Omega_{\bar{1},+1}-2iF_{+1 \bar{1} q}{}^q)
+ifg(\Omega_{\bar{1},p}{}^p- \Omega_{1,p}{}^p + \Omega_{1,-+}+\Omega_{\bar{1},-+})=0~.
\eea

{}From the real part of this expression we get

\beq
\label{con23x}
g^2 (\Omega_{1,-\bar{1}}+\Omega_{\bar{1},-1})-f^2(\Omega_{1,+\bar{1}}+\Omega_{\bar{1},+1})=0~.
\eeq

Next take $ig$(\ref{x12x})-$2f$(\ref{x18x}). By taking the duals of
(\ref{x21x}) and (\ref{x27x}), one can eliminate the $G$-flux dependence from
this combination to obtain

\bea
\label{con24x}
i (g \partial_p f - f \partial_p g) -g^2(\Omega_{p,1-}+\Omega_{1,p-}-2iF_{-1pq}{}^q)
\cr
-f^2(\Omega_{p,+1}+\Omega_{1,+p}+2iF_{+1 \bar{1} {\bar{q}}_1 {\bar{q}}_2} \epsilon^{{\bar{q}}_1 {\bar{q}}_2}{}_p)
\cr
+ifg(\Omega_{p,q}{}^q + \Omega_{p,-+}+\Omega_{1,{\bar{q}}_1 {\bar{q}}_2} \epsilon^{{\bar{q}}_1 {\bar{q}}_2}{}_p
-\Omega_{1,1p}-\Omega_{1,p\bar{1}})=0~.
\eea

Also take the complex conjugate of
$ig$[(\ref{x31x})-(\ref{x38x})]-$2f$[(\ref{x33x})+(\ref{x40x})] to obtain

\bea
\label{con25x}
i (f \partial_p g - g \partial_p f)-g^2(\Omega_{p,1-}+\Omega_{1,p-}+2iF_{-1pq}{}^q)
\cr
-f^2 (\Omega_{p,+1}+\Omega_{1,+p}-2iF_{+1 \bar{1} {\bar{q}}_1 {\bar{q}}_2} \epsilon^{{\bar{q}}_1 {\bar{q}}_2}{}_p)
\cr
-ifg (\Omega_{p,q}{}^q + \Omega_{p,-+}-\Omega_{1,p \bar{1}}- \Omega_{1,1p}+ \Omega_{1, {\bar{q}}_1 {\bar{q}}_2}
\epsilon^{{\bar{q}}_1 {\bar{q}}_2}{}_p) =0~.
\eea

Considering the sum of (\ref{con24x}) and (\ref{con25x}) we get

\beq
\label{con26x}
g^2 ( \Omega_{p,-1}+\Omega_{1,-p})-f^2(\Omega_{p,+1}+\Omega_{1,+p})=0~.
\eeq

Next take $ig$(\ref{x15x})-$2f$(\ref{x17x}). By taking the dual of
(\ref{x37x}) and (\ref{x43x}) one can remove the $G$-flux dependence from this
combination to obtain

\bea
\label{con27x}
i(g \partial_p f - f \partial_p g) -g^2(\Omega_{p, \bar{1}-}+\Omega_{\bar{1},p-}
-2iF_{-1 \bar{1} {\bar{q}}_1 {\bar{q}}_2}  \epsilon^{{\bar{q}}_1 {\bar{q}}_2}{}_p)
\cr
-f^2(\Omega_{p,+\bar{1}}+\Omega_{\bar{1},+p}+2iF_{+ \bar{1} pq}{}^q)
\cr
+ifg(-\Omega_{p,q}{}^q + \Omega_{p,-+}-\Omega_{\bar{1},1p}-\Omega_{\bar{1},p \bar{1}}
+\Omega_{\bar{1} {\bar{q}}_1 {\bar{q}}_2}  \epsilon^{{\bar{q}}_1 {\bar{q}}_2}{}_p)=0~.
\eea

Also take the complex conjugate of $ig$[(\ref{x28x})-(\ref{x22x})]-$2f$[(\ref{x34x})+(\ref{x24x})]
to get

\bea
\label{con28x}
-i(g \partial_p f - f \partial_p g) -g^2(\Omega_{p, \bar{1}-}+\Omega_{\bar{1},p-}
+2iF_{-1 \bar{1} {\bar{q}}_1 {\bar{q}}_2}  \epsilon^{{\bar{q}}_1 {\bar{q}}_2}{}_p)
\cr
-f^2(\Omega_{p,+\bar{1}}+\Omega_{\bar{1},+p}-2iF_{+ \bar{1} pq}{}^q)
\cr
-ifg(-\Omega_{p,q}{}^q + \Omega_{p,-+}-\Omega_{\bar{1},1p}-\Omega_{\bar{1},p\bar{1}}+
\Omega_{\bar{1} {\bar{q}}_1 {\bar{q}}_2}  \epsilon^{{\bar{q}}_1 {\bar{q}}_2}{}_p)=0~.
\eea

Taking the sum of (\ref{con27x}) and (\ref{con28x}), we find

\beq
\label{con29x}
g^2 (\Omega_{p,-\bar{1}}+\Omega_{\bar{1},-p})-f^2(\Omega_{p,+\bar{1}}+\Omega_{\bar{1},+p})=0~.
\eeq

Next take $f$(\ref{x19x})+${ig \over 2}$(\ref{x13x}). Take the dual and then symmetrize over the
two free indices to obtain

\bea
\label{con30x}
f^2(-\Omega_{p,+q}-\Omega_{q,+p}+iF_{+ \bar{1} p {\bar{q}}_1 {\bar{q}}_2} \epsilon^{{\bar{q}}_1 {\bar{q}}_2}{}_q
+iF_{+ \bar{1} q {\bar{q}}_1 {\bar{q}}_2} \epsilon^{{\bar{q}}_1 {\bar{q}}_2}{}_p)
\cr
-g^2(\Omega_{p,q-}+\Omega_{q,p-}-i F_{-1p {\bar{q}}_1 {\bar{q}}_2}\epsilon^{{\bar{q}}_1 {\bar{q}}_2}{}_q
-i F_{-1q {\bar{q}}_1 {\bar{q}}_2}\epsilon^{{\bar{q}}_1 {\bar{q}}_2}{}_p)
\cr
+ifg(-\Omega_{p,q \bar{1}}-\Omega_{q,p \bar{1}}-\Omega_{p,1q}-\Omega_{q,1p}
+\Omega_{p {\bar{q}}_1 {\bar{q}}_2} \epsilon^{{\bar{q}}_1 {\bar{q}}_2}{}_q
+\Omega_{q {\bar{q}}_1 {\bar{q}}_2} \epsilon^{{\bar{q}}_1 {\bar{q}}_2}{}_p)=0~.
\eea

Also take the complex conjugate of -$ig$(\ref{x30x})-$2f$(\ref{x32x}) and symmetrize over the two free indices to get

\bea
\label{con31x}
-g^2(\Omega_{p,q-}+\Omega_{q,p-}+iF_{-1p {\bar{q}}_1 {\bar{q}}_2}  \epsilon^{{\bar{q}}_1 {\bar{q}}_2}{}_q
+iF_{-1q {\bar{q}}_1 {\bar{q}}_2}  \epsilon^{{\bar{q}}_1 {\bar{q}}_2}{}_p)
\cr
+f^2(-\Omega_{p,+q}-\Omega_{q,+p}-iF_{+ \bar{1} p {\bar{q}}_1 {\bar{q}}_2} \epsilon^{{\bar{q}}_1 {\bar{q}}_2}{}_q
-iF_{+ \bar{1} q {\bar{q}}_1 {\bar{q}}_2} \epsilon^{{\bar{q}}_1 {\bar{q}}_2}{}_p)
\cr
+ifg(\Omega_{p,1q}+\Omega_{q,1p}+\Omega_{p,q \bar{1}}+\Omega_{q,p \bar{1}}
-\Omega_{p {\bar{q}}_1 {\bar{q}}_2} \epsilon^{{\bar{q}}_1 {\bar{q}}_2}{}_q
-\Omega_{q {\bar{q}}_1 {\bar{q}}_2} \epsilon^{{\bar{q}}_1 {\bar{q}}_2}{}_p)=0~.
\eea

Taking the sum of (\ref{con30x}) and (\ref{con31x}), we find

\beq
\label{con32x}
g^2 (\Omega_{p,-q}+\Omega_{q,-p})-f^2(\Omega_{p,+q}+\Omega_{q,+p})=0~.
\eeq

Next consider the combination $ig$(\ref{x14x})+$2f$(\ref{x16x}).
By comparing with the dual of $ig$(\ref{x29x})+$2f$(\ref{x35x}), the
$G$-flux terms can be removed to obtain

\bea
\label{con33x}
g^2(\Omega_{p , \bar{q}-}+\Omega_{\bar{q}, p-}+2i g_{p \bar{q}} F_{-1 \bar{1} r}{}^r -4iF_{-1 \bar{1}p \bar{q}})
\cr
+f^2(\Omega_{p,+\bar{q}}+\Omega_{\bar{q},+p}+2ig_{p \bar{q}}F_{+1 \bar{1} r}{}^r -4iF_{+1 \bar{1}p \bar{q}})
\cr
+ifg(\Omega_{p, \bar{q}1}+\Omega_{p, \bar{1} \bar{q}}+\Omega_{\bar{q},1p}+\Omega_{\bar{q},p \bar{1}}
-\Omega_{p, r_1 r_2}\epsilon^{r_1 r_2}{}_{\bar{q}} - \Omega_{\bar{q}, {\bar{r}}_1 {\bar{r}}_2}
\epsilon^{{\bar{r}}_1 {\bar{r}}_2}{}_p) =0~.
\eea

Take the complex conjugate of (\ref{con33x}), swap $p,q$ and add the result to (\ref{con33x})
to obtain

\beq
\label{con34x}
f^2 (\Omega_{p,+ \bar{q}}+\Omega_{\bar{q},+p})-g^2(\Omega_{p,-\bar{q}}+\Omega_{\bar{q},-p})=0~.
\eeq
This completes the first phase of solving the linear system.

\subsection{Fluxes and the solution to the linear system}

There are eleven types of $G$-flux terms with one free
holomorphic index: $G_{-+p}$, $G_{-1p}$, $G_{- \bar{1} p}$, $G_{+1p}$,
$G_{+ \bar{1} p}$, $G_{1 \bar{1} p}$, $G_{p q}{}^q$, $G_{- {\bar{q}}_1 {\bar{q}}_2}\epsilon^{{\bar{q}}_1 {\bar{q}}_2}{}_p$,
$G_{+ {\bar{q}}_1 {\bar{q}}_2}\epsilon^{{\bar{q}}_1 {\bar{q}}_2}{}_p$,
$G_{1 {\bar{q}}_1 {\bar{q}}_2}\epsilon^{{\bar{q}}_1 {\bar{q}}_2}{}_p$,
and $G_{\bar{1} {\bar{q}}_1 {\bar{q}}_2}\epsilon^{{\bar{q}}_1 {\bar{q}}_2}{}_p$.
These terms are constrained by sixteen equations ({\ref{y6y}}), ({\ref{y10y}}), ({\ref{x12x}}),
({\ref{x15x}}), ({\ref{x17x}}), ({\ref{x18x}}), ({\ref{x21x}}), ({\ref{x27x}}),
({\ref{x37x}}), ({\ref{x43x}}), ({\ref{x45x}}), ({\ref{x51x}}), ({\ref{x53x}}),
({\ref{x59x}}) and also the dualized antisymmetric parts of ({\ref{x30x}}) and ({\ref{x32x}}).
Only ten of these types of $G$-flux terms are constrained by these equations, one
must remain arbitrary. There are also six $G$-independent constraints.

It is straightforward, but tedious, to show that, taking  $G_{1 \bar{1} p}$ to be arbitrary, then

\bea
\label{g1eq}
\xone &=&2g^2 f^{-2} \ethree -i f^{-1} (g \eone +2g \afive -2g \etwo +2g \afour -4i \df)
\cr
&-& P_p-2 \athree -2i \ytwo -2i \yone -2i \yeight~,
\eea

\bea
\label{g2eq}
\xtwo &=& ig f^{-1} \ethree -{1 \over 2}g^{-1} \big(2g \afive +2g \afour
-4ig \ynine -4i \df
\cr
&-& 2g \cthree +4ig \yfour -g \eone +2g \etwo -4ig \ythree \big)
\cr
&-&{i \over 2} f g^{-1} \big(-6i \yone -2i \yeight -2i \ytwo +P_p -2 \athree
\cr
&-& \xsix +2 \cone -2 \ctwo \big)~,
\eea

\bea
\label{g3eq}
\xthree &=& -i g f^{-1} \ethree -{1 \over 2}g^{-1} (-2g \cthree +6g \afive +2g \afour +4ig \yfour
\cr
&-&8i \df +g \eone -2g \etwo -4g \dthree )
- {i \over 2} f g^{-1}(-2i \yone -6i \ytwo
\cr
&-&6i\yeight +2 \cone -4 \athree +2 \aone -2 \ctwo
+\xsix +P_p
\cr
&-&4 \dtwo +2 \done +2 \atwo )~,
\eea

\bea
\label{g4eq}
\xfour &=& ig^3 f^{-3} \ethree -{1 \over 2} g f^{-2} (-8ig \ythree
+2g \afive -4i \df +2g \afour +4g \btwo
\cr
&-&2g \cthree +4ig \yfour -g \eone
+2g \etwo -4ig \ynine)
\cr
&+&{i \over 2} f^{-1} (-4 \dg +4g \athree +2ig \ytwo +g \xsix + gP_p
+2g \aone +2g \bone
\cr
&-&4g \bfive -2g \atwo +6ig \yone +2ig \yeight
-2g \cone +2g \ctwo)
\cr
&-&2 \aseven -2i\yten~,
\eea

\bea
\label{g5eq}
\xfive &=& -i g^3 f^{-3} -{1 \over 2} g f^{-2}(-4ig \ynine +2g \afour +6g \afive +g \eone
\cr
&-&2g \etwo -2g \cthree +4ig \yfour -8i \df)
+{i \over 2} f^{-1}(-4 \dg +6g \athree
\cr
&+& g P_p +2ig \yone +6ig \ytwo
+6ig \yeight -g \xsix -2g \cone
\cr
&+&2g \ctwo -4g \aone)
-2i \ysix -2 \asix~,
\eea

\bea
\label{g7eq}
\xseven &=&  2i f^{-1} (-2ig \ynine +2g \afive -2i \df -g \cthree +2ig \yfour)
\cr
&+&4i \yeight +2 \athree -2 \atwo +4i \yone -2 \aone -2 \cone
\cr
&+&2 \ctwo~,
\eea

\bea
\label{g8eq}
\xeight &=& -2ig f^{-1} \ethree - g^{-1}(g \eone -2g \etwo +2g \afour +2g \afive
\cr
&-&2g \cthree
+4ig \yfour -4i \df)
-i f g^{-1} (-6i \yone
\cr
&-&2i \yeight -2i \ytwo +P_p -2 \athree -\xsix +2 \cone
\cr
&-&2 \ctwo)~,
\eea

\bea
\label{g9eq}
\xnine &=& -2i g^3 f^{-3} \ethree  - g^2 f^{-2} (-2 \afive +2 \afour +2 \cthree -4i \yfour
\cr
&+&\eone -2 \etwo +4i \ynine)
-i f^{-1} (-2g \athree
\cr
&-&2g \cone -2ig \ytwo +2ig \yeight +2ig \yone +g P_p
\cr
&-&g \xsix
+2g \ctwo +4ig \yseven -4g \aone +4 \dg)
\cr
&-&4 \aseven -4i \yten~,
\eea

\bea
\label{g10eq}
\xten &=&  -2 g^2 f^{-2} \ethree -i f^{-1} (-8ig \ythree
+2g \afive -4i \df +2g \afour
\cr
&+&4g \btwo -2g \cthree +4ig \yfour
-g \eone +2g \etwo
\cr
&-&4ig \ynine)
-2 \athree +2 \cone -2 \bone -2i \ytwo
\cr
&-&2i \yeight
-6i \yone +P_p -\xsix -2 \ctwo +4 \bfive
\cr
&-&4i \yseven~,
\eea

\bea
\label{g11eq}
\xeleven &=& -2 g^2 f^{-2} \ethree +i f^{-1} (-4g \dthree +g \eone -2g \etwo
+2g \afour
\cr
&+&4ig \ynine +2g \afive - 2g \cthree + 4ig \yfour -4i \df)
\cr
&+&6i \ytwo +2i \yeight -2 \done +2i \yone +P_p
\cr
&+&2 \athree +4 \dtwo - \xsix
-2 \cone +2 \ctwo~.
\eea

We also get with the $G$-independent constraints

\bea
\label{gg1con}
-gf^{-1} (g \afour +g \btwo -2ig \ythree -i \df) +i (g \aone +g \athree -g \bfour - \dg
\cr
+g \bone -g \bfive) +f(-\aseven +\bthree -2i \yten) =0~,~~~~
\eea

\bea
\label{gg2con}
-{1 \over 2} g f^{-1} (-g \cthree +2ig \yfour +2g \afive -2ig \ynine -2i \df)
\cr
-{i \over 2} (-2 \dg -g \cfive -2ig \yeight +2g \aone
\cr
-2ig \yseven
-4ig \yone -2g \ctwo +g \cone)
\cr
+f(-{1 \over 2} \cfour
+i \yten -i \yfive + \aseven) =0~,
\eea

\bea
\label{gg3con}
-g f^{-1} (g \dthree +g \afive -2ig \ynine -i \df)
\cr
-i(g \dfive +g \aone -g \athree -g \done +g \dtwo +\dg)
\cr
+f(-\asix + \dfour -2i \ysix)=0~,
\eea

\bea
\label{gg4con}
-g^2 f^{-1} \ethree +i(-2i \df +g \afive +g \afour
\cr
 -g \etwo +g \eone -g \efive)
\cr
+f(\athree -\efour +2i \yeight +2i \ytwo +2i \yone)=0~,
\eea

\bea
\label{gg5con}
-2ig^3 f^{-2} \ethree -g f^{-1} (2g \afour +2g \afive +g \eone -2g \etwo -4i \df)
\cr
+2ig (\athree + \tthree + i \yeight +i \yseven + 2i \yone)
\cr
+ f( \tone -2 \ttwo) = 0~,~
\eea

\bea
\label{gg6con}
-g^4 f^{-3} \ethree +{i \over 2} g^2 f^{-2} (2g \afour +2g \afive +g \eone -2g \etwo
-4i \df)
\cr
+ g f^{-1} (ig \yeight +2ig \ytwo +2g \athree -2 \dg -ig \yseven)
\cr
+{i \over 2} g (2 \asix +2 \tfive + \tone +2 \aseven) -f \tfour =0.~
\eea

 The conditions ({\ref{gg1con}}), ({\ref{gg2con}}), ({\ref{gg3con}}), ({\ref{gg4con}}),
({\ref{gg5con}}), ({\ref{gg6con}}) imply ({\ref{con13x}}), ({\ref{con16x}}),
({\ref{con24x}}) and ({\ref{con27x}}).

There are also eleven types of $G$-flux terms with one free
anti-holomorphic index: $G_{-+{\bar{p}}}$, $G_{-1{\bar{p}}}$, $G_{- \bar{1} {\bar{p}}}$, $G_{+1{\bar{p}}}$,
$G_{+ \bar{1} {\bar{p}}}$, $G_{1 \bar{1} {\bar{p}}}$, $G_{{\bar{p}} q}{}^q$, $G_{- q_1 q_2}\epsilon^{q_1 q_2}{}_{\bar{p}}$,
$G_{+ q_1 q_2}\epsilon^{q_1 q_2}{}_{\bar{p}}$,
$G_{1 q_1 q_2}\epsilon^{q_1 q_2}{}_{\bar{p}}$,
and $G_{\bar{1} q_1 q_2}\epsilon^{q_1 q_2}{}_{\bar{p}}$.
These terms are constrained by sixteen equations ({\ref{y4y}}), ({\ref{y11y}}), ({\ref{x22x}}),
({\ref{x24x}}), ({\ref{x28x}}), ({\ref{x31x}}), ({\ref{x33x}}), ({\ref{x34x}}),
({\ref{x38x}}), ({\ref{x40x}}), ({\ref{x46x}}), ({\ref{x48x}}), ({\ref{x54x}}),
({\ref{x56x}}) and also the dualized antisymmetric parts of ({\ref{x13x}}) and ({\ref{x19x}}).
Only ten of these types of $G$-flux terms are constrained by these equations, one
must remain unfixed. There are also six $G$-independent constraints.

It is straightforward, but tedious, to show that, taking  $G_{1 \bar{1} {\bar{p}}}$ to be arbitrary, then

\bea
\label{g1eqxx}
\xonexx &=& 2 g^2 f^{-2} \ethreexx -i f^{-1} (2g \afivexx +g \eonexx +2g \afourxx
-2g \etwoxx-4i \dfxx)
\cr
&-&P_{\bar{p}} -2 \athreexx -2i \ysevenxx +2i \yonexx +2i \ytwoxx~,
\eea

\bea
\label{g2eqxx}
\xtwoxx &=& -ig f^{-1} \ethreexx +{1 \over 2}g^{-1}(2g\cthreexx -6g \afivexx
-2g \afourxx +4ig \ythreexx
\cr
&+&8i \dfxx +4g \dthreexx +2g \etwoxx -g \eonexx)
\cr
&-&{i \over 2} f g^{-1} (2i \yonexx -6i \ysevenxx +6i \ytwoxx +P_{\bar{p}}
+2 \donexx
\cr
&-&2 \aonexx -4 \athreexx +2 \cfivexx - \xsixxx -2 \ctwoxx
\cr
&-&2 \atwoxx -4 \dtwoxx)~,
\eea

\bea
\label{g3eqxx}
\xthreexx &=& - g f^{-1} \ethree  +{1 \over 2}g^{-1}(-2g \afourxx -2g \etwoxx
+4ig \ythreexx +g \eonexx
\cr
 &-&2g \afivexx +2g \cthreexx -4ig \yninexx +4i \dfxx
-4ig \yfourxx)
\cr
&-&{i \over 2} f g^{-1}(-2i \ysevenxx +6i \yonexx +2i \ytwoxx +P_{\bar{p}}+\xsixxx
\cr
&-&2 \ctwoxx -2 \athreexx +2 \cfivexx)~,
\eea

\bea
\label{g4eqxx}
\xfourxx &=& -ig^3 f^{-3} \ethreexx +{1 \over 2} g f^{-2}(-6g \afivexx -2g \afourxx
+2g \etwoxx +4ig \ythreexx
\cr
&-&g \eonexx -4ig \yninexx +8i \dfxx +2g \cthreexx)
\cr
&+& {i \over 2}f^{-1} (-2ig \yonexx -6ig \ytwoxx +g \xsixxx +6ig \ysevenxx
+g P_{\bar{p}}
\cr
&+&2g \ctwoxx +6g \athreexx -2g \cfivexx -4 \dgxx +4g \aonexx)
\cr
&+&2i \yfivexx -2 \asixxx~,
\eea

\bea
\label{g5eqxx}
\xfivexx &=&- g^3 f^{-3} \ethreexx +{1 \over 2} g f^{-2} (-2g \afourxx
-2g \etwoxx +4ig \ythreexx +g \eonexx
\cr
&-&2g \afivexx +2g \cthreexx
-4ig \yninexx +4i \dfxx -4g \btwoxx -8ig \yfourxx)
\cr
&+& {i \over 2} f^{-1}(2ig \ysevenxx -2ig \ytwoxx -6ig \yonexx +2g \ctwoxx -2g \cfivexx
\cr
&-&4 \dgxx -4g \bfivexx +2g \bonexx +2g \atwoxx -g \xsixxx  +g P_{\bar{p}}
+4g \athreexx -2g \aonexx)
\cr
&-&2 \asevenxx +2i \ytenxx~,
\eea

\bea
\label{g7eqxx}
\xsevenxx &=& 2i f^{-1}(-2g \afivexx +g \cthreexx +2i \dfxx +2ig \ythreexx -2ig \yninexx)
\cr
&-&2 \athreexx -4i \ysevenxx +4i \yonexx -2 \ctwoxx +2 \cfivexx
\cr
&-&2 \aonexx -2 \atwoxx~,
\eea

\bea
\label{g8eqxx}
\xeightxx &=& -2ig f^{-1} \ethreexx +g^{-1}(-2g \afourxx +2g \etwoxx +4ig \ythreexx -g \eonexx
\cr
&-&2g \afivexx +2g \cthreexx +4i \dfxx)
\cr
&-&i f g^{-1}(2i \ytwoxx +P_{\bar{p}}-2i \ysevenxx + \xsixxx +6i \yonexx
\cr
&-&2 \ctwoxx
-2 \athreexx +2 \cfivexx)~,
\eea

\bea
\label{g9eqxx}
\xninexx &=& -2i g^3 f^{-3} \ethreexx - g^2 f^{-2}(2 \afourxx -2 \etwoxx +4i \ythreexx
+\eonexx
\cr
&-&2 \afivexx +2 \cthreexx -4i \yninexx)
\cr
&-&i f^{-1} (g P_{\bar{p}} -2g \athreexx +4 \dgxx +g \xsixxx +2ig \ysevenxx
+2ig \ytwoxx
\cr
&+&2g \ctwoxx -2g \cfivexx -2ig \yonexx +4ig \yeightxx +4g \aonexx)
\cr
&-&4 \asevenxx +4i \ytenxx~,
\eea

\bea
\label{g10eqxx}
\xtenxx &=& -2 g^2 f^{-2} \ethreexx -i f^{-1} (-2g \afourxx +2g \etwoxx +4g \dthreexx
+4ig \ythreexx
\cr
&-&g \eonexx -2g \afivexx +2g \cthreexx +4ig \yninexx +4i \dfxx)
\cr
&+& 2i \ysevenxx +P_{\bar{p}} -2 \donexx + \xsixxx -2i \yonexx
\cr
&+&2 \ctwoxx
+2 \athreexx -6i \ytwoxx -2 \cfivexx +4 \dtwoxx~,
\eea

\bea
\label{g11eqxx}
\xelevenxx &=& -2 g^2 f^{-2} \ethreexx +i f^{-1}(-2g \afourxx -2g \etwoxx
+4ig \ythreexx +g \eonexx
\cr
&-&2g \afivexx +2g \cthreexx -4ig \yninexx
+4i \dfxx  -4g \btwoxx -8ig \yfourxx)
\cr
&-&2 \athreexx -2i \ysevenxx +4 \bfivexx -2 \bonexx +2i \ytwoxx
\cr
&-&2 \ctwoxx +2 \cfivexx
+P_{\bar{p}} + \xsixxx +6i \yonexx -4i \yeightxx~.
\eea

In addition, we have  the $G$-independent constraints

\bea
\label{gg1conxx}
{1 \over 2} g f^{-1} (2ig \yfourxx + g \btwoxx +g \afourxx -i \dfxx)
\cr
+{i \over 2} (\dgxx +g \aonexx +g \bfivexx -g \bonexx +g \bfourxx -g \athreexx)
\cr
+f (-{1 \over 2} \bthreexx +{1 \over 2} \asevenxx -i \ytenxx)=0~,
\eea

\bea
\label{gg2conxx}
{1 \over 2} g f^{-1} (-2g \afivexx +g \cthreexx +2i \dfxx +2ig \ythreexx -2ig \yninexx)
\cr
+{i \over 2} (2ig \ysevenxx +2g \ctwoxx +2 \dgxx +2ig \yeightxx
\cr
 -4ig \yonexx
+2g \aonexx +g \conexx -g \cfivexx)
\cr
+f (i \ysixxx -i \ytenxx +\asevenxx -{1 \over 2}\cfourxx) =0~,
\eea

\bea
\label{gg3conxx}
{1 \over 2} g f^{-1} (2ig \yninexx +g \afivexx -i \dfxx +g \dthreexx )
\cr
+{i \over 2} (-g \athreexx +g \dfivexx +g \dtwoxx -g \donexx -g \aonexx + \dgxx)
\cr
+f ({1 \over 2} \asixxx -{1 \over 2} \dfourxx -i \yfivexx) =0~,
\eea

\bea
\label{gg4conxx}
{1 \over 2} g^2 f^{-1} \ethreexx -{i \over 2} (-2i \dfxx + g \afivexx
-g \etwoxx
\cr
+g \eonexx +g \afourxx -g \efivexx)
\cr
+{i \over 2}f (-i \efourxx +2 \ytwoxx +2 \yonexx +i \athreexx -2 \ysevenxx)=0~,
\eea

\bea
\label{gg5conxx}
i g^3 f^{-2} \ethreexx + {1 \over 2} g f^{-1} (2g \afivexx
+2g \afourxx -2g \etwoxx +g \eonexx -4i \dfxx)
\cr
-g(-\ysevenxx - \yeightxx +i \athreexx +i \tthreexx +2 \yonexx)
\cr
 + f(-{1 \over 2} \tonexx + \ttwoxx) =0~,
\eea

\bea
\label{gg6conxx}
{1 \over 2} g^4 f^{-3} \ethreexx -{i \over 4} g^2 f^{-2} (2g \afivexx +2g \afourxx
-2g \etwoxx +g \eonexx -4 i \dfxx)
\cr
+{i \over 2} g f^{-1} (2g \ytwoxx -g \ysevenxx +2ig \athreexx -2i \dgxx +g \yeightxx)
\cr
-{i \over 4}g (2 \asevenxx + \tonexx +2 \asixxx +2 \tfivexx) +{1 \over 2}f \tfourxx =0~.
\eea

To proceed, we note that the constraints ({\ref{gg1con}}), ({\ref{gg2con}}),
({\ref{gg3con}}), ({\ref{gg4con}}), ({\ref{gg5con}}), ({\ref{gg6con}})
and the complex conjugates of ({\ref{gg1conxx}}), ({\ref{gg2conxx}}),
({\ref{gg3conxx}}), ({\ref{gg4conxx}}), ({\ref{gg5conxx}}), ({\ref{gg6conxx}})
give twelve constraints on the ten types of $F$-flux term with a single
free holomorphic index (${F_{-+1 \bar{1} p}}$,
${F_{-+pq}{}^q}$, ${F_{-1pq}{}^q}$, ${F_{- \bar{1} pq}{}^q}$, ${F_{+1pq}{}^q}$,
${F_{+ \bar{1} pq}{}^q}$, ${F_{-+1 {\bar{q}}_1 {\bar{q}}_2} \epsilon^{{\bar{q}}_1 {\bar{q}}_2}{}_p}$,
${F_{-+\bar{1} {\bar{q}}_1 {\bar{q}}_2} \epsilon^{{\bar{q}}_1 {\bar{q}}_2}{}_p}$,
${F_{-1 \bar{1} {\bar{q}}_1 {\bar{q}}_2} \epsilon^{{\bar{q}}_1 {\bar{q}}_2}{}_p}$, and
${F_{+ 1 \bar{1} {\bar{q}}_1 {\bar{q}}_2} \epsilon^{{\bar{q}}_1 {\bar{q}}_2}{}_p}$).

These equations do not fix these $F$-flux terms uniquely: five of
the terms remain arbitrary. So, taking $\ytwo$, $\yfive$,
$\yseven$, $\yeight$, $\ynine$ and $\yten$ to be arbitrary, one has

\bea
\label{yycon1}
\ythree &=& -{g \over 4} f^{-1} \ethree -{i \over 2}(\btwo+\afour)-{f \over 2} g^{-1}(\aone+\bone
-\bfour -\bfive)
\cr
&-&{i \over 2} f^2 g^{-2}(2i \yten - \bthree + \aseven)-{f^3 \over 4} g^{-3} \tfour~,
\eea

\bea
\label{yycon2}
\yfour &=& {g \over 2} f^{-1} \ethree + \ynine +i(-\afour + \etwo -{1 \over 2} \eone
\cr
&-&{1 \over 2} \cthree)
+{f \over 2} g^{-1}(-2 \tthree +2 \ctwo -2 \aone + \cfive
\cr
&-& \cone)
-{f^2 \over 2} g^{-2} (2 \yfive -2 \yten +i(2 \ttwo +2 \aseven
\cr
&-& \tone - \cfour)) -{f^3 \over 2} g^{-3} \tfour~,
\eea

\bea
\label{yycon3}
\ysix &=& {g^3 \over 4} f^{-3} \ethree  + f^{-2}g^2(\ynine +{i \over 2}(\afive + \dthree))
\cr
&-&{g \over 2} f^{-1} (\aone - \done + \dfive + \dtwo) +{i \over 2}(\asix - \dfour)
\cr
&+&{f \over 4} g^{-1} \tfour~,
\eea

\bea
\label{yycon4}
\yone &=& -{g \over 2} f^{-1} (\afour - \efive + \afive - \etwo + \eone)
\cr
&+&{1 \over 2}Q_p - \ytwo - \yeight~,
\eea

\bea
\label{yycon5}
\yseven &=& -{g \over 2} f^{-1} (2 \efive - \eone) +\yeight +2 \ytwo
\cr
&+&{f \over 2} g^{-1}
(\tone -2 \ttwo)~.
\eea

There are in addition seven $F$-independent geometric constraints:

\be
\label{geocon1}
\efour + \tthree =0~,
\ee

\be
\label{geocon2}
\partial_p f = {g^2 \over 2} f^{-1} \ethree -{f \over 2} (\tthree+ \athree)
\ee

\be
\label{geocon3}
\partial_p g = -{f^2 \over 2} g^{-1} \tfour +{g \over 2}(\athree - \tthree)~,
\ee

\be
\label{geocon4}
g^2(- \efive - \afour + \etwo - \afive) - f^2(\asix + \tfive + \aseven + \ttwo) =0~,
\ee

\be
\label{geocon5}
g^2 (\btwo + \afour)  + f^2 (\aseven - \bthree) =0~,
\ee

\bea
\label{geocon6}
g^2 (2 \afour + \cthree -2 \etwo + \eone)
\cr
+ f^2 (2 \ttwo - \tone - \cfour +2 \aseven)=0~,
\eea

\be
\label{geocon7}
g^2 ( \dthree + \afive)+ f^2(\asix - \dfour)=0~.
\ee

Next, we shall investigate those equations constraining $G$ and $F$ flux terms
with no free holomorphic index $p$ (or antiholomorphic index ${\bar{p}}$).
There are 24 such equations: ({\ref{y5y}}), ({\ref{y7y}}),
({\ref{y8y}}), ({\ref{y9y}}), ({\ref{x20x}}),
 ({\ref{x23x}}),  ({\ref{x25x}}),  ({\ref{x26x}}),
 ({\ref{x36x}}),  ({\ref{x39x}}),  ({\ref{x41x}}),  ({\ref{x42x}}),
 ({\ref{x44x}}),  ({\ref{x47x}}),  ({\ref{x49x}}),  ({\ref{x50x}}),
 ({\ref{x52x}}),  ({\ref{x55x}}),  ({\ref{x57x}}),  ({\ref{x58x}})
together with the traces of  ({\ref{x14x}}), ({\ref{x16x}}),
 ({\ref{x29x}}) and  ({\ref{x35x}}). There are ten such $G$-flux terms
($\vone$, $\vtwo$, $\vthree$, $\vfour$, $\vfive$, $\vsix$, $\vseven$,
$\veight$, $\vnine$ and $\vten$) and ten such $F$-flux terms
($\zone$, $\zonec$, $\ztwo$, $\zthree$, $\zfour$, $\zfourc$, $\zfive$,
$\zfivec$, $\zsix$, $\zsixc$). The $G$-flux terms are fixed
in terms of the $F$-flux terms and the geometry via

\bea
\label{gnofree1}
\vone &=& f^{-1} (-2i \DGtwo + g(i \Ptwo +i \sone -i \stwo +i \sthree))
\cr
&+&2i \zone -4i \zfivec -2 \sfive - \Pthree~,
\eea

\bea
\label{gnofree2}
\vtwo &=& f^{-1} (-2i \DGtwo +g (i \Ptwo-i \sone +i \stwo +i \sthree))
\cr
&-&2i \zonec -4i \zfive -2 \sfivec - \Pfour~,
\eea

\bea
\label{gnofree3}
\vthree &=& (2i \DGtwo -4 \DFthree +g(-i \Ptwo -i \sone +i \stwo -8 \zfour -i \sthree -4i \mthree))
\cr
&-&2 \mone -6i \zone -2 \mfour +4i \zfivec -2 \mtwo +2 \sfive + \Pthree~,
\eea

\bea
\label{gnofree4}
\vfour &=& f^{-1}(-2i \DGtwo +4 \DFfour +g (8 \zfourc +i \Ptwo -i \sone +i \stwo +i \sthree
\cr
&+&4i \mthreec))
-4i \zfive - \Pfour -6i \zonec -2 \monec -2 \mtwoc
\cr
&+&2 \mfourc -2 \sfivec~,
\eea

\bea
\label{gnofree5}
\vfive &=& f^{-1} (4 \DFtwo +4ig \sfourc) +g^{-1}(4 \DGtwo +2i \DFthree
+g(8i \zfourc +6i \ztwo
\cr
&-&4 \sone -2 \mthree +4i \zfour + \Ptwo +2 \hthreec))
+f g^{-1}(i \mfour +i \Pthree
\cr
&+&8 \zfivec +i \mtwo +i \mone -4i \sfive
-4 \zone
\cr
&-&2i \honec -i \hfourc)~,
\eea

\bea
\label{gnofree6}
\vsix &=& -2ig^2 f^{-3} (i \DFtwo -g \sfourc)
-ig f^{-2}(-4i \DGtwo +2 \DFthree +g(-i \stwo
\cr
&+&2 i \mthree + 3i \sthree -2i \hthree +4 \zfour
-4 \zfourc -i \sone -3i \Ptwo))
\cr
&+&ig f^{-1}(-\mfour - \Pthree +4i \zfivec - \mtwo +4i \zfive - \mone +2 \sfivec
\cr
&+&2 \hone + \hfour +2 \sfive -4i \zone) +i g^{-1}(2i \DGone +g(4 \zsixc
\cr
&+&2 \zthree
+i \qone -2i \Pone -i \qtwo -i \qthree)) +2i f g^{-1} \qfivec~,
\eea

\bea
\label{gnofree7}
\vseven &=& 2ig f^{-2}(\DFtwo +ig \sfourc) + f^{-1}(-2i \DGtwo +g(-4 \zfourc -2i \sone
\cr
&+&2i \sthree -2i \hthree)) - \hfour -2 \hone
\cr
&-&2 \sfivec -4i \zfive~,
\eea

\bea
\label{gnofree8}
\veight &=& -2ig f^{-2} (\DFtwo +i g \sfourc) + f^{-1}(-2i \DGtwo +g(4 \zfourc
+4 \ztwo
\cr
 &+&2i \sone -2i \hthreec)) -4i \zfivec -2 \honec
 \cr
 &-&\hfourc -2 \sfive~,
\eea

\bea
\label{gnofree9}
\vnine &=& -2i g^2 f^{-3} (i \DFtwo -g \sfourc) -ig f^{-2}(-4i \DGtwo
+2 \DFthree +g(-i \stwo
\cr
&+&2i \mthree +3i \sthree -2i \hthree +4 \zfour -4 \zfourc
-i \sone -3i \Ptwo))
\cr
&+&ig f^{-1}(-\mfour - \Pthree +4i \zfivec -\mtwo +4i \zfive - \mone +2 \sfivec
\cr
&+&2 \hone
+ \hfour +2 \sfive -4i \zone) -i g^{-1}(2i \DGone +g (-i \qtwo
\cr
&+&i \qone +4 \zsixc
-i \qthree +2 \zthree +2i \Pone)) -2i f g^{-1} \qfivec~,
\eea

\bea
\label{gnofree10}
\vten &=& -g^{-1}(2i \DFthree +4 \DGtwo+g(-2 \sone -2 \sthree +2 \stwo -2 \mthree
\cr
&+&4i \zfour +2i \ztwo + \Ptwo +2 \hthreec))
+f g^{-1}(i \mfour +i \Pthree
\cr
&+&8 \zfivec +i \mtwo +i \mone -4i \sfive
-4 \zone
\cr
&-&2i \honec -i \hfourc)~.
\eea

There are also fourteen $G$-independent constraints

\bea
\label{fnofree1}
i \DGone +g(\zthree -{i \over 2} \qthree + \zsix + \zsixc)+{f \over 2}(\qfive+ \qfivec)=0~,
\eea

\bea
\label{fnofree2}
2i \DFtwo -g(\sfour + \sfourc) +f(-2 \ztwo -2 \zfour +i \sthree -2 \zfourc)=0~,
\eea

\bea
\label{fnofree3}
\DGtwo -{g \over 2} \sthree +f g^{-1}(-\DFone -{ig \over 2}(\qfour + \qfourc
\cr
+ \sfive + \sfivec))-{f^2 \over 2} g^{-1} \qthree =0~,
\eea

\bea
\label{fnofree4}
-i \DFone +2 \DGthree +g(-\mfour +\qfour +2i \zone -4i \zfive)
\cr
+g^{-1}(i \DGone +g(i \mseven +i \msix -i \qthree -i \qone))+f^2 g^{-1} \qfive=0~,
\eea

\bea
\label{fnofree5}
-\DFone -2i \DGfour +g(-4 \zfivec -2 \zonec -i \qfourc +i \mfourc)
\cr
+i f g^{-1}(\DGone +g(\msixc + \qone - \qthree + \msevenc))+f^2 g^{-1} \qfivec=0~,
\eea

\bea
\label{fnofree6}
i \DFtwo -g \sfour +f g^{-1}(-i \DGtwo +2 \DFthree +g(i \mfive +i \mthree +i \sone
\cr
+i \sthree))+ f^2 g^{-1}(2i \zone -4i \zfivec + \mfour - \sfive)=0~,
\eea

\bea
\label{fnofree7}
i \DFtwo -g \sfourc +i f g^{-1}(-\DGtwo -2i \DFfour +g(\mthreec + \sthree - \sone
\cr
+ \mfivec)) -i f^2 g^{-1}(2 \zonec +i \mfourc +4 \zfive -i \sfivec)=0~,
\eea

\bea
\label{fnofree8}
-{i \over 2} \DFfour +g({1 \over 2} \mthreec -i \zfourc)+{i \over 2}f g^{-1}
(\DGfour +g(\monec - \mfourc))
\cr
-f^2 g^{-1}({1 \over 2} \msixc +i \zsixc)=0~,
\eea

\bea
\label{fnofree9}
-{i \over 2} \DFthree +g({1 \over 2} \mthree - \zfour) +{i \over 2} f g^{-1}
(\DGthree -g(\mone + \mfour))
\cr
-f^2 g^{-1} ({1 \over 2} \msix +i \zsix)=0~,
\eea

\bea
\label{fnofree10}
2 \DFfour -8i \DGtwo +2 \DFthree +g(2i \mthreec -2i \hthree +4 \ztwo +2i \mthree +4 \zfourc
\cr
-2i \Ptwo
+4i \sthree +4 \zfour -2i \hthreec) +f(4i \zone + \Pfour
-4 \sfivec
+ \mtwo
\cr
+\mfour -2 \hone - \hfour - \hfourc
-8i \zfivec
-4 \sfive + \Pthree
\cr
+ \mone - \monec - \mtwoc
+ \mfourc -8i \zfive
-4i \zonec -2 \honec)=0~,
\eea

\bea
\label{fnofree11}
f^2 \Pone -g^2 \Ptwo =0~,
\eea

\bea
\label{fnofree12}
{i \over 2} g^2 f^{-2}(i \DFtwo -g \sfourc) +{i \over 4}g^2 f^{-1}(-2i \hthree
+i \sthree +i \stwo
\cr
-4 \zfourc +i \Ptwo -3i \sone)
+{i \over 2}g(- \hfour -2i \zfive + \hfive
\cr
- \sfivec +2i \zfivec - \hone + \sfive)
+{i \over 4} f g^{-1}(2i \DGone +g(-i \qtwo
\cr
+2i \htwo -i \Pone +i \qone -i \qthree +4 \zthree
+4 \zsixc)
+{i \over 2} f^2 g^{-1} \qfivec=0~,
\eea

\bea
\label{fnofree13}
-{i \over 2} g^2 f^{-1} (i \hthree -2 \ztwo +i \hthreec)-{i \over 2 }g(\hone + \honec
+ \hfour
\cr
+ \hfourc
- \hfive - \hfivec)+{i \over 2}f(2 \zthree +i \htwo +i \htwoc)=0~,
\eea

\bea
\label{fnofree14}
f^{-1} (- \DGtwo +{1 \over 2}g(\stwo - \sone + \sthree + \Ptwo))+g^{-1}(\DFone
+g(i \qfourc
\cr
-2 \zfive +2 \zfivec+i \sfivec))+{1 \over 1}f g^{-1}(- \qtwo +
\qthree - \qone - \Pone)=0.
\eea

There are ten real degrees of freedom in the $F$-flux terms which appear in
({\ref{fnofree1}})-({\ref{fnofree14}}), but only eight of them are fixed by
these equations. In particular, taking the real parts of $\zfive$ and $\zsix$
to be arbitrary we find that:

\bea
\label{fnofreesol1}
\zone &=& f^{-1}({1 \over 2} \partial_- g +g(-{1 \over 4} \sone
+{1 \over 4} \stwo-{1 \over 4} \sthree)
\cr
&+&{1 \over 2}Q_1
-{1 \over 4}f g^{-1}(- \qone + \qtwo +2 \msix +2 \mseven)
\cr
&+& (\zfive+\zfivec)~,
 \eea

\bea
\label{fnofreesol2}
\ztwo &=& -{1 \over 4}gf^{-1}(\sfour+\sfourc) +{1 \over 2}Q_-
\cr
&+&{1 \over 4} f g^{-1}(\qfour + \qfourc+ \sfive +\sfivec +2 \mone -2 \monec)
\cr
&+&{1 \over 4} f^3 g^{-3}(\qfive + \qfivec) + f^2 g^{-2}(\zsix+\zsixc)~,
\eea

\be
\label{fnofreesol3}
\zthree = -{1 \over 2}Q_+ -{1 \over 2}f g^{-1}(\qfive+\qfivec)-(\zsix+\zsixc)~,
\ee

\bea
\label{fnofreesol4}
\zfour &=& {1 \over 16} g f^{-1}(\sfourc-5\sfour)
-{1 \over 16}f g^{-1}(-2 \hfivec +6 \mone +7 \sfive
-3 \sfivec
\cr
&+& \qfourc -2 \monec +2 \hfive +3 \qfour
+2 \honec-2 \hone-2 \hfour
\cr
&+&2 \hfourc)
-{1 \over 16}f^3 g^{-3}(\qfive+3 \qfivec)
\cr
&-&{1 \over 2} f^2 g^{-2}(\zsix+\zsixc)~,
\eea

\bea
\label{fnofreesol5}
{1 \over 2}(\zfive - \zfivec)&=&-{1 \over 8}g f^{-1} (\sone - \stwo)-{1 \over 8}f g^{-1}(\qone + \qtwo)~,
\eea

\bea
\label{fnofreesixsol}
{1 \over 2}(\zsix-\zsixc) &=& -{1 \over 16}g^3 f^{-3}(\sfourc- \sfour)
+{1 \over 16}f^{-1}g(-2 \hfivec+2 \hfive
\cr
&+&2 \honec -2 \hone
+2 \hfourc -2 \hfour
\cr
&+& \qfourc - \qfour
+3 \sfive -3 \sfivec
-2 \mone -2 \monec)
\cr
&+&{3 \over 16}f g^{-1}(\qfivec-\qfive)~,
\eea

We also obtain the following constraints on the $P_M$
terms

\be
\label{nofreeP1}
f^2 \Pone -g^2 \Ptwo =0~,
\ee

\bea
\label{nofreeP2}
\Ptwo &=& {i \over 4} g f^{-1} (\sfour + \sfourc)
+g^{-1} (-2 \partial_- g +g (\sthree + \sone - \hthreec
+i Q_-))
\cr
&+&{1 \over 4}f g^{-1}(-2iP_1 -2i P_{\bar{1}}
+2Q_1 +2 Q_{\bar{1}}
+2i \mtwoc +i \sfivec -5i \qfourc +2i \hfivec
\cr
&+&2i \hone +3i \qfour -2i \mtwo +2i \hfive +2i \honec+9i \sfive)
\cr
&-&f^2 g^{-2}(\msevenc + \qone + \mseven + \htwo)
-{i \over 4} f^3 g^{-3}(\qfive + \qfivec)~,
\eea

together with the geometric constraints

\be
\label{nofreecon1}
g^2 \mthree -f^2 \msix =0~,
\ee

\be
\label{nofreecon2}
f^2(\qone + \htwo)-g^2(\hthree+\sone)=0~,
\ee

\be
\label{nofreecon3}
f^2(\mseven +\qtwo)-g^2(\mfive+\stwo)=0~,
\ee

\bea
\label{nofreecon4}
{1 \over 2} g^2 f^{-2}(\sfour + \sfourc)
+ \hone + \honec + \hfour + \hfourc
\cr
- \hfive - \hfivec + \monec - \mone
+{1 \over 2}(\sfivec + \sfive + \qfour + \qfourc)
\cr
+{1 \over 2} f^2 g^{-2}(\qfive + \qfivec)=0~,
\eea

\be
\label{nofreecon5}
\partial_+f = -{1 \over 2} g f^{-1}(-2 \partial_- g +g \sthree)-{1 \over 2}f \qthree~,
\ee

\be
\label{nofreecon6}
\partial_- f = -{1 \over 2} \sthree f~,
\ee

\be
\label{nofreecon7}
\partial_1 f = {1 \over 2}g^2 f^{-1} \sfour +{1 \over 2}f(\sfive - \mfour)~,
\ee

\be
\label{nofreecon8}
\partial_+g = {1 \over 2} \qthree g~,
\ee

\be
\label{nofreecon9}
\partial_1 g = -{1 \over 2}g(\qfour - \mfour) -{1 \over 2} f^2 g^{-1} \qfive~,
\ee

\be
\label{nofreecon10}
f^2 Q_+ - g^2 Q_- =2fg (\qfour+\sfive)~,
\ee

This completes our analysis. The conditions on the geometry are summarized in section two.

\newsection{Killing spinor equations for maximally supersymmetric $Spin(7)\ltimes \bR^8$-backgrounds}

\subsection{A linear system}
In this appendix, we give the details of the derivation of the
conditions on the geometry and the fluxes implied by the Killing
spinor equations for maximally supersymmetric $Spin(7)\ltimes
\bR^8$-backgrounds. These conditions have already been summarized in
section four. The linear system derived from the Killing spinor
equations of IIB supergravity associated with the maximally
supersymmetric $Spin(7)\ltimes \bR^8$-backgrounds can be easily
derived using the results of \cite{gju},  (\ref{maxspins}) and the
properties of the spinors in appendix A. In fact, one can easily
derive the conditions below from the results of \cite{gju} without
any further computation. We can simplify this linear system  by
using the
 self-duality condition of the five-form field strength $F_{M_1\dots M_5}=- {1\over 5!}
\epsilon_{M_1\dots M_5}{}^{N_1\dots N_5}
F_{N_1\dots N_5}$. This condition written in $SU(5)$ representations   relates the  components of $F$ as

\bea
F_{\a_1\a_2\a_3\a_4\bar\b}&=&-{1\over6} \epsilon_{\a_1\a_2\a_3\a_4}
\epsilon_{\bar\b}{}^{\g_1\g_2\g_3} F_{-+\g_1\g_2\g_3}~,
\cr
F_{\a_1\a_2\a_3\bar\b_1\bar\b_2}&=&-{1\over2} \epsilon_{\a_1\a_2\a_3}{}^{\bar\g_1}
\epsilon_{\bar\b_1\bar\b_2}{}^{\g_2\g_3} F_{-+\bar\g_1\g_2\g_3}~,
\cr
F_{+\a_1\a_2\a_3\a_4}&=&0~,
\cr
F_{+\a_1\a_2\a_3\bar\b}&=&-{1\over6} \epsilon_{\bar\b}{}^{\b_1\b_2\b_3}
\epsilon_{\a_1\a_2\a_3}{}^{\bar\g} F_{+\b_1\b_2\b_3\bar\g}~,
\cr
F_{+\bar\b_1\bar\b_2\a_1\a_2}&=&-{1\over4} \epsilon_{\bar\b_1\bar\b_2}{}^{\d_1\d_2}
\epsilon_{\a_1\a_2}{}^{\bar\g_1\bar\g_2} F_{+\bar\g_1\bar\g_2\d_1\d_2}~,
\cr
F_{-\a_1\a_2\a_3\a_4}&=&{1\over4!} \epsilon_{\a_1\a_2\a_3\a_4}
\epsilon^{\b_1\b_2\b_3\b_4} F_{-\b_1\b_2\b_3\b_4}~,
\cr
F_{-\a_1\a_2\a_3\bar\b}&=&{1\over6} \epsilon_{\bar\b}{}^{\b_1\b_2\b_3}
\epsilon_{\a_1\a_2\a_3}{}^{\bar\g} F_{-\b_1\b_2\b_3\bar\g}~,
\cr
F_{-\bar\b_1\bar\b_2\a_1\a_2}&=&{1\over4} \epsilon_{\bar\b_1\bar\b_2}{}^{\d_1\d_2}
\epsilon_{\a_1\a_2}{}^{\bar\g_1\bar\g_2} F_{-\bar\g_1\bar\g_2\d_1\d_2}~,
\la{selfdualsu}
\eea
where $\a,\b,\g=1,\dots,5$ and $\epsilon$ is the standard $SU(5)$-invariant (5,0)-form.
The above relations imply the following conditions

\bea
F_{\a_1\a_2\a_3\d}{}^\d&=&F_{-+\a_1\a_2\a_3}~,
\cr
F_{\a_1\a_2\bar\b\d}{}^\d&=&-F_{-+\a_1\a_2\bar\b}-2 g_{\bar\b[\a_1} F_{\a_2]-+\d}{}^\d~,
\cr
F_{\a\b}{}^\b{}_{\g}{}^\g&=&2 F_{-+\a\d}{}^\d~,
\cr
F_{+\a\b\d}{}^\d&=&0~,
\cr
F_{+\a_1\a_2\a_3[\bar\b} \epsilon_{\bar\g]}{}^{\a_1\a_2\a_3}&=&0~,
\cr
F_{+\a}{}^\a{}_\b{}^\b&=&0~.
\la{trselfdualsu}
\eea

As the self-duality conditions above and the case of backgrounds with one $Spin(7)\ltimes \bR^8$-invariant
spinor investigated in \cite{gju}, the linear system is written after decomposing the fluxes and geometry
in $SU(5)$ representation. We shall demonstrate that the solution of the linear system can be rewritten in terms of
$Spin(7)\ltimes \bR^8$ representations as one may have been expected because the structure
group of the spacetime reduces from $Spin(9,1)\times U(1)$ to $Spin(7)\ltimes \bR^8$.
The algebraic Killing spinor equations give
\bea
 P_{\bar{\alpha}}=0~,~~~  G_{-+\bar{\alpha}}
+G_{\bar{\alpha}\b}{}^\b
+{1 \over 3} \epsilon_{\bar{\alpha}}{}^{\b_1 \b_2 \b_3} G_{\b_1\b_2\b_3} =0~,
\la{k3k}
\eea

\bea
P_\alpha =0~,~~~ G_{-+\alpha}
-G_{\alpha \b}{}^\b
+{1 \over 3}  \epsilon_\alpha{}^{{\bar{\b_1}} {\bar{\b_2}} {\bar{\b_3}}}
G_{{\bar{\b_1}} {\bar{\b_2}} {\bar{\b_3}}} =0~,
\la{k4k}
\eea

\bea
P_+=0~,~~~  G_{+ \alpha}{}^\a =0~,
\la{k5k}
\eea

and

\be
 G_{+{\bar{\alpha}} {\bar{\beta}}}-{1 \over 2}
\epsilon_{{\bar{\alpha}}
{\bar{\beta}}}{}^{\g \d} G_{+ \g \d}=0~.
\la{k6k}
\ee

Next we turn into the Killing spinor equation associated
with the supercovariant derivative (\ref{kseqna}).
In particular the conditions arise from the ${\cal D}_{\a}$ component of the supercovariant derivative are

\bea
{1\over2}[(f^{-1}D_\a f)_{11}+(f^{-1}D_\a f)_{22}+ i(f^{-1}D_\a f)_{12} -i (f^{-1}D_\a f)_{21}]
+{1\over2}\Omega_{\a,\b}{}^\b
\cr
+
{1\over2} \Omega_{\a,-+}
+i F_{\a-+\b}{}^\b=0~,
\cr
\la{k7k}
\eea

\bea
(f^{-1}D_\a f)_{11}-(f^{-1}D_\a f)_{22}+ i(f^{-1}D_\a f)_{12} +i (f^{-1}D_\a f)_{21}
\cr
+{1\over2}G_{\a\b}{}^\b
+{1\over2} G_{\a-+}=0~,
\la{k8k}
\eea

\bea
 \Omega_{\a,\bar\b_1\bar\b_2}
  +2i F_{\a -+\bar\b_1\bar\b_2}-2i \delta_{\alpha [{\bar{\beta}}_1}F_{{\bar{\beta}}_2]-+\delta}{}^\delta
 -{1\over2} \Omega_{\a,\g_1\g_2} \epsilon^{\g_1\g_2}{}_{\bar\b_1\bar\b_2}=0~,
 \cr
 \la{k9k}
 \eea

 \bea
G_{\a\bar\b_1\bar\b_2}-
 {1\over2} G_{\a\g_1\g_2}\epsilon^{\g_1\g_2}{}_{\bar\b_1\bar\b_2}=0~,
 \la{k10k}
 \eea

 \bea
 {1\over2}[(f^{-1}D_\a f)_{11}+(f^{-1}D_\a f)_{22}+ i(f^{-1}D_\a f)_{12} -i (f^{-1}D_\a f)_{21}]
 \cr
  -{1\over2} \Omega_{\a,\b}{}^\b+{1\over2} \Omega_{\a,-+}
-{i\over3} \epsilon_\alpha{}^{{\bar{\beta}}_1 {\bar{\beta}}_2 {\bar{\beta}}_3}
F_{-+ {\bar{\beta}}_1 {\bar{\beta}}_2 {\bar{\beta}}_3}  =0~,
 \la{k11k}
\eea

\bea
(f^{-1}D_{\a} f)_{11}-(f^{-1}D_{\a} f)_{22}+ i(f^{-1}D_{\a} f)_{12} +i (f^{-1}D_{\a} f)_{21}
\cr
-{1\over2} G_{\a\b}{}^\b+{1\over2} G_{\a-+}=0~,
\la{k12k}
\eea

\bea
 \Omega_{\a,+\bar\b}+i F_{\a+\bar\b\g}{}^\g=0~,
\la{k13k}
\eea

\bea
G_{\a+\bar\b}=0~,
\la{k14k}
\eea

\bea
i F_{\a+\bar\b_1\bar\b_2\bar\b_3}
+{1\over2}
 \Omega_{\a,+\g} \epsilon^\g{}_{\bar\b_1\bar\b_2\bar\b_3} =0~
\la{k15k}
\eea
and
\bea
G_{\a+\g}=0~.
\la{k16k}
\eea

The conditions arise from the ${\cal D}_{\bar\a}$ component of the supercovariant derivative are

\bea
{1\over2}[(f^{-1}D_{\bar\a} f)_{11}+(f^{-1}D_{\bar\a} f)_{22}+ i(f^{-1}D_{\bar\a} f)_{12} -i (f^{-1}D_{\bar\a} f)_{21}]
\cr
+{1\over2} \Omega_{\bar\a,\b}{}^\b+{1\over2} \Omega_{\bar\a, -+}
-{i\over3}  \epsilon_{\bar{\alpha}}{}^{\gamma_1 \gamma_2 \gamma_3}F_{-+\gamma_1 \gamma_2 \gamma_3}
=0~,
\la{k17k}
\eea

\bea
(f^{-1}D_{\bar\a} f)_{11}-(f^{-1}D_{\bar\a} f)_{22}+ i(f^{-1}D_{\bar\a} f)_{12}+i (f^{-1}D_{\bar\a} f)_{21}
\cr
+{1\over2} G_{\bar\a\b}{}^\b+{1\over2}G_{\bar\a -+}=0~,
\la{k18k}
\eea

\bea
\Omega_{\bar\a, \bar\b_1\bar\b_2}
-[{1\over2}\Omega_{\bar\a,\g_1\g_2}
+i F_{\bar\a-+\g_1\g_2}+i \delta_{{\bar{\alpha}} [\gamma_1}F_{\gamma_2] -+ \delta}{}^\delta]
\epsilon^{\g_1\g_2}{}_{\bar\b_1\bar\b_2}=0~,
\la{k19k}
\eea

\bea
G_{\bar\a \bar\b_1\bar\b_2}-{1\over2}G_{\bar\a\g_1\g_2}\epsilon^{\g_1\g_2}{}_{\bar\b_1\bar\b_2}=0~,
\la{k20k}
\eea

\bea
{1\over2}[(f^{-1}D_{\bar\a} f)_{11}+(f^{-1}D_{\bar\a} f)_{22}+ i(f^{-1}D_{\bar\a} f)_{12} -i (f^{-1}D_{\bar\a} f)_{21}]
\cr
-{1\over2} \Omega_{\bar\a,\g}{}^\g+{1\over2} \Omega_{\bar\a,-+}
-i F_{\bar\a-+\g}{}^\g=0~,
\la{k21k}
\eea

\bea
(f^{-1}D_{\bar\a} f)_{11}-(f^{-1}D_{\bar\a} f)_{22}+ i(f^{-1}D_{\bar\a} f)_{12}+i (f^{-1}D_{\bar\a} f)_{21}
\cr
-{1\over2} G_{\bar\a\g}{}^\g+{1\over2} G_{\bar\a-+}=0~,
\la{k22k}
\eea

\bea
 \Omega_{\bar\a,+\bar\b}
-{i\over3}  F_{\bar\a+\g_1\g_2\g_3} \epsilon^{\g_1\g_2\g_3}{}_{\bar\b}=0~,
\la{k23k}
\eea

\bea
G_{\bar\a+\bar\b}=0~,
\la{k24k}
\eea
and
\bea
\Omega_{\bar\a,+\g}
-i F_{\bar\a+\g\d}{}^\d=0~.
\la{k25k}
\eea

Similarly, the conditions arise from the ${\cal D}_{-}$ component of the supercovariant derivative are

\bea
{1\over2}[(f^{-1}D_{-} f)_{11}+(f^{-1}D_{-} f)_{22}+ i(f^{-1}D_{-} f)_{12} -i (f^{-1}D_{-} f)_{21}]
\cr
+{1\over2} \Omega_{-,\g}{}^\g+{1\over2} \Omega_{-,-+}
+{i\over4} F_{-\g}{}^\g{}_\d{}^\d
+{i\over12}  F_{-\g_1\g_2\g_3\g_4} \epsilon^{\g_1\g_2\g_3\g_4}=0~,
\la{k26k}
\eea

\bea
(f^{-1}D_- f)_{11}-(f^{-1}D_- f)_{22}+ i(f^{-1}D_- f)_{12}+i (f^{-1}D_{\bar\a} f)_{21}+{1\over2} G_{-\g}{}^\g=0~,
\la{k27k}
\eea

\bea
\Omega_{-,\bar\b_1\bar\b_2}+i F_{-\bar\b_1\bar\b_2\g}{}^\g
- [{1\over2} \Omega_{-,\g_1\g_2}
-{i\over2} F_{-\g_1\g_2\d}{}^\d] \epsilon^{\g_1\g_2}{}_{\bar\b_1\bar\b_2}=0~,
\la{k28k}
\eea

\bea
G_{-\bar\b_1\bar\b_2}-{1\over2} G_{-\g_1\g_2}\epsilon^{\g_1\g_2}{}_{\bar\b_1\bar\b_2}=0~,
\la{k29k}
\eea

\bea
{1\over2}[(f^{-1}D_{-} f)_{11}+(f^{-1}D_{-} f)_{22}+ i(f^{-1}D_{-} f)_{12} -i (f^{-1}D_{-} f)_{21}]
\cr
-{1\over2} \Omega_{-,\g}{}^\g+{1\over2}\Omega_{-,-+}
+{i\over4} F_{-\g}{}^\g{}_\d{}^\d
+{i\over 12} F_{-\bar\b_1\bar\b_2\bar\b_3\bar\b_4}
\epsilon^{\bar\b_1\bar\b_2\bar\b_3\bar\b_4}=0~,
\la{k30k}
\eea

\bea
(f^{-1}D_- f)_{11}-(f^{-1}D_- f)_{22}+ i(f^{-1}D_- f)_{12}+i (f^{-1}D_{\bar\a} f)_{21}-{1\over2} G_{-\g}{}^\g=0~,
\la{k31k}
\eea

\bea
{1\over2} \Omega_{-,+\bar\b}+{i\over2} F_{-+\bar\b\g}{}^\g
-{i\over6}  F_{-+\g_1\g_2\g_3} \epsilon^{\g_1\g_2\g_3}{}_{\bar\b}=0~,
\la{k32k}
\eea

\bea
G_{-+\bar\b}=0~,
\la{k33k}
\eea

\bea
i F_{-+\bar\b_1\bar\b_2\bar\b_3}
+
[{1\over2} \Omega_{-,+\g}-{i\over2}
F_{-+\g\d}{}^\d] \epsilon^\g{}_{\bar\b_1\bar\b_2\bar\b_3}=0~,
\la{k34k}
\eea
and
\bea
G_{-+\g}=0~.
\la{k35k}
\eea

Finally, the conditions arise from the ${\cal D}_{+}$ component of the supercovariant derivative are

\bea
{1\over2}[(f^{-1}D_{+} f)_{11}+(f^{-1}D_{+} f)_{22}+ i(f^{-1}D_{+} f)_{12} -i (f^{-1}D_{+} f)_{21}]
\cr
+{1\over2} \Omega_{+,\g}{}^\g
+{1\over2} \Omega_{+,-+}=0~,
\la{k36k}
\eea

\bea
(f^{-1}D_+ f)_{11}-(f^{-1}D_+ f)_{22}+ i(f^{-1}D_+ f)_{12}+i (f^{-1}D_{\bar\a} f)_{21}+{1\over2} G_{+\g}{}^\g=0~,
\la{k37k}
\eea

\bea
\Omega_{+,\bar\b_1\bar\b_2}
-{1\over2} \Omega_{+,\g_1\g_2} \epsilon^{\g_1\g_2}{}_{\bar\b_1\bar\b_2}
=0~,
\la{k38k}
\eea

\bea
{1\over2}[(f^{-1}D_{+} f)_{11}+(f^{-1}D_{+} f)_{22}+ i(f^{-1}D_{+} f)_{12} -i (f^{-1}D_{+} f)_{21}]
\cr
-{1\over2} \Omega_{+,\g}{}^\g+{1\over2} \Omega_{+,-+}=0~,
 \la{k39k}
 \eea

 \bea
 (f^{-1}D_+ f)_{11}-(f^{-1}D_+ f)_{22}+ i(f^{-1}D_+ f)_{12}+i (f^{-1}D_{\bar\a} f)_{21}-{1\over2} G_{+\g}{}^\g=0~,
 \la{k40k}
 \eea

 and

 \bea
 \Omega_{+,+\a}=\Omega_{+,+\bar\a}=0~.
 \la{k41k}
 \eea

The above equations are all the restrictions that the Killing spinor equations put on the fluxes
and geometry of maximally supersymmetric $Spin(7)\ltimes \bR^8$-backgrounds. As we shall see, the linear
system can be easily solved to express some of the fluxes in terms of the geometry of spacetime.

\subsection{The solution of the linear system}

The solution of the linear system given in the previous section is straightforward, so we shall not elaborate.
Since the linear system factorizes in equations for $G,P$ and $\Omega, F$ we shall solve the two sets
of equations separately.

It is straightforward to observe that the equations for $P, G$ imply that
all the components of $P$ and $G$ vanish apart from
$P_-$ and $G_{-\a\bar\b}, G_{-\a\b}$ and $G_{-\bar\a\bar\b}$ which in addition satisfy the relations
\bea
G_{-\a}{}^\a=0~,~~~~G_{-\bar\a\bar\b}-{1\over2} \epsilon_{\bar\a\bar\b}{}^{\g\d} G_{-\g\d}~.
\eea
The component $P_-$ is unconstrained. The conditions on $G$ imply that $G_{-AB}$ takes values
in the Lie algebra $spin(7)\otimes \bC$.
In addition to these conditions on the fluxes, the functions that determine the spinors are constrained
as
\bea
(f^{-1}\partial_A f)_{11}-(f^{-1}\partial_A f)_{22}= (f^{-1}\partial_A f)_{12}+ (f^{-1}\partial_Af)_{21}=0~,~~~
A=-,+, \a, \bar\a
\eea

To solve of remaining equations of the linear system we first consider the equations derived
from ${\cal D}_+$ component of the supercovariant connection. After some straightforward analysis,
we find that these imply
\bea
\Omega_{+,+\a}=0~,~~~\Omega_{+,\a}{}^\a=0~,~~~
\cr
\Omega_{+,\bar\a\bar\b}-{1\over2} \epsilon_{\bar\a\bar\b}{}^{\g\d} \Omega_{+,\g\d}=0
\cr
(f^{-1}\partial_+ f)_{11}+(f^{-1}\partial_+ f)_{22}+\Omega_{+,-+}=0
\cr
(f^{-1}\partial_+ f)_{12}-(f^{-1}\partial_+ f)_{21}-Q_+=0~.
\eea
Next we turn to the conditions derived from ${\cal D}_-$ component of the supercovariant connection and after
a similar analysis, we find that
\bea
\Omega_{-,+\a}=0~,~~~~\Omega_{-,\a}{}^\a=0~,
\cr
\Omega_{-,\a\b}-{1\over2} \epsilon_{\a\b}{}^{\bar\g\bar\d} \Omega_{-,\bar\g\bar\d}=0
\cr
F_{-+\bar\a\d}{}^\d-{1\over3} F_{-+\g_1\g_2\g_3} \epsilon^{\g_1\g_2\g_3}{}_{\bar\a}=0
\cr
F_{-\bar\a_1\bar\a_2\bar\a_3\bar\a_4} \epsilon^{\bar\a_1\bar\a_2\bar\a_3\bar\a_4}-
F_{-\a_1\a_2\a_3\a_4} \epsilon^{\a_1\a_2\a_3\a_4}=0~,
\cr
F_{-\a_1\a_2\d}{}^\d+{1\over2} \epsilon_{\a_1\a_2}{}^{\bar\b_1\bar\b_2} F_{-\bar\b_1\bar\b_2\d}{}^\d=0~,
\cr
(f^{-1}\partial_- f)_{11}+(f^{-1}\partial_- f)_{22}+\Omega_{-,-+}=0
\cr
(f^{-1}\partial_- f)_{12}-(f^{-1}\partial_- f)_{21}- Q_-+{1\over2} F_{-\g}{}^\g{}_\d{}^\d+{1\over6}
F_{-\a_1\a_2\a_3\a_4} \epsilon^{\a_1\a_2\a_3\a_4}=0~.
\eea
Similarly, we investigate the conditions that arise from the ${\cal D}_\a$ and ${\cal D}_{\bar\a}$
components of the supercovariant derivative. We find that
\bea
\Omega_{\bar\a,+\b}=0~,~~~\Omega_{\a,+\b}=0
\cr
F_{+\bar\a\b_1\b_2\b_3}=0~,~~~F_{+\a\bar\b_1\bar\b_2\bar\b_3}=0~,~~~F_{-+\b_1\b_2\bar\a}=0~,
\cr
\Omega_{\bar\a, \b_1\b_2}-{1\over2} \epsilon_{\b_1\b_2}{}^{\bar\g_1\bar\g_2} \Omega_{\bar\a, \bar\g_1\bar\g_2}=0
\cr
F_{\bar\a\b_1\b_2\b_3\b_4}=0~,~~~F_{-+\b_1\b_2\b_3}=0~,~~~F_{+\a\bar\b\d}{}^\d=0~,
\cr
\Omega_{\a,\b}{}^\b=0
\cr
\cr
(f^{-1}\partial_\a f)_{11}+(f^{-1}\partial_\a f)_{22}+\Omega_{\a,-+}=0
\cr
(f^{-1}\partial_\a f)_{12}-(f^{-1}\partial_\a f)_{21}-Q_\a=0~.
\eea
The conditions on the geometry and the fluxes are summarized in section four. In the same section,
the geometry of the spacetime of maximally supersymmetric $Spin(7)\ltimes\bR^8$-backgrounds is also
investigated.

\newsection{Maximally supersymmetric $SU(4)\ltimes\bR^8$-backgrounds}

\subsection{A linear system}

In this appendix, we give the details of the derivation of the
conditions on the geometry and the fluxes implied by the Killing
spinor equations for maximally supersymmetric
$SU(4)\ltimes\bR^8$-backgrounds. These conditions have already been
summarized in section five. The linear system can be easily derived
using (\ref{maxsuf}), the results of \cite{gju} and appendix A. In
particular, the algebraic part of the Killing spinor equations in
({\ref{maxsuf}}) can be rewritten as

\be
\label{algsu4i}
P_A \Gamma^A 1 =0 , \qquad P_A \Gamma^A  e_{1234}=0
\ee

and

\be
\label{algsu4ii}
G_{A_1 A_2 A_3} \Gamma^{A_1 A_2 A_3} 1 =0 , \qquad G_{A_1 A_2 A_3} \Gamma^{A_1 A_2 A_3} e_{1234} =0~.
\ee

To investigate the rest of the Killing spinor equations, it is most convenient to define the one-forms

\bea
\label{xidents}
X^1 &=& {1 \over 2} \big[(f^{-1}Df)_{11}+(f^{-1}Df)_{33}+i(f^{-1}Df)_{13}-i(f^{-1}Df)_{31}\big]~,
\cr
X^2 &=& {1 \over 2} \big[(f^{-1}Df)_{12}+(f^{-1}Df)_{34}+i(f^{-1}Df)_{14}-i(f^{-1}Df)_{32}\big]~,
\cr
X^3 &=& {1 \over 2} \big[(f^{-1}Df)_{21}+(f^{-1}Df)_{43}+i(f^{-1}Df)_{23}-i(f^{-1}Df)_{41}\big]~,
\cr
X^4 &=& {1 \over 2} \big[(f^{-1}Df)_{22}+(f^{-1}Df)_{44}+i(f^{-1}Df)_{24}-i(f^{-1}Df)_{42}\big]~,
\cr
X^5 &=&  (f^{-1}Df)_{11}-(f^{-1}Df)_{33}+i(f^{-1}Df)_{13}+i(f^{-1}Df)_{31}~,
\cr
X^6 &=&  (f^{-1}Df)_{12}-(f^{-1}Df)_{34}+i(f^{-1}Df)_{14}+i(f^{-1}Df)_{32}~,
\cr
X^7 &=&  (f^{-1}Df)_{21}-(f^{-1}Df)_{43}+i(f^{-1}Df)_{23}+i(f^{-1}Df)_{41}~,
\cr
X^8 &=&  (f^{-1}Df)_{22}-(f^{-1}Df)_{44}+i(f^{-1}Df)_{24}+i(f^{-1}Df)_{42}~,
\eea
which  are constructed by taking the $D$ derivative on the functions $f$.

The remaining equations  of ({\ref{maxsuf}})  can be rewritten as

\bea
\label{puresp1}
{1 \over 2} (X^1-X^4+iX^3+iX^2)_M ~1 +{1 \over 2} (X^1+X^4+iX^3-iX^2)_M ~e_{1234}
\cr
+{1 \over 4} \Omega_{MAB} \Gamma^{AB} ~e_{1234}
+{i \over 48} \Gamma^{B_1 B_2 B_3 B_4} F_{M B_1 B_2 B_3 B_4} ~e_{1234}&=&0~,
\cr
{1 \over 2} (X^5-X^8+iX^7+iX^6)_M ~1 +{1 \over 2} (X^5+X^8+iX^7-iX^6)_M ~e_{1234}
\cr
+{1 \over 4} G_{MAB} \Gamma^{AB} ~e_{1234}&=&0~,
\eea

and

\bea
\label{puresp2}
{1 \over 2} (X^1-X^4-iX^2-iX^3)_M ~ e_{1234} +{1 \over 2} (X^1+X^4+iX^2-iX^3)_M ~1
\cr
+{1 \over 4} \Omega_{MAB} \Gamma^{AB} ~1
+{i \over 48} \Gamma^{B_1 B_2 B_3 B_4} F_{M B_1 B_2 B_3 B_4} ~1&=&0~,
\cr
{1 \over 2} (X^5-X^8-iX^6-iX^7)_M ~e_{1234} +{1 \over 2} (X^5+X^8+iX^6-iX^7) ~1
\cr
+{1 \over 4} G_{MAB} \Gamma^{AB} ~1&=&0~.
\eea

Rewriting the equations in this way allows for an extremely simple comparison
with the Killing spinor equations evaluated on pure spinors, which have
been already been computed in \cite{gju}. However, one must be careful to
include the  $X$-terms which are not present in the pure spinor case.

\subsection{The solution of the linear system}

The solution of the  algebraic Killing spinor equations, ({\ref{algsu4i}}),
  is

\be
\label{su4psol}
P_+=0 , \quad P_\alpha=0 , \quad P_{\bar{\alpha}}=0
\ee

and $P_-$ is {\it unconstrained}. From ({\ref{algsu4ii}}),
we also find that

\bea
\label{su4gsoli}
G_{+ \alpha}{}^\alpha &=&0~,
\cr
G_{+ \alpha \beta} &=& 0~,
\cr
G_{+ {\bar{\alpha}} {\bar{\beta}}}&=&0~,
\cr
G_{\alpha_1 \alpha_2 \alpha_3}&=&0~,
\cr
G_{ {\bar{\alpha}}_1 {\bar{\alpha}}_2 {\bar{\alpha}}_3} &=&0~,
\cr
G_{-+ \alpha}-G_{\alpha \beta}{}^\beta &=&0~,
\cr
G_{-+ {\bar{\alpha}}}+ G_{{\bar{\alpha}} \beta}{}^\beta &=&0~.
\eea

Next consider the $G$ dependent parts of ({\ref{puresp1}}) and ({\ref{puresp1}}).
In particular, the $\alpha$ components of these equations imply the conditions

\bea
\label{alphagcon1}
(X^5-X^8+iX^7+iX^6)_\alpha &=&0~,
\cr
G_{\alpha_1 \alpha_2 \alpha_3} &=&0~,
\cr
(X^5+X^8+iX^7-iX^6)_\alpha -G_{\alpha \beta}{}^\beta + G_{-+\alpha} &=&0~,
\cr
G_{+\alpha \beta} &=&0~,
\cr
(X^5+X^8+iX^6 -iX^7)_{\alpha} +G_{\alpha \beta}{}^\beta +G_{-+\alpha} &=&0~,
\cr
G_{\alpha {\bar{\beta}}_1 {\bar{\beta}}_2} &=&0~,
\cr
(X^5-X^8-iX^6-iX^7)_\alpha &=&0~,
\cr
G_{+ \alpha {\bar{\beta}}} &=&0~.
\eea

Similarly, for  the ${\bar{\alpha}}$ components,  we find
\bea
\label{alphagcon2}
(X^5-X^8+iX^7+iX^6)_{\bar{\alpha}} &=&0~,
\cr
G_{{\bar{\alpha}} \beta_1 \beta_2} &=&0~,
\cr
(X^5+X^8+iX^7-iX^6)_{\bar{\alpha}} -G_{{\bar{\alpha}} \beta}{}^\beta + G_{-+{\bar{\alpha}}} &=&0~,
\cr
G_{+{\bar{\alpha}} \beta} &=&0~,
\cr
(X^5+X^8+iX^6 -iX^7)_{{\bar{\alpha}}} +G_{{\bar{\alpha}} \beta}{}^\beta +G_{-+{\bar{\alpha}}} &=&0~,
\cr
G_{ {\bar{\alpha}}_1 {\bar{\alpha}}_2 {\bar{\alpha}}_3 } &=&0~,
\cr
(X^5-X^8-iX^6-iX^7)_{\bar{\alpha}} &=&0~,
\cr
G_{+ {\bar{\alpha}}_1 {\bar{\alpha}}_2 } &=&0~.
\eea

The $-$ components, we find  the conditions

\bea
\label{alphagcon3}
(X^5-X^8+iX^7+iX^6)_- &=&0~,
\cr
G_{-\gamma_1 \gamma_2}&=&0~,
\cr
(X^5+X^8+iX^7-iX^6)_- - G_{- \beta}{}^\beta &=&0~,
\cr
G_{-+\gamma} &=&0~,
\cr
(X^5+X^8+iX^6-iX^7)_- + G_{- \beta}{}^\beta &=&0~,
\cr
G_{- {\bar{\alpha}}_1 {\bar{\alpha}}_2} &=&0~,
\cr
(X^5-X^8-iX^6-iX^7)_- &=&0~,
\cr
G_{-+ {\bar{\gamma}}} &=&0~,
\eea

and from the $+$ components, we obtain

\bea
\label{alphagcon4}
(X^5-X^8+iX^7+iX^6)_+ &=&0~,
\cr
G_{+ {\bar{\alpha}}_1 {\bar{\alpha}}_2} &=&0~,
\cr
(X^5+X^8+iX^7-iX^6)_+ - G_{+ \beta}{}^\beta &=&0~,
\cr
(X^5+X^8+iX^6-iX^7)_+ + G_{+ \beta}{}^\beta &=&0~,
\cr
G_{+ {\bar{\alpha}}_1 {\bar{\alpha}}_2} &=&0~,
\cr
(X^5-X^8-iX^6-iX^7)_+ &=&0~.
\eea

Combining these equations with ({\ref{su4gsoli}}),  we see that all components of $G$ vanish,
except for $G_{-\alpha {\bar{\beta}}}$,  the trace satisfies
\be
\label{gmmincon}
G_{- \beta}{}^\beta = 2i ( (f^{-1}\partial_- f)_{21}- (f^{-1}\partial_- f)_{43}
+i (f^{-1}\partial_- f)_{23} +i (f^{-1}\partial_- f)_{41})~,
\ee
and  the following conditions on $f$:

\bea
\label{ffsu4con1}
(f^{-1} \partial_A f)_{11}&=&(f^{-1} \partial_A f)_{33}~,
\cr
(f^{-1} \partial_A f)_{13}+(f^{-1} \partial_A f)_{31} &=&0~,
\cr
(f^{-1} \partial_A f)_{12} &=& (f^{-1} \partial_A f)_{34}~,
\cr
(f^{-1} \partial_A f)_{14} +(f^{-1} \partial_A f)_{32} &=&0~,
\cr
(f^{-1} \partial_A f)_{21} &=& (f^{-1} \partial_A f)_{43}~,
\cr
(f^{-1} \partial_A f)_{23}+(f^{-1} \partial_A f)_{41} &=&0~,
\cr
(f^{-1} \partial_A f)_{22} &=& (f^{-1} \partial_A f)_{44}~,
\cr
(f^{-1} \partial_A f)_{24} + (f^{-1} \partial_A f)_{42} &=&0~,
\eea
for $A=+, \alpha, {\bar{\alpha}}$, and

\bea
\label{ffsu4con2}
(f^{-1} \partial_- f)_{11} &=& (f^{-1} \partial_- f)_{33}~,
\cr
(f^{-1} \partial_- f)_{13}+(f^{-1} \partial_- f)_{31} &=& 0~,
\cr
(f^{-1} \partial_- f)_{22} &=& (f^{-1} \partial_- f)_{44}~,
\cr
(f^{-1} \partial_- f)_{24}+(f^{-1} \partial_- f)_{42} &=& 0~,
\cr
(f^{-1} \partial_- f)_{12} -(f^{-1} \partial_- f)_{34}+(f^{-1} \partial_- f)_{21}-(f^{-1} \partial_- f)_{43} &=& 0~,
\cr
(f^{-1} \partial_- f)_{14} +(f^{-1} \partial_- f)_{41}+(f^{-1} \partial_- f)_{23}+(f^{-1} \partial_- f)_{32} &=& 0~.
\eea

We next consider the $F$-dependent parts of the Killing spinor
equation. From $\alpha$ component, we obtain

\bea
\label{su4Feqn1}
(X^1-X^4+iX^3+iX^2)_\alpha &=&0~,
\cr
(X^1+X^4+iX^3-iX^2)_\alpha - \Omega_{\alpha, \beta}{}^\beta +\Omega_{\alpha, -+} &=&0~,
\cr
\Omega_{\alpha_1 , \alpha_2 \alpha_3} &=&0~,
\cr
\Omega_{\alpha,+ \beta} &=&0~,
\cr
(X^1+X^4+iX^2-iX^3)_\alpha + \Omega_{\alpha, \beta}{}^\beta+ \Omega_{\alpha,-+}
+2F_{-+\alpha \beta}{}^\beta &=&0~,
\cr
\Omega_{\alpha, {\bar{\beta}}_1 {\bar{\beta}}_2}+2i F_{-+\alpha {\bar{\beta}}_1 {\bar{\beta}}_2}
-2i \g_{\alpha [{\bar{\beta}}_1}F_{{\bar{\beta}}_2]-+ \gamma}{}^\gamma &=&0~,
\cr
(X^1-X^4-iX^2-iX^3)_\alpha-{2i \over 3} \epsilon_{\alpha}{}^{{\bar{\beta}}_1
{\bar{\beta}}_2 {\bar{\beta}}_3} F_{-+{\bar{\beta}}_1
{\bar{\beta}}_2 {\bar{\beta}}_3} &=&0~,
\cr
\Omega_{\alpha, + {\bar{\beta}}}+i F_{\alpha + {\bar{\beta}} \gamma}{}^\gamma &=&0
\cr
F_{\alpha + {\bar{\beta}}_1
{\bar{\beta}}_2 {\bar{\beta}}_3} &=&0~.
\eea

Similarly, from the ${\bar{\alpha}}$ component, we find

\bea
\label{su4Feqn2}
(X^1-X^4+iX^3+iX^2)_{\bar{\alpha}} -{2i \over 3} \epsilon_{\bar{\alpha}}{}^{\beta_1 \beta_2 \beta_3}
F_{-+ \beta_1 \beta_2 \beta_3} &=&0~,
\cr
\Omega_{{\bar{\alpha}}, \beta_1 \beta_2}+2iF_{-+{\bar{\alpha}} \beta_1 \beta_2}
+2ig_{{\bar{\alpha}} [\beta_1}F_{\beta_2] -+\gamma}{}^\gamma &=&0~,
\cr
(X^1+X^4+iX^3-iX^2)_{\bar{\alpha}}-\Omega_{{\bar{\alpha}}, \beta}{}^\beta + \Omega_{{\bar{\alpha}},-+}
-2iF_{-+ {\bar{\alpha}} \beta}{}^\beta &=&0~,
\cr
F_{{\bar{\alpha}}+\beta_1 \beta_2 \beta_3} &=&0~,
\cr
\Omega_{{\bar{\alpha}},+ \beta} -i F_{{\bar{\alpha}}+ \beta \gamma}{}^\gamma &=&0~,
\cr
(X^1+X^4+iX^2-iX^3)_{\bar{\alpha}} + \Omega_{{\bar{\alpha}}, \beta}{}^\beta + \Omega_{{\bar{\alpha}},-+}&=&0~,
\cr
\Omega_{{\bar{\alpha}}_1, {\bar{\alpha}}_2 {\bar{\alpha}}_3} &=&0~,
\cr
(X^1-X^4-iX^2-iX^3)_{\bar{\alpha}} &=&0~,
\cr
\Omega_{\bar{\alpha}, +{\bar{\beta}}}&=&0~.
\eea

Next from the $-$ component,  we obtain

\bea
\label{su4Feqn3}
(X^1-X^4+iX^3+iX^2)_- +{i \over 6} F_{-\alpha_1 \alpha_2 \alpha_3 \alpha_4}
\epsilon^{\alpha_1 \alpha_2 \alpha_3 \alpha_4} &=&0~,
\cr
\Omega_{-,\beta_1 \beta_2}-iF_{- \beta_1 \beta_2 \gamma}{}^\gamma &=&0~,
\cr
(X^1+X^4+iX^3-iX^2)_- -\Omega_{-, \beta}{}^\beta +\Omega_{-,-+}+{i \over 2}
F_{-\beta}{}^\beta{}_\gamma{}^\gamma &=&0~,
\cr
F_{-+\beta_1 \beta_2 \beta_3}&=&0~,
\cr
\Omega_{-,+\alpha}-iF_{-+\alpha \beta}{}^\beta &=&0~,
\cr
(X^1+X^4+iX^2-iX^3)_- + \Omega_{-,\beta}{}^\beta + \Omega_{-,-+}+{i \over 2}
F_{-\beta}{}^\beta{}_\gamma{}^\gamma &=&0~,
\cr
\Omega_{-, {\bar{\beta}}_1 {\bar{\beta}}_2}+iF_{- {\bar{\beta}}_1 {\bar{\beta}}_2 \gamma}{}^\gamma &=&0~,
\cr
(X^1-X^4-iX^2-iX^3)_- +{i \over 6}F_{- {\bar{\beta}}_1 {\bar{\beta}}_2
{\bar{\beta}}_3 {\bar{\beta}}_4} \epsilon^{{\bar{\beta}}_1 {\bar{\beta}}_2
{\bar{\beta}}_3 {\bar{\beta}}_4} &=&0~,
\cr
\Omega_{-,+{\bar{\alpha}}}+iF_{-+{\bar{\alpha}} \gamma}{}^\gamma &=&0~,
\cr
F_{-+{\bar{\beta}}_1 {\bar{\beta}}_2 {\bar{\beta}}_3} &=&0~,
\eea

and from the $+$ component,  we find

\bea
\label{su4Feqn4}
(X^1-X^4+iX^3+iX^2)_+ &=&0~,
\cr
\Omega_{+,\beta_1 \beta_2} &=&0~,
\cr
(X^1+X^4+iX^3-iX^2)_+ - \Omega_{+, \beta}{}^\beta + \Omega_{+,-+} &=&0~,
\cr
\Omega_{+,+\alpha} &=&0~,
\cr
\Omega_{+,+{\bar{\alpha}}} &=&0~,
\cr
(X^1+X^4+iX^2-iX^3)_+ + \Omega_{+, \beta}{}^\beta + \Omega_{+,-+} &=&0~,
\cr
\Omega_{+,{\bar{\beta}}_1 {\bar{\beta}}_2} &=&0~,
\cr
(X^1-X^4-iX^2-iX^3)_+ &=&0~.
\eea

Next we turn our attention to the $X$-independent parts of the Killing spinor equations to  obtain

\bea
\label{xindep1}
F_{+ \alpha {\bar{\beta}}_1 {\bar{\beta}}_2 {\bar{\beta}}_3} &=&0~,
\cr
F_{-+ \beta_1 \beta_2 \beta_3} &=&0~,
\cr
F_{- \beta_1 \beta_2 \gamma}{}^\gamma &=&0~,
\cr
F_{+\alpha {\bar{\beta}} \gamma}{}^\gamma &=&0~,
\cr
F_{-+ \alpha  {\bar{\beta}}_1 {\bar{\beta}}_2 } &=&0~,
\eea

on the flux $F$,  and

\bea
\label{xindep2}
\Omega_{\alpha_1, \alpha_2 \alpha_3} &=&0~,
\cr
\Omega_{{\bar{\alpha}} , \beta_1 \beta_2} &=&0~,
\cr
\Omega_{\alpha, + {\bar{\beta}}} &=&0~,
\cr
\Omega_{-,\alpha_1 \alpha_2} &=&0~,
\cr
\Omega_{-,+\alpha} &=&0~,
\cr
\Omega_{+, \beta_1 \beta_2}&=&0~,
\cr
\Omega_{\alpha_1 , + \alpha_2} &=&0~,
\eea
on the geometry of spacetime.

 Lastly, we consider
the remaining equations, together with ({\ref{ffsu4con1}})
and ({\ref{ffsu4con1}}). From the $+$, $\alpha$ and ${\bar{\alpha}}$ components,
we find

\bea
\label{finalsu4fconstri}
(f^{-1} \partial_A f)_{11} = (f^{-1} \partial_A f)_{22} = (f^{-1} \partial_A f)_{33} = (f^{-1} \partial_A f)_{44}~,
\cr
(f^{-1} \partial_A f)_{23} = (f^{-1} \partial_A f)_{32} = (f^{-1} \partial_A f)_{24} = (f^{-1} \partial_A f)_{41} =0~,
\cr
(f^{-1} \partial_A f)_{12} = - (f^{-1} \partial_A f)_{21} = (f^{-1} \partial_A f)_{34} = -(f^{-1} \partial_A f)_{43}~,
\cr
(f^{-1} \partial_A f)_{13} = - (f^{-1} \partial_A f)_{31} = (f^{-1} \partial_A f)_{24} = - (f^{-1} \partial_A f)_{42}~,
\eea
and
\bea
\label{finalsu4fconstrii}
(f^{-1} \partial_A f)_{11} + {1 \over 2} \Omega_{A,-+} &=&0
\cr
2 (f^{-1} \partial_A f)_{13}-Q_A &=& 0
\cr
2i (f^{-1} \partial_A f)_{21}-\Omega_{A, \beta}{}^\beta &=& 0
\eea
for $A=+,\alpha, {\bar{\alpha}}$. From the $-$ component, we obtain
\be
\label{finalsu4fconstr3}
(f^{-1} \partial_- f)_{11} = (f^{-1} \partial_- f)_{22} = (f^{-1} \partial_- f)_{33} = (f^{-1} \partial_- f)_{44}~,
\ee
and
\be
\label{finalsu4fconstr4}
(f^{-1} \partial_- f)_{ij}+(f^{-1} \partial_- f)_{ji} =0~,
\ee
for $i,j=1,2,3,4$ and $i \neq j$, together with the constraints

\bea
\label{finalsu4fconstr5}
(f^{-1} \partial_- f)_{11}+{1 \over 2} \Omega_{-,-+} &=&0
\cr
  (f^{-1} \partial_- f)_{13} + (f^{-1} \partial_- f)_{24} - Q_-+{1 \over 2} F_{- \gamma}{}^\gamma{}_\beta{}^\beta&=&0~,
\cr
i ((f^{-1} \partial_- f)_{21}+(f^{-1} \partial_- f)_{43})-\Omega_{-,\beta}{}^\beta &=& 0~,
\eea

and

\bea
(f^{-1} \partial_- f)_{13} - (f^{-1} \partial_- f)_{24}
+ {1 \over 12}(F_{- \alpha_1 \alpha_2
\alpha_3 \alpha_4} \epsilon^{\alpha_1 \alpha_2
\alpha_3 \alpha_4}
\qquad &&
\cr
+ F_{- {\bar{\alpha}}_1 {\bar{\alpha}}_2 {\bar{\alpha}}_3 {\bar{\alpha}}_4}
\epsilon^{{\bar{\alpha}}_1 {\bar{\alpha}}_2 {\bar{\alpha}}_3 {\bar{\alpha}}_4})&=&0~,
\cr
(f^{-1} \partial_- f)_{14}-(f^{-1} \partial_- f)_{32}
 -{i \over 12}
(F_{- \alpha_1 \alpha_2
\alpha_3 \alpha_4} \epsilon^{\alpha_1 \alpha_2
\alpha_3 \alpha_4}
\qquad &&
\cr
 - F_{- {\bar{\alpha}}_1 {\bar{\alpha}}_2 {\bar{\alpha}}_3 {\bar{\alpha}}_4}
\epsilon^{{\bar{\alpha}}_1 {\bar{\alpha}}_2 {\bar{\alpha}}_3 {\bar{\alpha}}_4})&=&0~.
\label{finalsu4fconstr6}
\eea
The conditions on the geometry and fluxes derived in this appendix are summarized in section five. In this section,
the geometry of maximally supersymmetric $SU(4)\ltimes \bR^8$ backgrounds is also investigated.

\newsection{Maximally supersymmetric $G_2$-backgrounds}

\subsection{Simplifying the Linear System}

As in the previous cases, in this appendix, we shall give the detail derivation of the conditions
for the fluxes and geometry for maximally supersymmetric $G_2$-backgrounds. A summary
of these conditions has already appeared in section six.
In order to find the solution of the Killing spinor equations of the maximally supersymmetric $G_2$ backgrounds
in the most efficient manner,
it is convenient to deal with the Killing spinor equations given
in ({\ref{maxsuf}}) in a particular order. First, from the algebraic Killing spinor equations
\be
P_A \Gamma^A \eta_1=0~,~~~P_A\Gamma^A \eta_2=0~,
\ee
we find that
\be
P =0~,
\ee
i.e. the $P$ flux vanishes and the scalars are constant.

Next consider the algebraic Killing spinor equations involving $G$.
In particular,  from
\be
\label{algg2con1}
 G_{N_1N_2N_3} \Gamma^{N_1N_2N_3} \eta_1=0~,~~~G_{N_1N_2N_3} \Gamma^{N_1N_2N_3} \eta_2=0~,
\ee
we find
\bea
\label{algg2con2}
G_{-+\bar{p}}+G_{\bar{p}q}{}^q +G_{1 \bar{1} \bar{p}} - \epsilon_{\bar{p}}{}^{q_1 q_2}
G_{1 q_1 q_2} &=&0~,
\cr
G_{-+ \bar{1}} +G_{\bar{1} q}{}^q +2 G_{234} &=&0
\cr
G_{-+p}-G_{pq}{}^q -G_{1 \bar{1} p}-\epsilon_p{}^{{\bar{q}}_1 {\bar{q}}_2} G_{\bar{1}
{\bar{q}}_1 {\bar{q}}_2} &=&0~,
\cr
G_{-+1}-G_{1q}{}^q +2 G_{\bar{2} \bar{3} \bar{4}} &=&0~,
\cr
G_{+q}{}^q +G_{+ 1 \bar{1}} &=&0~,
\cr
G_{+\bar{p} \bar{q}}+\epsilon_{\bar{p} \bar{q}}{}^r G_{+r1} &=&0~,
\cr
G_{+ \bar{1} \bar{p}} -{1 \over 2} \epsilon_{\bar{p}}{}^{q_1 q_2}G_{+ q_1 q_2} &=&0~,
\cr
G_{\bar{p} 1-}-{1 \over 2} \epsilon_{\bar{p}}^{q_1 q_2} G_{- q_1 q_2} &=&0~,
\cr
G_{-q}{}^q -G_{-1 \bar{1}} &=&0~,
\cr
G_{p \bar{1} -} -{1 \over 2} \epsilon_p{}^{{\bar{q}}_1 {\bar{q}}_2}G_{- {\bar{q}}_1 {\bar{q}}_2}
&=&0~,
\cr
G_{-+1}-G_{1q}{}^q +2 G_{234} &=&0~,
\cr
G_{-+\bar{1}}+G_{\bar{1}q}{}^q +2 G_{\bar{2} \bar{3} \bar{4}} &=&0~,
\cr
G_{pq}{}^q -G_{p 1 \bar{1}} +G_{-+p} &=&0~,
\cr
G_{\bar{p} q}{}^q -G_{\bar{p} 1 \bar{1}}-G_{-+\bar{p}}+\epsilon_{\bar{p}}^{q_1 q_2}
G_{\bar{1} q_1 q_2} &=&0~.
\eea
These complete the list of conditions on $G$ required by the algebraic Killing spinor equations.

Next consider the equations
\bea
\label{algg2con3}
X^5{}_M \eta_1 + X^6{}_M \eta_2 + {1\over 4}  \Gamma^{N_1N_2} G_{MN_1N_2} \eta_1&=&0~,
\cr
X^7{}_M \eta_1 + X^8{}_M \eta_2 + {1\over 4}  \Gamma^{N_1N_2} G_{MN_1N_2} \eta_2&=&0~,
\eea
where we have found it useful to use the notation given in ({\ref{xidents}}).
Contracting with $\Gamma^M$ and using ({\ref{algg2con1}}), we obtain
\bea
\label{algg2con4}
X^5{}_p = X^5{}_{\bar{p}} &=&0~,
\cr
X^6{}_p = X^6{}_{\bar{p}} &=&0~,
\cr
X^7{}_p = X^7{}_{\bar{p}} &=&0~,
\cr
X^8{}_p = X^8{}_{\bar{p}} &=&0~,
\eea
and
\bea
\label{algg2con5}
X^5{}_1 &=& X^5{}_{\bar{1}}~,
\cr
X^6{}_1 &=& X^6{}_{\bar{1}}~,
\cr
X^7{}_1 &=& X^7{}_{\bar{1}}~,
\cr
X^8{}_1 &=& X^8{}_{\bar{1}}~.
\eea
Hence, from ({\ref{algg2con4}}), we must have
\bea
\label{simplermaxg21}
G_{p N_1 N_2} \Gamma^{N_1 N_2} \eta_1 &=&0~,
\cr
G_{\bar{p} N_1 N_2} \Gamma^{N_1 N_2} \eta_1 &=&0~,
\cr
G_{p N_1 N_2} \Gamma^{N_1 N_2} \eta_2 &=&0~,
\cr
G_{\bar{p} N_1 N_2} \Gamma^{N_1 N_2} \eta_2 &=&0~,
\eea

which imply the conditions
\bea
\label{simplermaxg22}
G_{pq}{}^q +G_{p 1 \bar{1}} +G_{-+p} &=&0~,
\cr
G_{p1-} &=&0~,
\cr
G_{p {\bar{q}}_1 {\bar{q}}_2}-\epsilon_{{\bar{q}}_1 {\bar{q}}_2}{}^r G_{p1r} &=&0~,
\cr
G_{-pq} &=&0~,
\cr
G_{p \bar{1} \bar{q}}-{1 \over 2} \epsilon_{\bar{q}}{}^{r_1 r_2}G_{p r_1 r_2} &=&0~,
\cr
G_{-p \bar{q}} &=&0~,
\cr
-G_{pq}{}^q -G_{p 1 \bar{1}}+G_{-+p} &=&0~,
\cr
G_{p \bar{1}-} &=&0~,
\cr
G_{+p \bar{q}} &=&0~,
\cr
G_{p \bar{q} 1} -{1 \over 2} \epsilon_{\bar{q}}{}^{r_1 r_2} G_{p r_1 r_2} &=&0~,
\cr
G_{p+\bar{1}} &=&0~,
\cr
-G_{p 1 \bar{1}}-G_{-+p}+G_{pq}{}^q &=&0~,
\cr
G_{p+1} &=&0~,
\cr
-G_{pq}{}^q +G_{p 1 \bar{1}} -G_{-+p} &=&0~,
\cr
G_{p+q}&=&0~,
\cr
G_{p {\bar{q}}_1 {\bar{q}}_2}- \epsilon_{{\bar{q}}_1 {\bar{q}}_2}{}^r G_{pr \bar{1}} &=&0~,
\eea
and
\bea
\label{simplermaxg23}
G_{{\bar{p}}q}{}^q +G_{{\bar{p}} 1 \bar{1}} +G_{-+{\bar{p}}} &=&0~,
\cr
G_{{\bar{p}}1-} &=&0~,
\cr
G_{{\bar{p}} {\bar{q}}_1 {\bar{q}}_2}-\epsilon_{{\bar{q}}_1 {\bar{q}}_2}{}^r G_{{\bar{p}}1r} &=&0~,
\cr
G_{-{\bar{p}}q} &=&0~,
\cr
G_{{\bar{p}} \bar{1} \bar{q}}-{1 \over 2} \epsilon_{\bar{q}}{}^{r_1 r_2}G_{{\bar{p}} r_1 r_2} &=&0~,
\cr
G_{-{\bar{p}} \bar{q}} &=&0~,
\cr
-G_{{\bar{p}}q}{}^q -G_{{\bar{p}} 1 \bar{1}}+G_{-+{\bar{p}}} &=&0~,
\cr
G_{{\bar{p}} \bar{1}-} &=&0~,
\cr
G_{+{\bar{p}} \bar{q}} &=&0~,
\cr
G_{{\bar{p}} \bar{q} 1} -{1 \over 2} \epsilon_{\bar{q}}{}^{r_1 r_2} G_{{\bar{p}} r_1 r_2} &=&0~,
\cr
G_{{\bar{p}}+\bar{1}} &=&0~,
\cr
-G_{{\bar{p}} 1 \bar{1}}-G_{-+{\bar{p}}}+G_{{\bar{p}}q}{}^q &=&0~,
\cr
G_{{\bar{p}}+1} &=&0~,
\cr
-G_{{\bar{p}}q}{}^q +G_{{\bar{p}} 1 \bar{1}} -G_{-+{\bar{p}}} &=&~,
\cr
G_{{\bar{p}}+q}&=&0~,
\cr
G_{{\bar{p}} {\bar{q}}_1 {\bar{q}}_2}- \epsilon_{{\bar{q}}_1 {\bar{q}}_2}{}^r G_{{\bar{p}}r \bar{1}} &=&0~.
\eea

Then, taking ({\ref{algg2con2}}) together with ({\ref{simplermaxg22}})
and ({\ref{simplermaxg23}}), it is straightforward to show that all components of
$G$ are constrained to vanish, with the exception of
$G_{-+1}$, $G_{-+\bar{1}}$,
$G_{1 p {\bar{q}}}$, $G_{\bar{1} p \bar{q}}$, $G_{234}$, $G_{\bar{2} \bar{3} \bar{4}}$
which are related via
\bea
\label{simplermaxg24}
G_{1 p \bar{q}} &=& g_{p \bar{q}} G_{234}~,
\cr
G_{\bar{1} p \bar{q}} &=& -  g_{p \bar{q}} G_{234}~,
\cr
G_{234} &=& G_{\bar{2} \bar{3} \bar{4}}~,
\cr
G_{-+1} &=& G_{234}~,
\cr
G_{-+\bar{1}} &=& G_{234}~.
\eea
These components also vanish. To see this take ({\ref{algg2con5}}), which implies
that
\be
(G_{1 N_1 N_2}-G_{\bar{1} N_1 N_2}) \Gamma^{N_1 N_2} \eta_1=0~,
\ee
which in turn gives
\be
G_{1q}{}^q = G_{\bar{1} q}{}^q~,
\ee
and hence all components of $G$ vanish. Then, from the vanishing of
$X^5$, $X^6$, $X^7$ and $X^8$, it follows that $f^{-1} Df$ is constrained via
\bea
\label{g2fdfcon1}
(f^{-1} \partial_M f)_{11} &=&(f^{-1} \partial_M f)_{33}~,
\cr
(f^{-1} \partial_M f)_{22} &=& (f^{-1} \partial_M f)_{44}~,
\cr
(f^{-1} \partial_M f)_{13}+(f^{-1} \partial_M f)_{31} &=&0~,
\cr
(f^{-1} \partial_M f)_{24}+(f^{-1} \partial_M f)_{42} &=&0~,
\cr
(f^{-1} \partial_M f)_{12}&=&(f^{-1} \partial_M f)_{34}~,
\cr
(f^{-1} \partial_M f)_{14}+(f^{-1} \partial_M f)_{32}&=&0~,
\cr
(f^{-1} \partial_M f)_{21}&=&(f^{-1} \partial_M f)_{43}~,
\cr
(f^{-1} \partial_M f)_{23}+(f^{-1} \partial_M f)_{41}&=&0~.
\eea

Next consider the portions of  ({\ref{maxsuf}}) which depend on
the spin connection and the flux $F$. It will turn out to be most convenient
to consider first those components which do not have any dependence on $f$,
i.e. to exclude the $1$, $e_{1234}$, $e_{51}$ and $e_{5234}$ components.
Then from the $+$ component, we obtain
\bea
\label{simplermaxg25}
\Omega_{+, {\bar{q}}_1 {\bar{q}}_2}- \epsilon_{ {\bar{q}}_1 {\bar{q}}_2}{}^r \Omega_{+,1r} &=&0~,
\cr
2i F_{-+1 {\bar{q}}_1 {\bar{q}}_2}+  \epsilon_{ {\bar{q}}_1 {\bar{q}}_2}{}^r
(\Omega_{+,r-}-i F_{-+rs}{}^s+iF_{-+1 \bar{1}r}) &=&0~,
\cr
-2i F_{-+ \bar{1} q_1 q_2} + \epsilon_{q_1 q_2}{}^{\bar{r}}
(- \Omega_{+,\bar{r}-}-iF_{-+\bar{r}s}{}^s+iF_{-+1 \bar{1} \bar{r}}) &=&0~,
\cr
\Omega_{+,+p} = \Omega_{+,+\bar{p}} &=&0~,
\cr
\Omega_{+,\bar{p}1}-iF_{+1 \bar{p}r}{}^r- \epsilon_{\bar{p}}{}^{q_1 q_2}
({1 \over 2}\Omega_{+,q_1 q_2}+iF_{+ 1 \bar{1} q_1 q_2}) &=&0~,
\cr
-\Omega_{+,p\bar{1}}-iF_{+ \bar{1} pr}{}^r + \epsilon_p{}^{{\bar{q}}_1 {\bar{q}}_2}
({1 \over 2} \Omega_{+,{\bar{q}}_1 {\bar{q}}_2}-iF_{+ 1 \bar{1} {\bar{q}}_1
{\bar{q}}_2}) &=&0~.
\eea
The $-$ component implies that
\bea
\label{simplermaxg26}
-\Omega_{-,1r}+iF_{-1rs}{}^s +{1 \over 2} \epsilon_r{}^{{\bar{q}}_1 {\bar{q}}_2}
({1 \over 2}\Omega_{-,{\bar{q}}_1 {\bar{q}}_2}+iF_{-1 \bar{1} {\bar{q}}_1 {\bar{q}}_2}) &=&0~,
\cr
\Omega_{-,\bar{1} \bar{r}}+iF_{- \bar{1} \bar{r}s}{}^s + \epsilon_{\bar{r}}{}^{q_1 q_2}
(-{1 \over 2} \Omega_{-,q_1 q_2}+iF_{-1 \bar{1} q_1 q_2}) &=&0~,
\cr
\Omega_{-,-p}=\Omega_{-,- \bar{p}} &=&0~,
\cr
\Omega_{-,+\bar{p}}+iF_{-+\bar{p}q}{}^q +iF_{-+ 1 \bar{1} \bar{p}}-i
\epsilon_{\bar{p}}{}^{q_1 q_2}F_{-+1 q_1 q_2} &=&0~,
\cr
\Omega_{-,\bar{p}1}-{1 \over 2} \epsilon_{\bar{p}}{}^{q_1 q_2}\Omega_{-,q_1 q_2} &=&0~,
\cr
-\Omega_{-,+p}+iF_{-+ps}{}^s+iF_{-+1 \bar{1}p}+i\epsilon_p{}^{{\bar{q}}_1 {\bar{q}}_2}
F_{-+\bar{1} {\bar{q}}_1 {\bar{q}}_2} &=&0~,
\cr
\Omega_{-,p\bar{1}}-{1 \over 2} \epsilon_p{}^{{\bar{q}}_1 {\bar{q}}_2} \Omega_{-,{\bar{q}}_1
{\bar{q}}_2} &=&0~.
\eea
The $1$ component implies that
\bea
\label{simplermaxg27}
\Omega_{1, {\bar{q}}_1 {\bar{q}}_2}-\Omega_{1,1\ell}\epsilon^\ell{}_{{\bar{q}}_1 {\bar{q}}_2}
+2iF_{-+1 {\bar{q}}_1 {\bar{q}}_2} &=&0~,
\cr
\Omega_{1, \ell -}-iF_{-1\ell s}{}^s &=&0~,
\cr
\Omega_{1, \bar{1} \bar{p}}-{1 \over 2} \Omega_{1 q_1 q_2} \epsilon_{\bar{p}}{}^{q_1 q_2}
-iF_{-+\bar{p} q}{}^q +iF_{-+1 \bar{1} \bar{p}} &=&0~,
\cr
\Omega_{1,\bar{p}-}+iF_{-1 \bar{1} q_1 q_2}\epsilon^{q_1 q_2}{}_{\bar{p}} &=&0~,
\cr
\Omega_{1,+\bar{p}}-iF_{+1 \bar{p}q}{}^q &=&0~,
\cr
\Omega_{1,\bar{p}1}-{1 \over 2}\Omega_{1 q_1 q_2} \epsilon^{q_1 q_2}{}_{\bar{p}}
+iF_{-+1 q_1 q_2}  \epsilon^{q_1 q_2}{}_{\bar{p}} &=&0~,
\cr
\Omega_{1,+p}+iF_{+1 \bar{1} {\bar{q}}_1 {\bar{q}}_2} \epsilon^{ {\bar{q}}_1 {\bar{q}}_2}{}_p
&=&0~,
\cr
\Omega_{1,{\bar{q}}_1 {\bar{q}}_2}-\Omega_{1,r\bar{1}}\epsilon^r{}_{{\bar{q}}_1 {\bar{q}}_2}
-iF_{-+rs}{}^s \epsilon^r{}_{{\bar{q}}_1 {\bar{q}}_2}-iF_{-+1 \bar{1} r}
\epsilon^r{}_{{\bar{q}}_1 {\bar{q}}_2} &=&0~,
\eea
and the ${\bar{1}}$ component implies that
\bea
\label{simplermaxg28}
-\Omega_{\bar{1},1p}-iF_{-+ps}{}^s +i F_{-+1 \bar{1} p}+{1 \over 2}
\Omega_{\bar{1},{\bar{q}}_1 {\bar{q}}_2} \epsilon^{{\bar{q}}_1 {\bar{q}}_2}{}_p &=&0~,
\cr
\Omega_{\bar{1},p-}-iF_{-1 \bar{1} {\bar{q}}_1 {\bar{q}}_2} \epsilon^{{\bar{q}}_1 {\bar{q}}_2}{}_p
&=&0~,
\cr
-\Omega_{\bar{1},\bar{p} \bar{1}}-{1 \over 2} \Omega_{\bar{1},q_1 q_2}
\epsilon^{q_1 q_2}{}_{\bar{p}}-iF_{-+\bar{1}q_1 q_2} \epsilon^{q_1 q_2}{}_{\bar{p}} &=&0~,
\cr
\Omega_{\bar{1},\bar{p}-}+iF_{-\bar{1} \bar{p}s}{}^s &=&0~,
\cr
\Omega_{\bar{1},+\bar{p}}-iF_{+ 1 \bar{1}q_1 q_2} \epsilon^{q_1 q_2}{}_{\bar{p}} &=&0
\cr
-\Omega_{\bar{1}, q_1 q_2}+\Omega_{\bar{1} \bar{\ell} 1}
\epsilon^{\bar{\ell}}{}_{q_1 q_2}-iF_{-+{\bar{\ell}}s}{}^s \epsilon^{\bar{\ell}}{}_{q_1 q_2}
-iF_{-+1 \bar{1} \bar{\ell}} \epsilon^{\bar{\ell}}{}_{q_1 q_2} &=&0~,
\cr
\Omega_{\bar{1},+p}+iF_{+ \bar{1} ps}{}^s &=&0~,
\cr
-\Omega_{\bar{1},p \bar{1}}+{1 \over 2}\Omega_{\bar{1}, {\bar{q}}_1 {\bar{q}}_2}
\epsilon^{{\bar{q}}_1 {\bar{q}}_2}{}_p-iF_{-+\bar{1} {\bar{q}}_1 {\bar{q}}_2}
\epsilon^{{\bar{q}}_1 {\bar{q}}_2}{}_p &=&0~.
\eea

{}From the above equations, it is straightforward but tedious to show that
\bea
\label{maxg2Fvan1}
F_{-1 \bar{1} q_1 q_2} = F_{+1 \bar{1} q_1 q_2} &=&0~,
\cr
F_{-1pq}{}^q = F_{+1 \bar{p}q}{}^q &=&0~,
\cr
F_{-+pq}{}^q&=&0~,
\cr
F_{-+1 q_1 q_2} = F_{-+ 1 {\bar{q}}_1 {\bar{q}}_2} &=&0~,
\cr
F_{-+1 \bar{1} p}&=&0~.
\eea

Now consider the $p$ and $\bar{p}$ components (again setting aside the
 $1$, $e_{1234}$, $e_{51}$ and $e_{5234}$ components for the moment).
On using ({\ref{maxg2Fvan1}}) there is some simplification. In particular,
from the $p$ component, we find
\bea
\label{simplermaxg29}
\Omega_{p,  {\bar{q}}_1 {\bar{q}}_2}+2iF_{-+p {\bar{q}}_1 {\bar{q}}_2}-\Omega_{p,1 \ell}
\epsilon^\ell{}_{ {\bar{q}}_1 {\bar{q}}_2} &=&0~,
\cr
-2iF_{-1p  {\bar{q}}_1 {\bar{q}}_2}+\Omega_{p,\ell -} \epsilon^\ell{}_{ {\bar{q}}_1 {\bar{q}}_2}
&=&0~,
\cr
\Omega_{p, \bar{1} \bar{q}}+i F_{-+ \bar{1} r}{}^r g_{p \bar{q}}-2iF_{-+ \bar{1}p
\bar{q}}-{1 \over 2} \Omega_{p , r_1 r_2}{}\epsilon^{r_1 r_2}{}_{\bar{q}} &=&0~,
\cr
-\Omega_{p, \bar{q}-}-iF_{-r}{}^r{}_{1 \bar{1}}g_{p \bar{q}}+2iF_{-1 \bar{1} p \bar{1}} &=&0~,
\cr
\Omega_{p,+ \bar{q}}+iF_{+1 \bar{1} r}{}^r g_{p \bar{q}}-iF_{-1 \bar{1} p \bar{q}} &=&0~,
\cr
\Omega_{p, \bar{q} 1}-2iF_{-+1p \bar{q}}+iF_{-+1r}{}^r g_{p \bar{q}}-{1 \over 2}
\Omega_{p,r_1 r_2} \epsilon^{r_1 r_2}{}_{\bar{q}} &=&0~,
\cr
F_{+ \bar{1} p  {\bar{q}}_1 {\bar{q}}_2}-{1 \over 2} \Omega_{p,+r} \epsilon^r{}_{ {\bar{q}}_1 {\bar{q}}_2} &=& 0~,
\cr
\Omega_{p,  {\bar{q}}_1 {\bar{q}}_2}-2iF_{-+p  {\bar{q}}_1 {\bar{q}}_2}-
\Omega_{p,r \bar{1}} \epsilon^r{}_{ {\bar{q}}_1 {\bar{q}}_2} &=&0~,
\eea
and from the ${\bar{p}}$ component, we obtain

\bea
\label{simplermaxg2x}
-\Omega_{\bar{p},1q}+i F_{-+1r}{}^r g_{\bar{p} q}+2iF_{-+1 \bar{p} q}+{1 \over 2} \Omega_{\bar{p},
{\bar{r}}_1 {\bar{r}}_2} \epsilon^{{\bar{r}}_1 {\bar{r}}_2}{}_q &=&0~,
\cr
\Omega_{\bar{p},q-}+iF_{-r}{}^r{}_{1 \bar{1}} g_{\bar{p} q}+2iF_{-1 \bar{1} \bar{p} q} &=&0~,
\cr
\Omega_{\bar{p}, \bar{1} \bar{\ell}}\epsilon^{\bar{\ell}}{}_{q_1 q_2}-\Omega_{\bar{p},q_1 q_2}
-2iF_{-+\bar{p}q_1 q_2} &=&0~,
\cr
-\Omega_{\bar{p}, \bar{\ell}-} \epsilon^{\bar{\ell}}{}_{q_1 q_2}+2i F_{- \bar{1} \bar{p} q_1 q_2} &=&0~,
\cr
\Omega_{\bar{p},+ \bar{\ell}} \epsilon^{\bar{\ell}}{}_{q_1 q_2}-2i F_{+1 \bar{p} q_1 q_2} &=&0~,
\cr
{1 \over 2} \Omega_{\bar{p}, \bar{\ell} 1}\epsilon^{\bar{\ell}}{}_{q_1 q_2}
-{1 \over 2} \Omega_{\bar{p}, q_1 q_2}+iF_{-+\bar{p} q_1 q_2} &=&0~,
\cr
-{1 \over 2} \Omega_{\bar{p},+q}-{i \over 2}F_{+1 \bar{1} r}{}^r g_{\bar{p} q}-iF_{+1 \bar{1}\bar{p}q} &=&0~,
\cr
{1 \over 2}\Omega_{\bar{p}, {\bar{r}}_1 {\bar{r}}_2} \epsilon^{ {\bar{r}}_1 {\bar{r}}_2}{}_q
-\Omega_{\bar{p}, q \bar{1}}+2iF_{-+ \bar{1} \bar{p} q}+iF_{-+\bar{1}r}{}^r g_{\bar{p} q} &=&0~.
\eea

It is straightforward to show that ({\ref{simplermaxg29}}) and ({\ref{simplermaxg2x}}) imply that
{\it all components of $F$ vanish}, with the exception of
$F_{-+234}$, $F_{+ \bar{1} 234}$ and $F_{-1234}$ and their complex conjugates, together with
those components related to these terms by the duality of $F$. These components have yet to be constrained.

To proceed, note that the difference of the $1$ and $e_{1234}$, and the $e_{51}$ and $e_{5234}$
components in  the $F$-dependent portions of ({\ref{maxsuf}}) does not contain any $f^{-1} Df$ dependence.
Consider these differences which arise in the $1$ and ${\bar{1}}$ components of these equations.
{}From the $1$ component, we find
\bea
\label{maxg2aux1}
\Omega_{1,p}{}^p + \Omega_{1,1 \bar{1}}+2iF_{-+ \bar{2} \bar{3} \bar{4}} &=&0~,
\cr
\Omega_{1,1-}-\Omega_{1,\bar{1}-}-2iF_{-1234} &=&0~,
\cr
\Omega_{1,+\bar{1}}-\Omega_{1,+1}+2iF_{+1 \bar{2} \bar{3} \bar{4}} &=&0~,
\cr
\Omega_{1,p}{}^p - \Omega_{1,1 \bar{1}}-2iF_{-+234} &=&0~,
\eea
and from the $\bar{1}$ component, we obtain
\bea
\label{maxg2aux2}
\Omega_{\bar{1},p}{}^p +\Omega_{\bar{1},1\bar{1}}-2iF_{-+234} &=&0~,
\cr
\Omega_{\bar{1},1-}-\Omega_{\bar{1}, \bar{1}-}+2iF_{- \bar{1} \bar{2} \bar{3} \bar{4}} &=&0~,
\cr
\Omega_{\bar{1},+\bar{1}}-\Omega_{\bar{1},+1}-2iF_{+\bar{1}234} &=&0~,
\cr
\Omega_{\bar{1},p}{}^p - \Omega_{\bar{1},1 \bar{1}}+2iF_{-+ \bar{2} \bar{3} \bar{4}} &=&0~.
\eea
These equations are sufficient to imply that $F_{-+234}$, $F_{+ \bar{1} 234}$ and $F_{-1234}$
vanish. Hence {\it the $F$-flux also vanishes}. It is then straightforward to read off the
constraints on the spin connection obtained from the above reasoning, together with
the remaining $+$, $-$, $p$ and $\bar{p}$ components of the differences between the
$1$ and $e_{1234}$, and the $e_{51}$ and $e_{5234}$ components. This exhausts all of the
content of the $f^{-1} Df$-independent part of the Killing spinor equation.
In particular, we find
\bea
\label{maxg2spin1}
\Omega_{M,1q}-{1 \over 2} \Omega_{M, {\bar{r}}_1 {\bar{r}}_2} \epsilon^{{\bar{r}}_1 {\bar{r}}_2}{}_q &=&0~,
\cr
\Omega_{M,\bar{1}q}+{1 \over 2} \Omega_{M, {\bar{r}}_1 {\bar{r}}_2} \epsilon^{{\bar{r}}_1 {\bar{r}}_2}{}_q &=&0~,
\cr
\Omega_{M,+q} &=&0~,
\cr
\Omega_{M,-q} &=&0~,
\cr
\Omega_{M,q}{}^q &=&0~,
\cr
\Omega_{M,1 \bar{1}} &=&0~,
\cr
\Omega_{M,-1}-\Omega_{M,-\bar{1}} &=&0~,
\cr
\Omega_{M,+1}-\Omega_{M,+\bar{1}} &=&0~,
\eea
for $M=+,-,1,\bar{1},p,\bar{p}$. Lastly, we examine the $1$ and $e_{15}$ components of the
Killing spinor equation ({\ref{maxsuf}}) in order to obtain the full set of constraints on
$f^{-1} Df$. We find
\bea
\label{maxg2fdf1}
(f^{-1} \partial_M f)_{11} = (f^{-1} \partial_M f)_{33}=-(f^{-1} \partial_M f)_{22}=-(f^{-1} \partial_M f)_{44}~,
\cr
(f^{-1} \partial_M f)_{13}=-(f^{-1} \partial_M f)_{31}=(f^{-1} \partial_M f)_{24}=-(f^{-1} \partial_M f)_{42}~,
\cr
(f^{-1} \partial_M f)_{23}=(f^{-1} \partial_M f)_{32}=(f^{-1} \partial_M f)_{14}=(f^{-1} \partial_M f)_{41}=0~,
\eea
and
\bea
\label{maxg2fdf2}
(f^{-1} \partial_M f)_{12}=(f^{-1} \partial_M f)_{34}=\Omega_{M,+\bar{1}}~,
\cr
(f^{-1} \partial_M f)_{21}=(f^{-1} \partial_M f)_{43}=\Omega_{M,1-}~,
\eea
and
\bea
\label{maxg2fdf3}
(f^{-1} \partial_M f)_{11} +{1 \over 2} \Omega_{M,-+}&=&0~,
\cr
2 (f^{-1} \partial_M f)_{13}-Q_M &=&0
\eea
for $M=+,-,1,\bar{1},p,\bar{p}$.
The conditions on the fluxes and geometry are summarized in section six. Despite the complexity of analyzing
the Killing spinor equations, it turns out that the spacetime in this case is a product
of a three-dimensional Minkowski space and a $G_2$ holonomy manifold. This may have been expected
because the Killing spinor equations imply that the fluxes vanish.


\begin{thebibliography}{00}
\addcontentsline{toc}{section}{References} \frenchspacing \small
\addtolength{\itemsep}{-4pt}

\bi{jfgpa}
 J.~Figueroa-O'Farrill and
G.~Papadopoulos,
``Maximally supersymmetric solutions of ten-dimensional
and eleven-dimensional supergravities,''
JHEP {\bf 0303} (2003) 048:
[arXiv:hep-th/0211089].

``Pluecker-type relations for orthogonal planes,''
[arXiv:math.ag/0211170].

\bibitem{schwarz}
J.~H.~Schwarz, ``Covariant Field Equations Of Chiral N=2 D = 10
Supergravity,'' Nucl.\ Phys.\ B {\bf 226} (1983) 269.

\bi{georgea}
M.~Blau, J.~Figueroa-O'Farrill, C.~Hull and G.~Papadopoulos,
``A new maximally supersymmetric background of IIB superstring theory,''
JHEP {\bf 0201} (2002) 047
[arXiv:hep-th/0110242].

``Penrose limits and maximal supersymmetry,''
Class.\ Quant.\ Grav.\  {\bf 19} (2002) L87
[arXiv:hep-th/0201081].

\bibitem{gju}
U.~Gran, J.~Gutowski and G.~Papadopoulos,
``The spinorial geometry of supersymmetric IIB backgrounds,''
arXiv:hep-th/0501177.

\bibitem{uggp}
J.~Gillard, U.~Gran and G.~Papadopoulos, ``The spinorial geometry of
supersymmetric backgrounds,'' arXiv:hep-th/0410155.

\bibitem{jones}
E.~J.~Hackett-Jones and D.~J.~Smith,
``Type IIB Killing spinors and calibrations,''
JHEP {\bf 0411} (2004) 029
[arXiv:hep-th/0405098].


\bibitem{grayb}
M. Fernandez, A. Gray, ``Riemannian manifolds with structure group $G_2$'', Ann. Mat. Pura Appl(4) 32 (1982), 19-45. 1.


\bi{ivanov} T. Friedrich and S. Ivanov, ``Killing spinor equations in dimension 7 and geometry of intergrable
$G_2$-manifolds,''
J. Geom. Phys. {\bf 48}  (2003), 1-11,
[math.DG/0112201].



\bi{salamonb} S. Chiossi and S. Salamon, ``The intrinsic torsion of $SU(3)$
 and $G_2$ structures'', Diff. Geom., Valencia 2001, World Sci. 2002, 115
 [arXiv:math.DG/0202282].

 \bibitem{penrose} R. Penrose, ``Any spacetime has a plane wave as a limit,''
Differential Geometry and Relativity, 271-275, Riedel Dordrecht,
(1976).



\bibitem{west}
J.~H.~Schwarz and P.~C.~West,
``Symmetries And Transformations Of Chiral N=2 D = 10 Supergravity,''
Phys.\ Lett.\ B {\bf 126} (1983) 301.

\bibitem{becker}
K.~Becker and L.~S.~Tseng, ``A note on fluxes in six-dimensional
string theory backgrounds,'' arXiv:hep-th/0410283.


\bibitem{howe}
P.~S.~Howe and P.~C.~West, ``The Complete N=2, D = 10
Supergravity,'' Nucl.\ Phys.\ B {\bf 238} (1984) 181.


\bi{gray}
A. Gray and L.M. Hervella, ``The sixteen classes of
 almost Hermitian manifolds and their linear invariants'',
 Ann.\ Mat.\ Pura\ e \ Appl.\ {\bf 282} (1980) 1.



\bibitem{amus}
 A.~Moroianu and U.~Semmelmann, ``Parallel spinors and
holonomy group,'' math.DG/9903062.

\bibitem{mcinnes} B. McInnes, {\em Spin holonomy of Einstein manifolds},
Commun. Math. Phys. {\bf 203} (1999) 349-364.


\bibitem{blau}
M.~Blau, J.~Figueroa-O'Farrill and G.~Papadopoulos, ``Penrose
limits, supergravity and brane dynamics,'' Class.\ Quant.\ Grav.\
{\bf 19} (2002) 4753 [arXiv:hep-th/0202111].

 M.~Blau, M.~Borunda, M.~O'Loughlin and G.~Papadopoulos,
  ``Penrose limits and spacetime singularities,''
  Class.\ Quant.\ Grav.\  {\bf 21} (2004) L43
  [arXiv:hep-th/0312029].

\bibitem{patricot}
  C.~Patricot,
  ``Kaigorodov spaces and their Penrose limits,''
  Class.\ Quant.\ Grav.\  {\bf 20} (2003) 2087
  [arXiv:hep-th/0302073].


\bibitem{klebanov}
  I.~R.~Klebanov and M.~J.~Strassler,
  ``Supergravity and a confining gauge theory: Duality cascades and
  chiSB-resolution of naked singularities,''
  JHEP {\bf 0008} (2000) 052
  [arXiv:hep-th/0007191].

\bibitem{volkov}
  A.~H.~Chamseddine and M.~S.~Volkov,
  ``Non-Abelian BPS monopoles in N = 4 gauged supergravity,''
  Phys.\ Rev.\ Lett.\  {\bf 79} (1997) 3343
  [arXiv:hep-th/9707176].

  \bibitem{nunez}
  J.~M.~Maldacena and C.~Nunez,
  ``Towards the large N limit of pure N = 1 super Yang Mills,''
  Phys.\ Rev.\ Lett.\  {\bf 86} (2001) 588
  [arXiv:hep-th/0008001].

\bibitem{aatgp}
  G.~Papadopoulos and A.~A.~Tseytlin,
  ``Complex geometry of conifolds and 5-brane wrapped on 2-sphere,''
  Class.\ Quant.\ Grav.\  {\bf 18} (2001) 1333
  [arXiv:hep-th/0012034].

\bibitem{minasian}
M.~Grana, R.~Minasian, M.~Petrini and A.~Tomasiello,
``Supersymmetric backgrounds from generalized Calabi-Yau
manifolds,'' JHEP {\bf 0408} (2004) 046 [arXiv:hep-th/0406137].

A.~Butti, M.~Grana, R.~Minasian, M.~Petrini and A.~Zaffaroni, ``The
baryonic branch of Klebanov-Strassler solution: A supersymmetric
family of SU(3) structure backgrounds,'' arXiv:hep-th/0412187.

\bibitem{cvetic}
  K.~Behrndt, M.~Cvetic and P.~Gao,
  ``General type IIB fluxes with SU(3) structures,''
  arXiv:hep-th/0502154.


\bibitem{warner}
  C.~N.~Gowdigere and N.~P.~Warner,
  ``Holographic Coulomb Branch Flows with N=1 Supersymmetry,''
  arXiv:hep-th/0505019.





\bibitem{wang} McKenzie Y. Wang,
 ``Parallel spinors and parallel forms'', Ann. Global Anal Geom.
Vol 7, No 1 (1989), 59.




\bibitem{lawson} H. Blaine Lawson and Marie-Louise Michelsohn, ``Spin geometry,'' Princeton
University Press (1989).


\bi{harvey} F. R. Harvey, ``Spinors and Calibrations,'' Academic
Press, London (1990).





\end{thebibliography}
\end{document}